\documentclass{article}

 \usepackage{hyperref}
\usepackage{graphicx}
\usepackage{bm}
\usepackage{mathtools}
\usepackage{subcaption}
\usepackage{amsfonts}
\usepackage{epstopdf} 

\begin{document}


\title{Robust probabilistic modeling of photoplethysmography signals with application to the classification of premature beats}

\author{M.\ Regis\textsuperscript{1,2}, L.M.\ Eerik\"{a}inen\textsuperscript{1,2}, R.\ Haakma\textsuperscript{2}, E.R.\ van den Heuvel\textsuperscript{1}, P.\ Serra\textsuperscript{1}}

\maketitle

\begin{abstract}
In this paper we propose a robust approach to model photoplethysmography (PPG) signals.
After decomposing the signal into two components, we focus the analysis on the pulsatile part, related to cardiac information. 
The goal is to enable a deeper understanding of the information contained in the pulse shape, together with that derived from the rhythm. 
Our approach combines functional data analysis with a state space representation and guarantees fitting robustness and flexibility on stationary signals, without imposing a priori information on the waveform and heart rhythm.
With a Bayesian approach, we learn the distribution of the parameters, used for understanding and monitoring PPG signals. The model can be used for data compression, for inferring medical parameters and to understand condition-related waveform characteristics. In particular, we detail a procedure for the detection of premature contractions based on the residuals of the fit. 
This method can handle both atrial and ventricular premature contractions, and classify the type by only using information from the model fit.
\end{abstract}

{\itshape PPG Signals, Premature Contraction Detection, Functional Data Analysis, Kalman Filter, Signal Synthesis, Spline Model}

\vfill
\vspace{1cm}
{\textsuperscript{1}Eindhoven University of Technology, Eindhoven, The Netherlands\\
	\textsuperscript{2}Philips Research, Eindhoven, The Netherlands}

\newpage5

\section{Introduction}
Photoplethysmography (PPG) is a light-based technology to measure relative peripheral blood volume variations~\cite{Allen2007}.
Nowadays PPG is commonly used and integrated in devices at different body locations (e.g.~wrist, finger, ear), since it can be used to derive physiological information that is normally measured via other measurement modalities, such as electrocardiogram (ECG).
Although this costs a loss of accuracy in the estimated values when compared to the same medical parameters, e.g.~heart rate from ECG, it comes with the advantage of unobtrusiveness and low costs.
Furthermore, it can provide a broader image on the health status, including for example information on the cardiovascular, respiratory and autonomic systems~\cite{Shelley2007, Allen2007, Peper2007, Elgendi2012, Charlton2018}.
Since it is nowadays well accepted that early diagnosis and treatment of threatening conditions can make a large difference both in terms of individuals' quality of life and public health costs, monitoring unobtrusively and on a large scale the population for cardiovascular problems at relatively small costs has become a priority in modern medicine.
This has motivated and boosted research on PPG signals to detect anomalies, monitor, and classify categories of patients.
However, the richness of the information combined with signal disruption, due to device misplacement and movement for example, results in an extremely complex signal.
In fact, a deep understanding of the PPG waveform is still a challenging open problem~\cite{Elgendi2012}.

Multiple attempts have been made to model the signal to disclose the underlying physiology~\cite{Shariati2005,Rubins2008,Couceiro2012, Wang2013, Martin2013, Baruch2014, Martin2015, Luo2016, He2017}. 
Shariati and Zahedi~\cite{Shariati2005} proposed a linear parametric model, and tested the fit with different choices for the correlation structure.
The conclusion among the tested possibilities was in favour of the autoregressive structure.
A more popular approach consists of decomposing the signal into a variable number of (log)normal components~\cite{Rubins2008,Couceiro2012, Wang2013, Martin2013, Baruch2014, Martin2015, Luo2016, He2017}. This method is motivated by the underlying physiology since each component has a physical explanation~\cite{Elgendi2012a}.
The major limitation of this approach is the fixed structure of the pulses, that requires a priori knowledge and reduces model robustness.
Most of the approaches listed above are in fact suited for modelling regular sinus rhythm.
A recent publication explores using a mixture of Gaussians and lognormals for model fitting and data simulation under atrial fibrillation~\cite{Solovsenko2017}. Also in this case, each patient is assigned a certain fixed pulse shape among five possible choices prior fitting.
This rigidity leads to the failure of fitting procedures every time the new encountered shape does not comply with the expected waveform, requiring the data to have limited within-subject variability, and the method necessitates some initial analyses to distinguish among the possible categories.
Also the fixed number of categories may not capture all the possible behaviours of the data and variability between individuals.
To overcome this limitation, curve registration has been proposed~\cite{Martin2013}. This way variability across pulses is extracted from the data, allowing separate analyses, and Gaussian mixture models can be fitted. 
In this publication then, the temporal evolution is captured by imposing an autoregressive structure on the fitted parameters.

We further enhance modelling flexibility, by combining curve registration with a spline model for the registered curves.
Since our focus is on the pulsatile component dependent on heart rate, we first exclude the low-frequency component~\cite{Peper2007}. This part contains information about the respiratory activity, and can be further modelled and analysed with the approaches introduced for extracting respiratory information~\cite{Addison2017, Mason2013, Karlen2013,Charlton2018}.
Then we segment the resulting signal component into pulses based on the detected local minima, define a functional data object for each pulse and model them iteratively.
We employ a state space representation to directly model the relationship between parameters, avoiding a second step to model the time evolution.
Estimation is performed with an extended Kalman filter.
The approach is flexible and is shown to fit data measured with different probes (i.e.~finger and wrist) and under various heart rhythms (e.g.~atrial fibrillation, presence of premature contractions).

Our main contribution to the understanding of the signal is on the waveform characteristics: instead of imposing a certain pulse shape that remains constant in time, we model it as random, and learn the posteriori distribution of the fitted parameters.
This enables multiple analyses on the health status such as heart rhythm disorders. Direct applications include model-based compression and summary of patient properties into a reduced set of parameters. Furthermore, clinical uses are foreseen by classifying the fitted parameters.
We detail the procedure for the detection of premature contractions.

The paper is organized as follows: in Section~\ref{sec:Data} we introduce the motivating example and dataset. Then in Section~\ref{sec:Methods} we illustrate the modelling procedure, detailing each step from the data processing until model fitting.
In the following sections we foresee possible clinical applications of the newly introduced model (Section~\ref{sec:Applications}), and show the results both from the fit and from the applications (Section~\ref{sec:Results}).
We end the paper with a short discussion on the findings and an outlook on possible future extensions and applications of the method presented here (in Section~\ref{sec:PPGDiscussion} and Section~\ref{sec:FutureWork} respectively).

\section{Data}\label{sec:Data}
In this section, we describe the motivating example and protocol defined for data collection.

The data motivating our study was originally collected to develop algorithms to detect atrial fibrillation from PPG.
Atrial fibrillation (AF) is an arrhythmia with 2\% incidence on the world population, with possible complications leading to stroke and heart failure among the others. Since it is so widely diffuse and often asymptomatic, it has motivated a wide spectrum of researches, trying to early detect this rhythm disorder via unobtrusive devices. 
Accurate detection of arrhythmias via photoplethysmography signals would in fact empower monitoring on a large scale, for long periods, unobtrusively.

The data was collected from 30 patients (63\% men, age 65 $\pm$ 14 years) assigned to a 24 hour Holter electrocardiography (ECG) examination at Catharina Hospital in Eindhoven, The Netherlands.
The study was approved by the local medical ethical committee and every patient provided written informed consent before participating.
The study protocol was the following: volunteers registered in the morning at the cardiology department of the hospital to have a 12-lead Holter monitor installed (H12+, Mortara, Milwaukee, WI, USA).
In parallel, patients received a wrist-wearable data logging device equipped with Philips Cardio and Motion Monitoring Module (CM3 Generation-3, Wearable Sensing Technologies, Philips, Eindhoven). The device was positioned on the non-dominant arm and recorded PPG and acceleration data. Sampling frequency of the raw data was $f_s = 128$ Hz and the dynamic range of the accelerometer was  $\pm$ 8 g.
The Holter monitor was attached to the patient following normal hospital procedures, and then a synchronization protocol was performed by simultaneously tapping the wrist-wearable device and pressing the event button on the Holter.
Patients were also asked to keep a diary of the taken medications, their activities and complaints for the whole monitoring period.
The protocol ended after 24 hours, when the patient would arrive to the hospital to return the Holter monitoring device, the diary and the wrist-wearable device. At that moment, the same synchronization procedure of the first day was repeated to allow time alignment of the collected data.
Patient information was obtained from medical records.
The ECG recordings were analyzed by trained analysts, supported by a software (Veritas, Mortara, Milwaukee, WI, USA) that automatically detects the time and type of the contraction. Every heart beat in the ECG was labelled either as sinus rhythm, atrial fibrillation (AF), premature supraventricular or ventricular contraction, artifact, or unknown. The output of the software was verified or corrected by analysts. In the 24-hour monitoring period, patients had either 100\% atrial fibrillation (8 participants, 26.67\%) or no atrial fibrillation (22 participants, 73.33\%). 
The raw PPG signal obtained from the wrist-worn device was pre with a 0.3 Hz high-pass filter and a 5 Hz low-pass filter, and downsampled to a frequency of 64 Hz.

The data of five patients has been excluded from the study for poor signal quality.
For each of the remaining patients, we selected a segment six minutes long from the two hours following the sleep onset.
Since the main goal of our research was modelling the physiology, we selected data under stable conditions, i.e.~segments of data from the night time with the least movement possible. This does not necessarily exclude position changes and device displacements.

\section{Methods}\label{sec:Methods}
\begin{figure}[h!]\centering
	\includegraphics[width=0.8\linewidth]{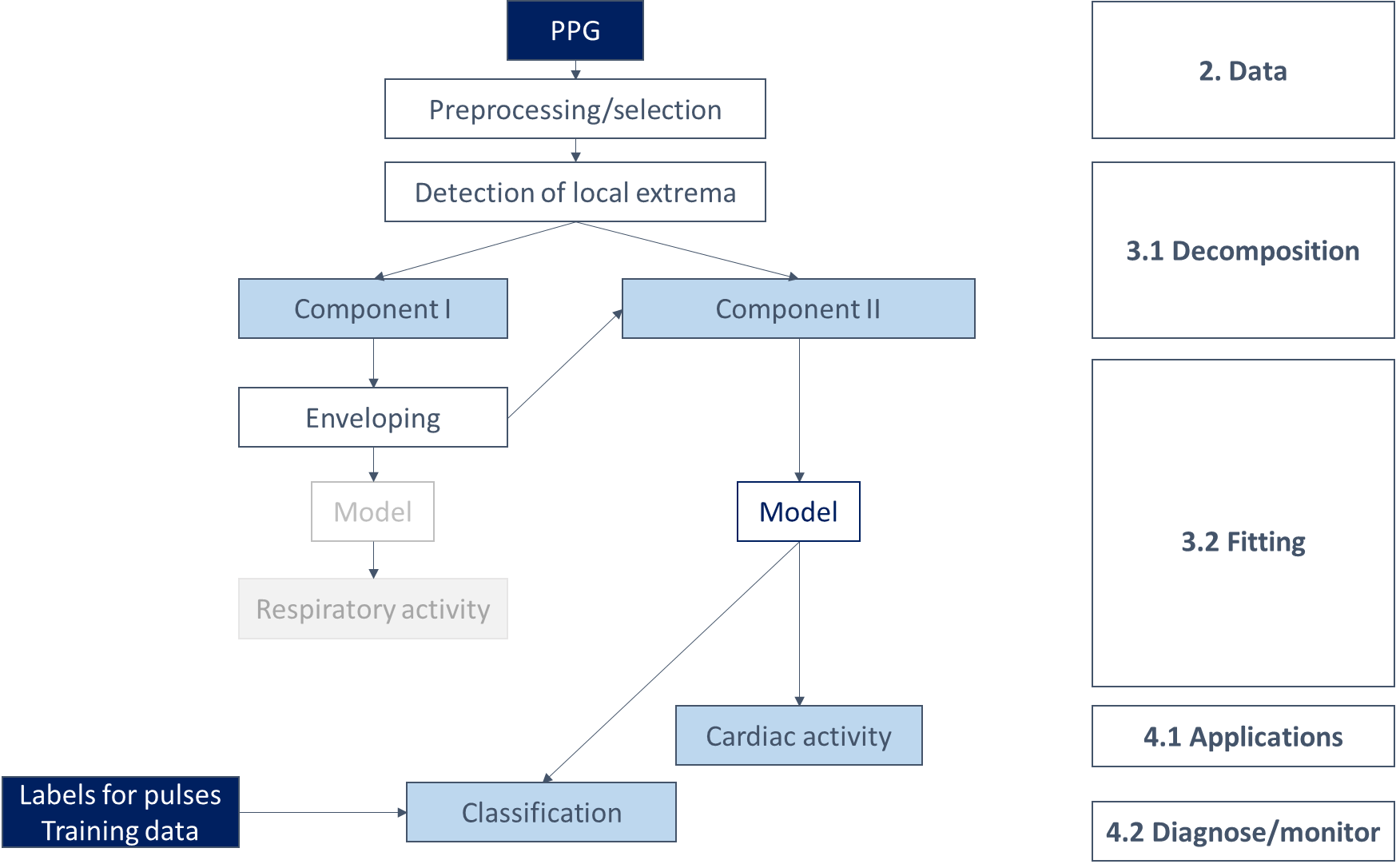}
	\caption{{\bfseries Flowchart of the algorithm.}
		Main steps in the algorithm presented sequentially and grouped into phases at the right hand side, where the numbers redirect to the corresponding section.
		Colour coding: blue = input; light blue = (intermediate) output; white = steps of the algorithm; grey = steps mentioned from literature. 		
	}
	\label{fig:flowchart}
\end{figure}

In this section we detail the modelling procedure, shown schematically in Figure~\ref{fig:flowchart}.

The boxes on the right hand side group the steps of the algorithm into main phases, and the number therein redirect to the corresponding section with the details on the implemented procedure.

The data motivating the study and used in the present work is described in Section~\ref{sec:Data}, together with the corresponding preprocessing procedure.
After that appropriate segments of signals are selected and processed, they are decomposed into two separate periodic components (Section~\ref{sec:decomposition}). Component I is the low-frequency component, captured by the local critical points. Component II shows pulsatile frequency, and is the original signal corrected for the envelope of the low-frequency component. 
The paper focuses on an in-depth analysis of Component II in Section~\ref{sec:fitting}. Component II undergoes first a transformation step, which includes segmentation into singular pulses and a check on abnormalities: pulses which are completely flat or with an outlying small length are excluded since for them no physiological explanation could be found. 
Then we scale each regular pulse to have unit amplitude, define a functional object and perform curve registration. In the current implementation, we show that using piecewise linear functions both for interpolation and for the time transformation already provides good fit, but these functions can be easily replaced by more complex transformations.
We propose a state space representation to model the evolution of the shape of the aligned and scaled PPG pulses. Since data collected under stable conditions does not have much residual variability after curve registration, we model the pulses as being constant on average and put a prior distribution on them.\\

In the following, $Z$ denotes the time series of observations and $[0,T]$ the time interval on which they were taken. 
The pair $(R_i,Z_i)$ represents a measurement $Z_i$ taken at time $R_i\in [0,T]$, where $R_i= (i-1)\Delta$ and $\Delta=1/{f_s}$ is the constant sampling period. 
Note that if the sampling frequency were not constant, $R_i$ would be more generally a sequence increasing in $i$.

\subsection{Decomposition}\label{sec:decomposition}

\paragraph{Detection of local extrema}\label{sec:decomposition:detection}
For detecting local extrema, we implement a greedy approach comparing the $i^{th}$ value $Z_i$ with the values in the neighborhood of radius $\rho$ centered in $R_i$. If it is larger (smaller) than or equal to all the considered points, then it is flagged as a maximum (minimum).
The critical points are then post-processed to keep only the first and the last of sequences of identical maxima (minima, respectively).

We denote the local minima by $\mathcal{L} = \{(T_i, L_i),\ i = 0, \dots, N\}$, with $N$ the total number of pulses, and assume that the first and last observations are local minima, i.e.~$(0,L_0)$ and $(T,L_N)$. Furthermore, with the term pulse we refer to the collection of observations between two consecutive local minima $T_{i-1}$ and $T_i$. These samples are denoted by $\bm V_i=(V_{i,1}, \dots, V_{i,n_i})$, and the corresponding observation times by $\bm R_i=(R_{i,1}, \dots, R_{i,n_i})$.
The length of the $i^{th}$ vector is $n_i$, with $n = \sum_{i=1}^N n_i$ the total number of observations by time $T$.
The duration of each pulse is denoted by $t_i=T_i-T_{i-1}$, with $\sum_{i=1}^N t_i = T$ and $T_N = T$. In case of constant sampling frequency, $T_i = n_i \Delta$.
Finally, the maximum in each pulse is denoted by $(S_i, U_i)$, with $\mathcal{U}=\{(S_i, U_i),\ i=1,\dots,N\}$.

\paragraph{Component I and Component II}\label{sec:components}
Component I consists of the union of $\mathcal{L}$ and $\mathcal{U}$, as shown in Figure~\ref{fig:componentI}.
Component II consists of the original signal normalized by the envelope of the local extrema (Figure~\ref{fig:componentII}).
To obtain the second component, we first subtract the piecewise linear interpolation of the local minima $\tilde{L}(t)$ to the original signal.
More precisely, $\tilde{L}(t): [0, T] \mapsto \mathbb{R}$ is the interpolating function connecting the \emph{extended} set of minima obtained by adding points iteratively
$$\text{if}\ \exists i\in 1,\dots,N,\ t\in[0,T]\ \text{s.t.}\ Z_i < \tilde{L}(t),\ \text{then}\ \tilde{\mathcal{L}} =  \tilde{\mathcal{L}} \cup (R_i, Z_i),$$
where initially $\tilde{\mathcal{L}}=\mathcal{L}$ and $\tilde{L}(t)=L(t)$ is the analogous function interpolating the initial set of minima $\mathcal{L}$.
This procedure is repeated until $Z_i \geq\tilde{ L}(t),\ \forall i,t$.
This way, $\tilde{L}(t)$ fully wraps the signal from below and
$$\tilde{V}_{i,j}=V_{i,j} - \tilde{L}(R_{i,j})\geq 0,\ \forall i=1,\dots,N,\ j=1,\dots,n_i.$$

Then each pulse with $\text{length}(\tilde{\bm V}_i)< l_0$ or amplitude $\tilde{U}_i=U_i - \tilde{L}(S_{i})=0$ is stored as a vector of missing values of matching length (both threshold can be set to exclude no pulse). The $N_k$ pulses $\tilde{\bm V}_i$ that pass these exclusion criteria are then normalized dividing by the shifted local maximum $\tilde{U}_i$,
$$\hat{V}_{i,j}=\frac{\tilde{V}_{i,j}}{\tilde{U}_i},\ i=1,\dots,N_k,\ j=1,\dots,n_i.$$
Component II is then segmented into pulses $\hat{\bm V}_{i},\ i=1,\dots,N_k$ and these are overlapped to have the same first observation as shown in Figure~\ref{fig:segmentation}.

\subsection{Model for Component II}\label{sec:fitting}
\paragraph{Curve registration}
We denote with $P_i(t)$ the functional version of $\hat{\bm V}_{i}$.
In the current implementation we perform linear interpolation such that $\hat{V}_{i,j} = P_i(R_{i,j})$, $i=1,\dots,N_k$, $j=1,\dots,n_i$, and evaluate the functions at the same $r$ equally spaced points in $[0,1]$, as shown in the three examples in Figure~\ref{fig:componentII_aligned}.
Then we perform curve registration, i.e.~we align pulse characteristics by composing the original curve with monotone functions of the independent argument (hereby referred to as warping functions).
The pulses can first be aligned to a target pulse, such as
\begin{equation}
M_i(t) = \text{median} (P_1(t), \dots, P_{i-1}(t))
\end{equation}
by composing the curves $P_i(t)$ with the corresponding warping functions
$$g_i(t):[0,1]\to [0,t_i],\ g_i(t) = arg\,min ||{M_i(t) - Q_i(t)}||_2 ,$$
where $Q_i(t) = P_i(g_i(t))$ and the difference is minimized with respect to the $L^2$ norm.
In the current implementation $g_i = \mathcal{I}\ \forall i$, where $\mathcal{I}(x)=x$ denotes the identity function.

Then we compose the aligned curves $Q_i(t)$ with piecewise linear warping functions $h_i(t):[0,1]\to [0,1]$ that interpolate the points $(0,0)$, $(m_i, {S}_i)$, and $(1,1)$ (shown in Figure~\ref{fig:timeTransf}). In particular, we choose
\begin{equation}
m_i = \text{median} ({S}_1, \dots, {S}_{i-1}).
\end{equation}
From this transformation the maximum of each pulse results aligned to $m_i$, as shown in Figure~\ref{fig:curveReg}. 
For each pulse we can thus define the composed warping function $w_i(t) = (h_i \circ g_i)(t)$ and denote the corresponding normalized, aligned functional object as $W_i(t) = P_i(w_i(t))$.

\paragraph{State space model}
Let $\bm X_i\in\mathbb{R}^p$, $p\in\mathbb{N}$ be state vectors and $\bm Y_i = (Y_{i,1}, \dots, Y_{i,r})$  the observation vectors with $Y_{i,j}=W_i((j-1)/r)$, $i=1,\dots,N_k$, $j=1,\dots,r$.
Consider also appropriate
design matrices $\bm H_i\in\mathbb{R}^{r\times p}$, $i=1,\dots,N_k$,
a Toeplitz matrix $\bm \Sigma\in\mathbb{R}^{p\times p}$, and
$\sigma_\epsilon^2, \sigma_\xi^2, \sigma^2 \ge 0$.
We model the observations $\bm Y_i$ as
\begin{equation}\label{eq:state_space_model}
\begin{aligned}
\bm Y_i &= \bm H_i \bm X_i + \bm\epsilon_i, \qquad
\bm\epsilon_i \sim N(\bm 0, \sigma^2\bm I_r + \sigma_\epsilon^2 \bm H_i \bm\Sigma\bm H_i^T),\\
\bm X_i &= \bm X_{i-1} + \bm\xi_i, \qquad
\bm\xi_i \sim N(\bm 0, \sigma_\xi^2\bm I_p),
\end{aligned}
\end{equation}
where the sequences $\bm\epsilon_i$ and $\bm\xi_i$ are mutually independent, $i = 1,\dots,N_k$.

The top equation in~\eqref{eq:state_space_model} models each pulse $\bm Y_i$ as a function of the state vector $\bm X_i$.
In fact we assume that $\bm Y = \bm H \bm\theta + \sigma\bm\eta$, with $\bm\theta = \bm X + \sigma_\epsilon\bm\zeta$, where $\bm\eta\sim N(\bm 0, \bm I_r)$ and $\bm\zeta\sim N(\bm 0, \bm\Sigma)$ are independent.
In other words, we assume that our observations are of the form $\bm H \bm\theta$, where the parameters $\bm\theta$ have mean $\bm X$.
The variance $\sigma^2$ represents the noise level of the observations $\bm Y_i$, $\sigma_\epsilon^2$ models the magnitude of the variability of the parameter $\bm\theta$ modelling the pulse shape, and $\bm\Sigma$ models dependencies between the coordinates of $\bm\theta$.
The bottom equation in~\eqref{eq:state_space_model} models the evolution of the means of the shape parameters as being constant in time.
Note that our modelling assumption still allows for variability around this mean.

Here we assume the entries of the design matrices $\bm H_i$ to be
\begin{equation}
\{\bm H_i\}_{k,j} = N_{j,d+1}\left(\frac{R_{i,k}-R_{i,1}}{t_i}\right),
\qquad k=1,\dots,r, \; j = 1, \dots, p,
\end{equation}
where $N_{1,d+1}, \dots, N_{p,d+1}$ form a B-spline basis of degree $d$ on $[0,1]$, with an appropriate number of knots.

\paragraph{Estimation}
The state space model described above includes parameters that have to be estimated from the data, namely the state vectors $\bm X_i$, the Toeplitz matrix $\bm\Sigma$, and the variances $\sigma_\epsilon^2$, $\sigma_\xi^2$, and $\sigma^2$.
These are estimated sequentially.

The Kalman filter equations can be used to estimate the states $\bm X_i$ of the system.
We initialize $\bm X_0 = \bm 0$, $\bm \Gamma_0 = \bm I_p$, $\hat{\bm\Sigma}_0 = \bm I_p$, $\hat\sigma_{\epsilon,0}^2 = 0.1$, $\hat\sigma_{\xi,0}^2 = 0.1$, and $\hat\sigma^2_0 = 0.1$, and iterate, for $i=1, \dots, N_k$, computing
\begin{equation}\label{eq:KF_equations}
\begin{aligned}
\bm \Theta_i &=
\hat\sigma^2_{i-1} \bm I_r + \bm H_i \left( \hat\sigma_{\epsilon,i-1}^2\hat{\bm\Sigma} + \hat\sigma^2_{\xi,i-1} \bm I_p + \bm\Gamma_{i-1} \right) \bm H_i^T,\\
\bm\Xi_i &=
\left( \bm\Gamma_{i-1} + \hat\sigma_{\xi,i-1}^2 \bm I_p \right) \bm H_i \bm \Theta_i^{-1},\\
\hat{\bm X}_i &=
\hat{\bm X}_{i-1} - \bm\Xi_i \left( \bm Y_i - \bm H_i \hat{\bm X}_{i-1} \right),\\
\bm\Gamma_i &=
\left(\bm I_p - \bm\Xi_i \bm H_i\right)\left( \bm\Gamma_{i-1} + \hat\sigma_{\xi,i-1}\bm I_p \right),
\end{aligned}
\end{equation}
followed by estimating $\hat{\bm\Sigma}_i$, $\hat\sigma^2_i$, $\hat\sigma_{\epsilon,i}^2$, and $\hat\sigma_{\xi,i}^2$, as minimizers of
\begin{equation}\label{eq:liielihood_KF}
\sum_{k=\max(1,i-s)}^i \left\{
\left(\bm Y_k - \bm H_k\hat{\bm X}_{k-1}\right)^T \bm\Theta_k^{-1} \left(\bm Y_k - \bm H_k\hat{\bm X}_{k-1}\right) + \log|\bm\Theta_k| 
\right\},
\end{equation}
for some $s\in\mathbb{N}$.
The expression in the previous display is, up to an additive constant, $-2$ times the log-likelihood of the most recent $s$ observations from the model.
Note that although the two terms $\sigma^2_{\xi}$ and $\sigma^2_{\epsilon}$ in $\bm \Theta_i$ are written separately, they are not identifiable. However, for our purpose it is sufficient to estimate their sum in order to account for the appropriate variability.

\section{Applications}\label{sec:Applications}
The model described in Section~\ref{sec:Methods} is motivated by and directed towards clinical applications.
Data compression is a direct consequence of model fitting, that reduces the data load as much as desired. The method can thus be used to reduce the memory load and increase the processing speed of various algorithms.
Also the distribution of the fitted parameters is directly obtained from the procedure, enabling analyses and inference. Examples include studying the covariance of the shape parameters.
In this section we discuss other possible uses of the output produced by the model, and detail a procedure for the detection of premature contractions.

\subsection{Inference}
The shape parameters from the model fit summarize the waveform as accurately as desired. 
An interesting application is expressing medical parameters as function of the fitted values.
Inter-beat-intervals (IBIs) and pulse amplitude are only few examples of the signal characteristics that can be expressed as functional of the shape parameters. Furthermore, new features can be derived beside the ones that are traditionally used in computational PPG analysis.
We illustrate the accuracy in deriving IBIs from our algorithm in Section~\ref{res:IBIs}.

\subsection{Detection of premature contractions}
The method introduced in Section~\ref{sec:Methods} relies on the assumption of underlying stationary process. 
As a consequence, disturbances and deviations from stationarity affect the magnitude of residuals. This characteristic can be used to diagnose and monitor patients.
We illustrate a procedure for the detection of premature contractions, both atrial and ventricular. In fact, the dataset was collected to study atrial fibrillation, but multiple arrhythmias are present in the data, and can easily be confused with atrial fibrillation.

The literature on the detection of rhythm anomalies in the PPG signal is broad.
Suzuki et al.~\cite{Suzuki2009} proposed a method for the detection of irregular pulses, irrespective of the cause, based on the changes in the pulse-to-pulse interval and the ratio pulse amplitude/pulse-to-pulse interval.
This method achieved 91.5\% sensitivity and 99.0\% specificity on data from three subjects during night, but did not allow classification of specific anomalies.
Specific methods for the detection of premature ventricular contractions have been proposed, in which pulse features are extracted and classified with various algorithms. Linear discriminant analysis classifier ($90.5\%$ sensitivity and 99.9 \% specificity, even in case of four-class classification based on the shape of the following pulse)~\cite{Gil2013}, artificial neural network (92.4\%-93.2\% sensitivity and 99.9\% specificity, also in case of bigemini)~\cite{Solovsenko2015} and k-nearest classification (90.4\% specificity)~\cite{Yousefi2018} are just a few examples.
These methods are all suited for the detection of premature ventricular contractions.
An approach for a more general detection of (sequences) of premature contractions exist~\cite{Polania2015}. The method, however, is segment-based and provides the classification for a time window of certain length (e.g.~15 seconds). This would cause problems in case different premature contractions happened in the same window.

Furthermore, all the approaches proposed so far focus on the extraction of features from each pulse and rely thus on some method detecting the pulses. In fact, signal segmentation in case of arrhythmias may not be accurate, as many authors report. This also makes difficult the detection of different types of premature contractions.
Recently more attention has been paid to the shape characteristics of the PPG signal~\cite{Papini2018}. In the study the focus was on developing a template matching that is robust to multiple types of arrhythmias, and led to detect premature contractions, irrespectively of the type.

We propose a model-based classification and a model-based threshold method to detect premature atrial and ventricular contractions, and classify them into the two types.
In fact premature contractions can be symptomatic and therefore cause discomfort for the patient. Understanding the burden (i.e.~the number of premature contractions) and the origin of the premature contractions is important since the decision whether to treat is influenced by these two factors.
The first classification approach is support vector machine (SVM) classification with features that are output of the model. In particular, we use the standardized residuals together with possibly the shape parameters and/or the data per pulse.
We explore different combinations of predictors and kernels, and train/test the model through leave-one-out cross-validation.
We consider both two and three class classification (regular vs premature atrial or ventricular contraction, and regular vs premature atrial vs
premature ventricular contraction, respectively).

The second approach concatenates the standardized residuals in a vector, losing the separation into pulses, smooths the absolute values and detects local peaks.
Since maxima in the smoothed absolute residuals are seen to align with anomalies in the PPG signal due to a mismatch between model prediction and observed data, we use a simple threshold to detect the premature contractions.

\section{Results}\label{sec:Results}
In this section we collect some results on the model fit and its applications. First we illustrate the procedure described in Section~\ref{sec:Methods} on different types of PPG segments. Then we focus on the application, showing the accuracy of the estimated inter-beat-intervals, and the performance of the methods for the detection of premature contractions.

\subsection{Fit}
In the following, we show the results of fitting the model to PPG signals, illustrating the whole procedure introduced in Section~\ref{sec:Methods}.
To prove the flexibility of the model, we fit data both from patients with atrial fibrillation (Patient 6) and without (Patient 5) from the internally collected dataset (cf. Section~\ref{sec:Data}).
Furthermore, we consider the publicly available~\emph{8-minute Data} from the CapnoBase database~\cite{Karlen2010, Karlen2013} and fit the data from patient 32 (Capno 32).
This way, we have data recorded with finger (CapnoBase data) and wrist (internally collected data) probes, either sinus rhythm or under atrial fibrillation.
The data from the CapnoBase dataset is firstly downsampled to reduce the sampling frequency to $64$ Hz.

\begin{figure}[h!]
	\centering
	\begin{subfigure}[b]{0.3\textwidth}
		\centering
		\includegraphics[width=\linewidth]{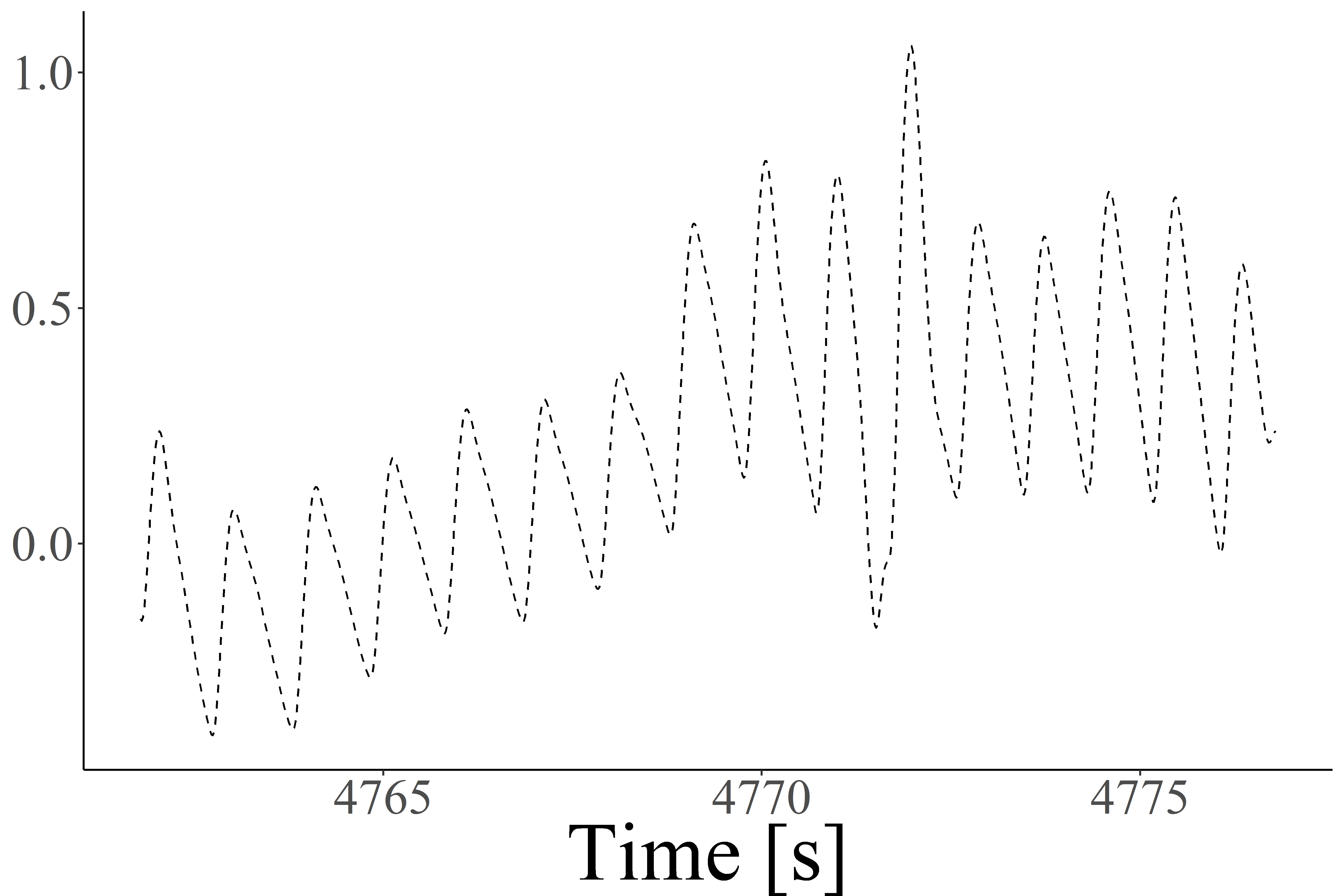}
		\subcaption{Patient 5}  		
	\end{subfigure}
	\begin{subfigure}[b]{0.3\textwidth}
		\centering
		\includegraphics[width=\linewidth]{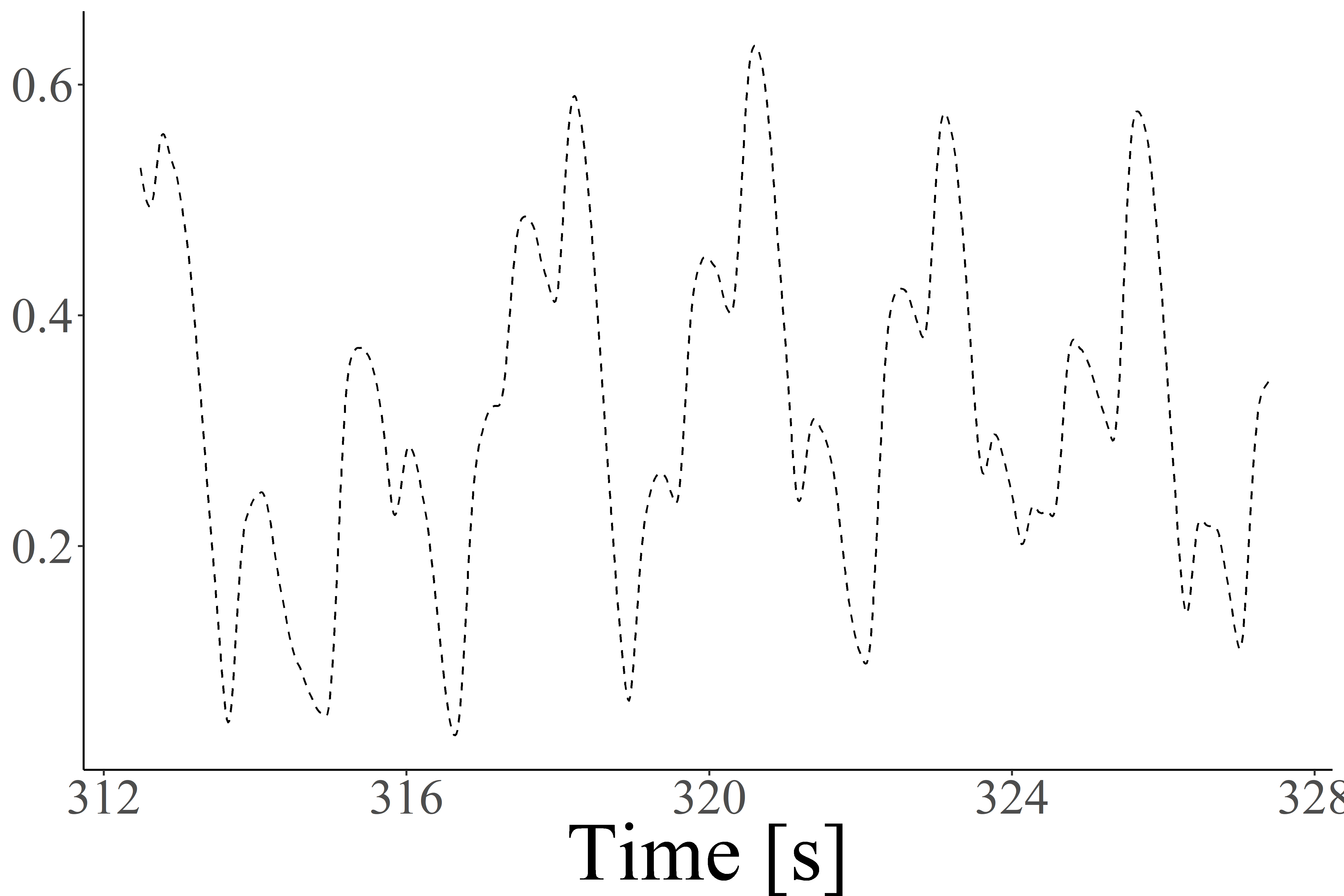}
		\subcaption{Patient 6}
	\end{subfigure}
	\begin{subfigure}[b]{0.3\textwidth}
		\centering
		\includegraphics[width=\linewidth]{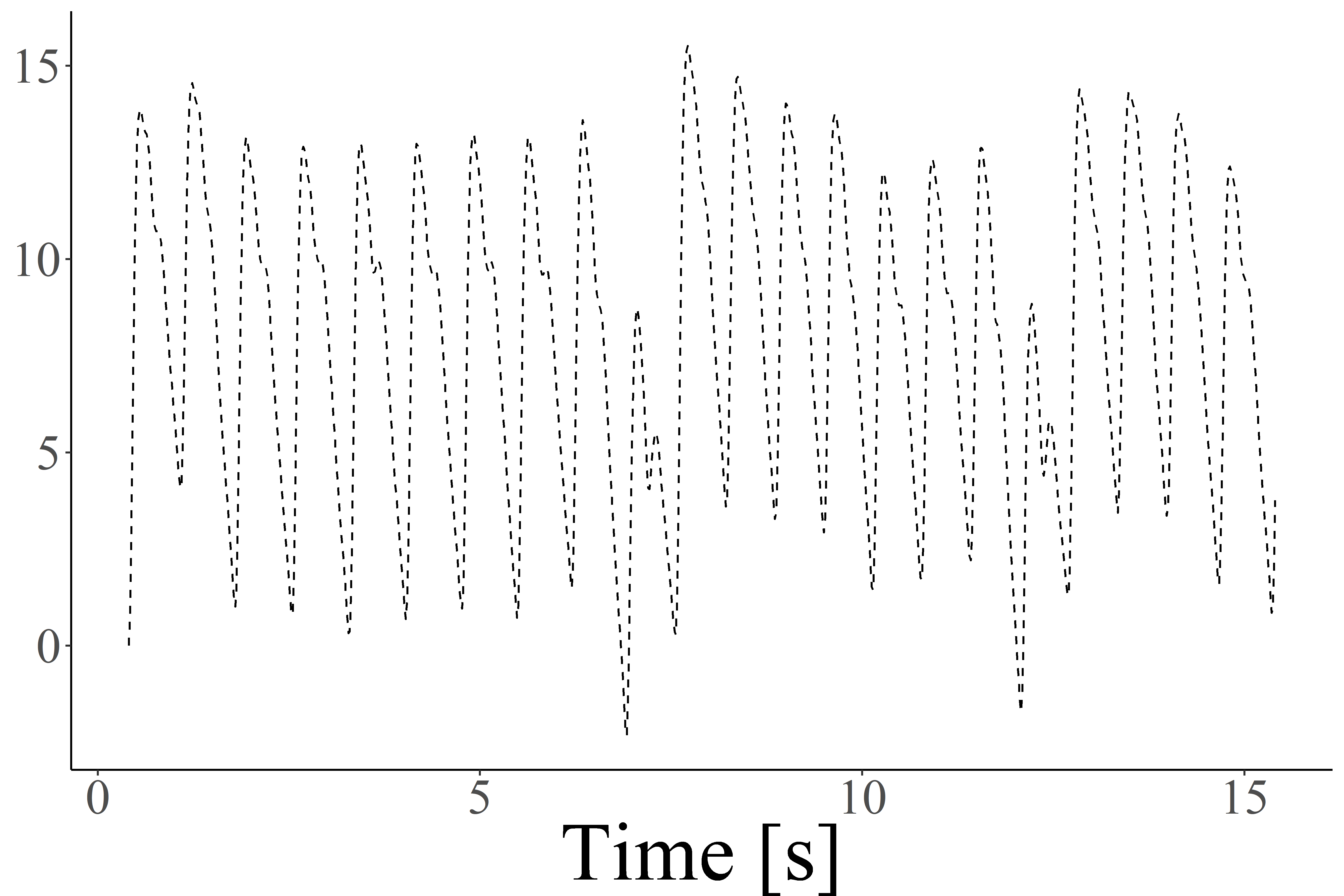}
		\subcaption{Capno 32}  		
	\end{subfigure}		
	\caption{{\bfseries Processed PPG signals.} Sample segments of about 15 seconds from Patient 5 (nonAF) and Patient 6 (AF) from the internally collected dataset, and from Patient Capno 32 from the CapnoBase dataset.}\label{fig:originalPPG}	
\end{figure}

\begin{figure}[h!]
	\centering
	\begin{subfigure}[b]{0.3\textwidth}
		\centering
		\includegraphics[width=\linewidth]{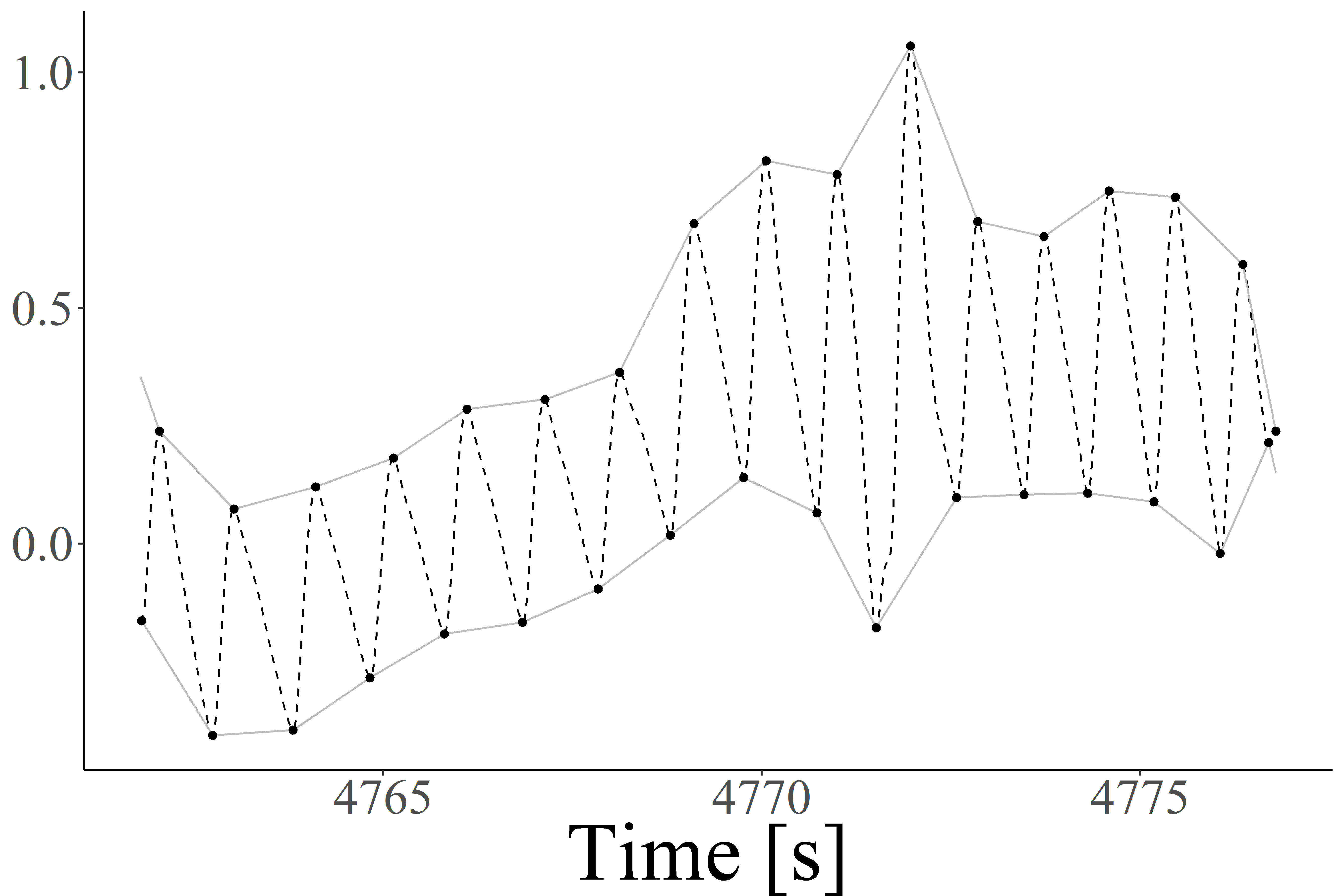}	   		
		\subcaption{Patient 5}  
	\end{subfigure}
	\begin{subfigure}[b]{0.3\textwidth}
		\centering
		\includegraphics[width=\linewidth]{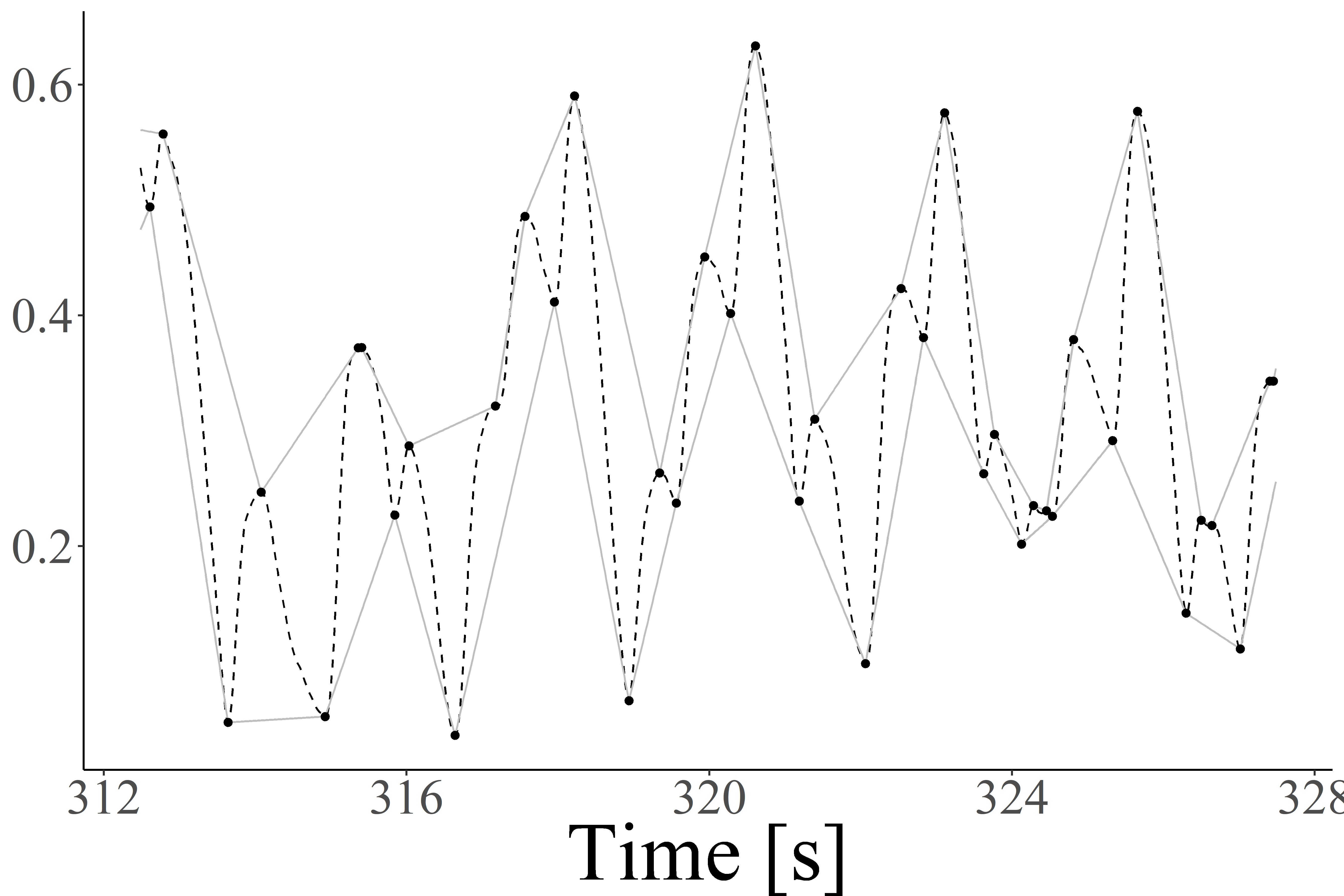}
		\subcaption{Patient 6}
	\end{subfigure}
	\begin{subfigure}[b]{0.3\textwidth}
		\centering
		\includegraphics[width=\linewidth]{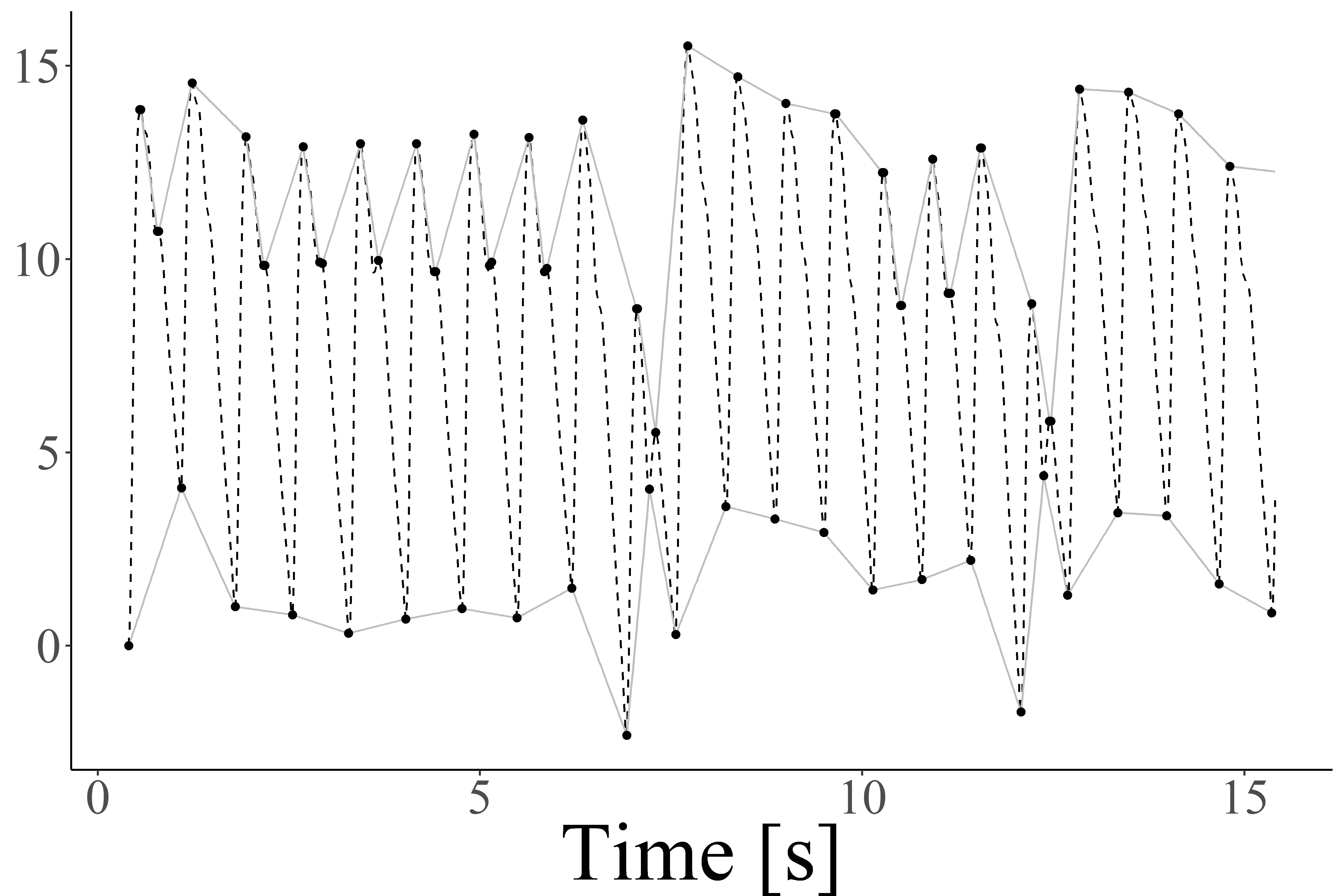}	   		
		\subcaption{Capno 32}  
	\end{subfigure}
	\caption{{\bfseries Decomposition - Component I.} Local maxima and minima constitute Component I, represented by the black dots. The dotted line is the original signal and the grey lines constitute the envelope, used to normalize the signal to obtain Component II. }\label{fig:componentI}	
\end{figure}

\begin{figure}[h!]
	\centering
	\begin{subfigure}[b]{0.3\textwidth}
		\centering
		\includegraphics[width=\linewidth]{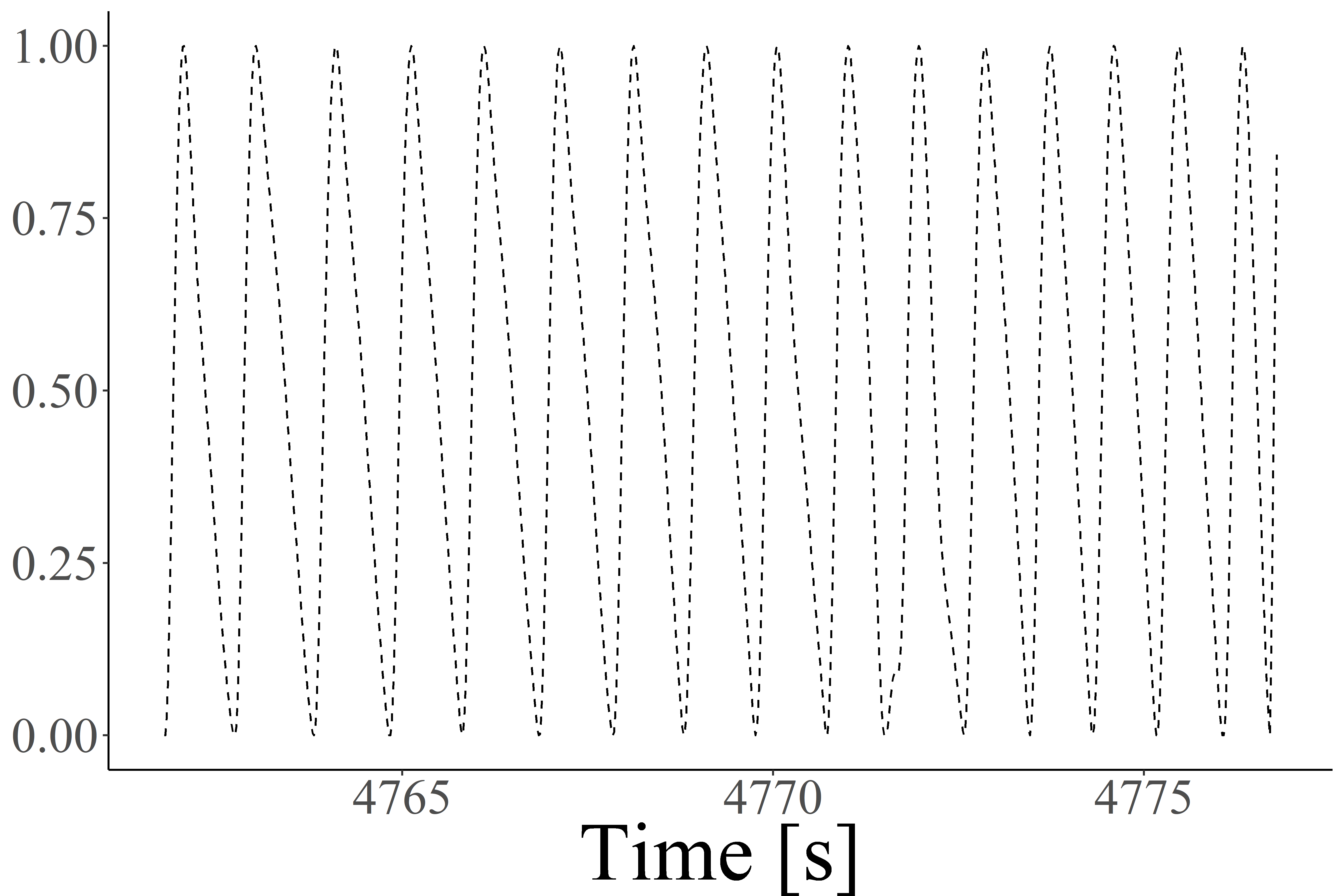}
		\subcaption{Patient 5}  
	\end{subfigure}
	\begin{subfigure}[b]{0.3\textwidth}
		\centering
		\includegraphics[width=\linewidth]{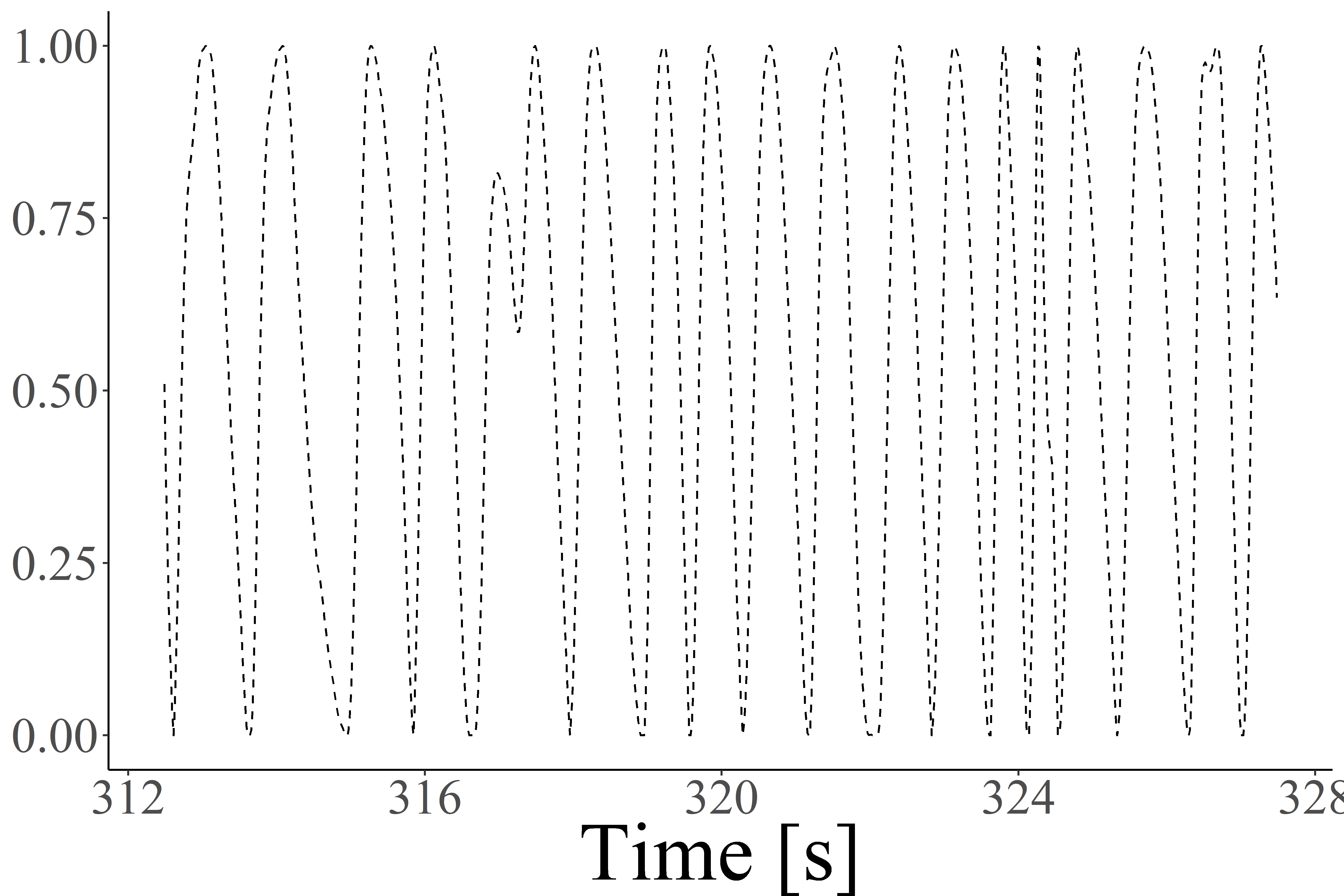}
		\subcaption{Patient 6}
	\end{subfigure}
	\begin{subfigure}[b]{0.3\textwidth}
		\centering
		\includegraphics[width=\linewidth]{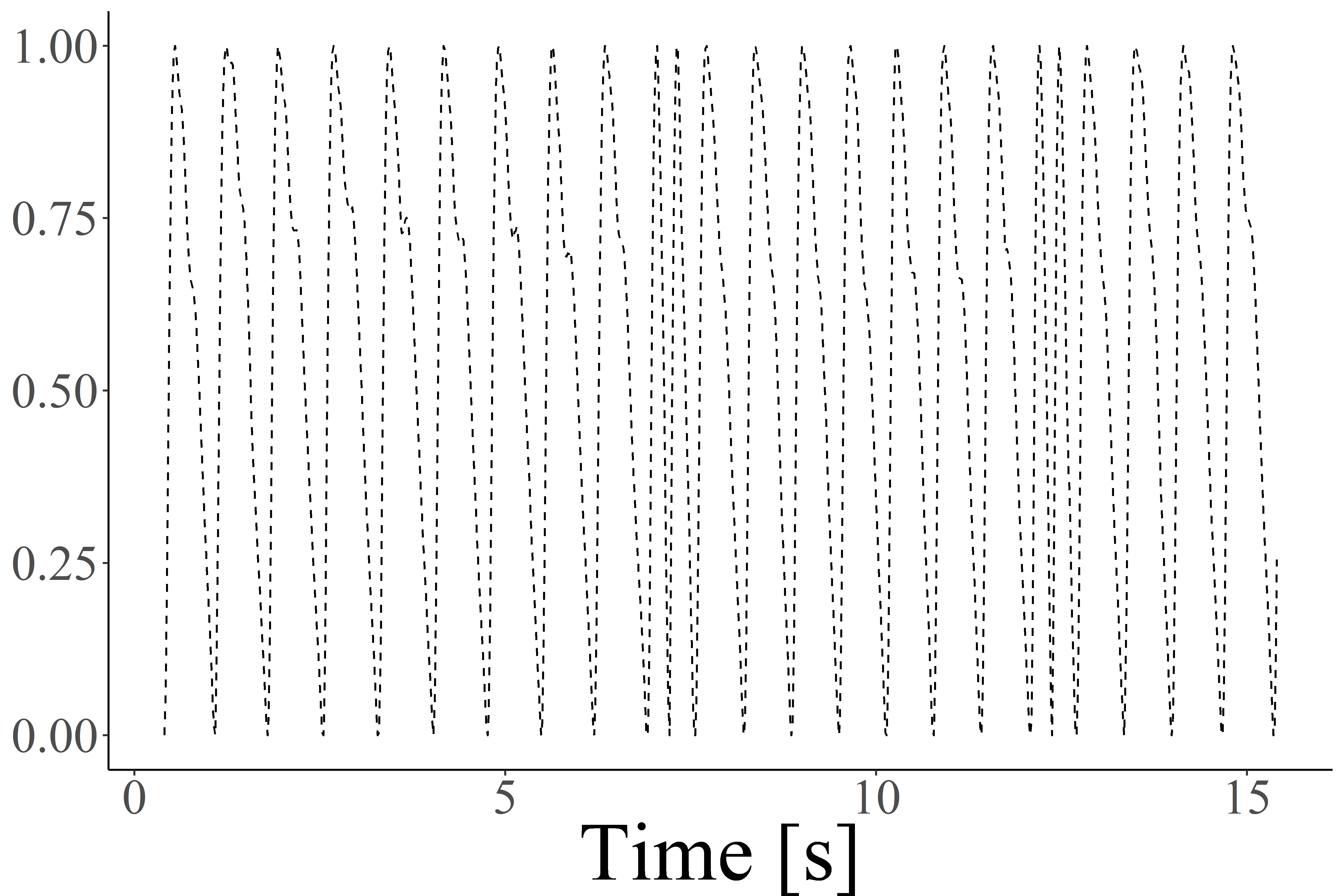}	   		
		\subcaption{Capno 32}  
	\end{subfigure}
	\caption{{\bfseries Decomposition - Component II.} Component II is the signal normalized by the envelope of Component I. Component II has thus unit amplitude. Dotted line for the observations.}\label{fig:componentII}	
\end{figure}

\begin{figure}[h!]
	\centering
	\begin{subfigure}[b]{0.3\textwidth}
		\centering
		\includegraphics[width=\linewidth]{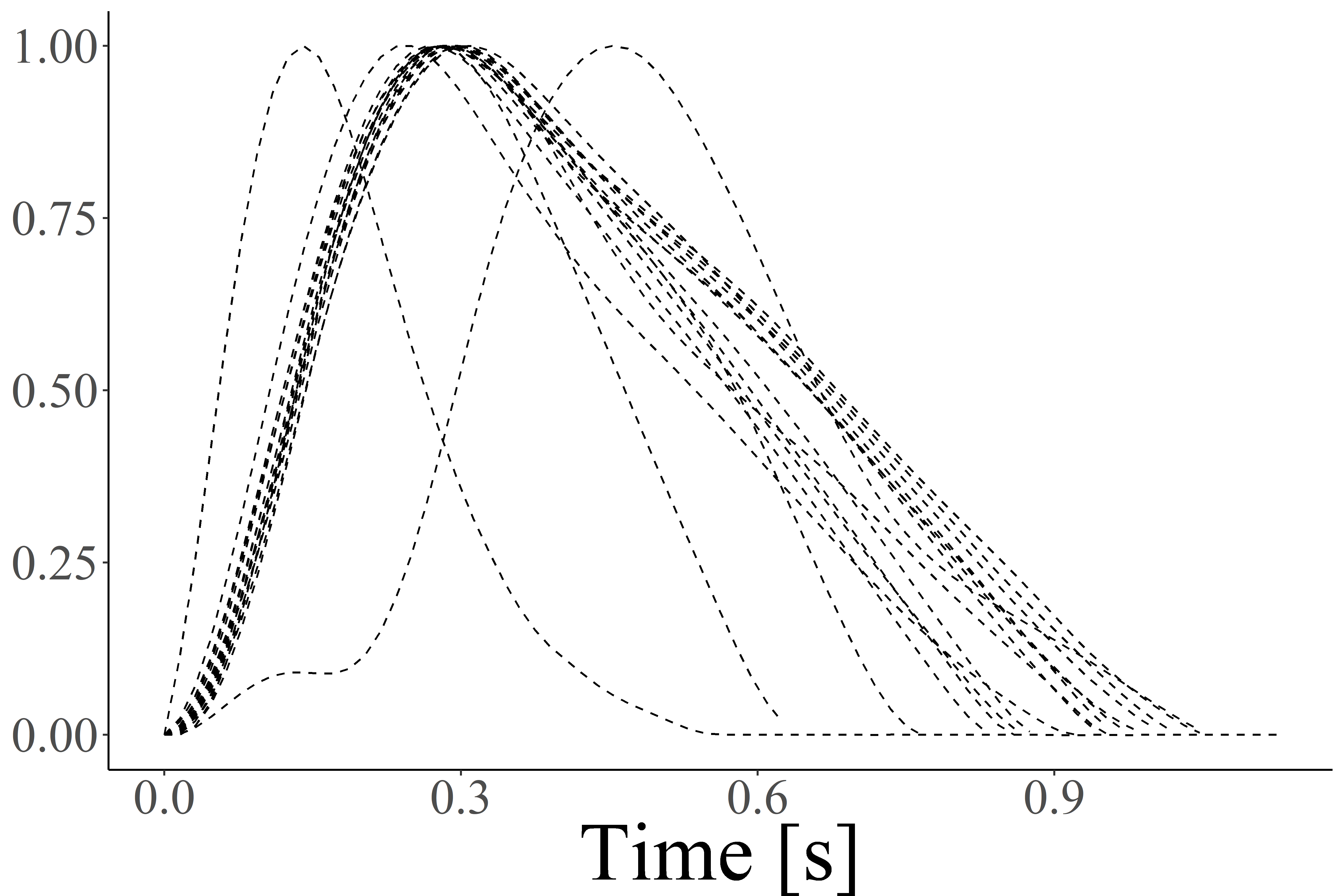}
		\subcaption{Patient 5}  
	\end{subfigure}
	\begin{subfigure}[b]{0.3\textwidth}
		\centering
		\includegraphics[width=\linewidth]{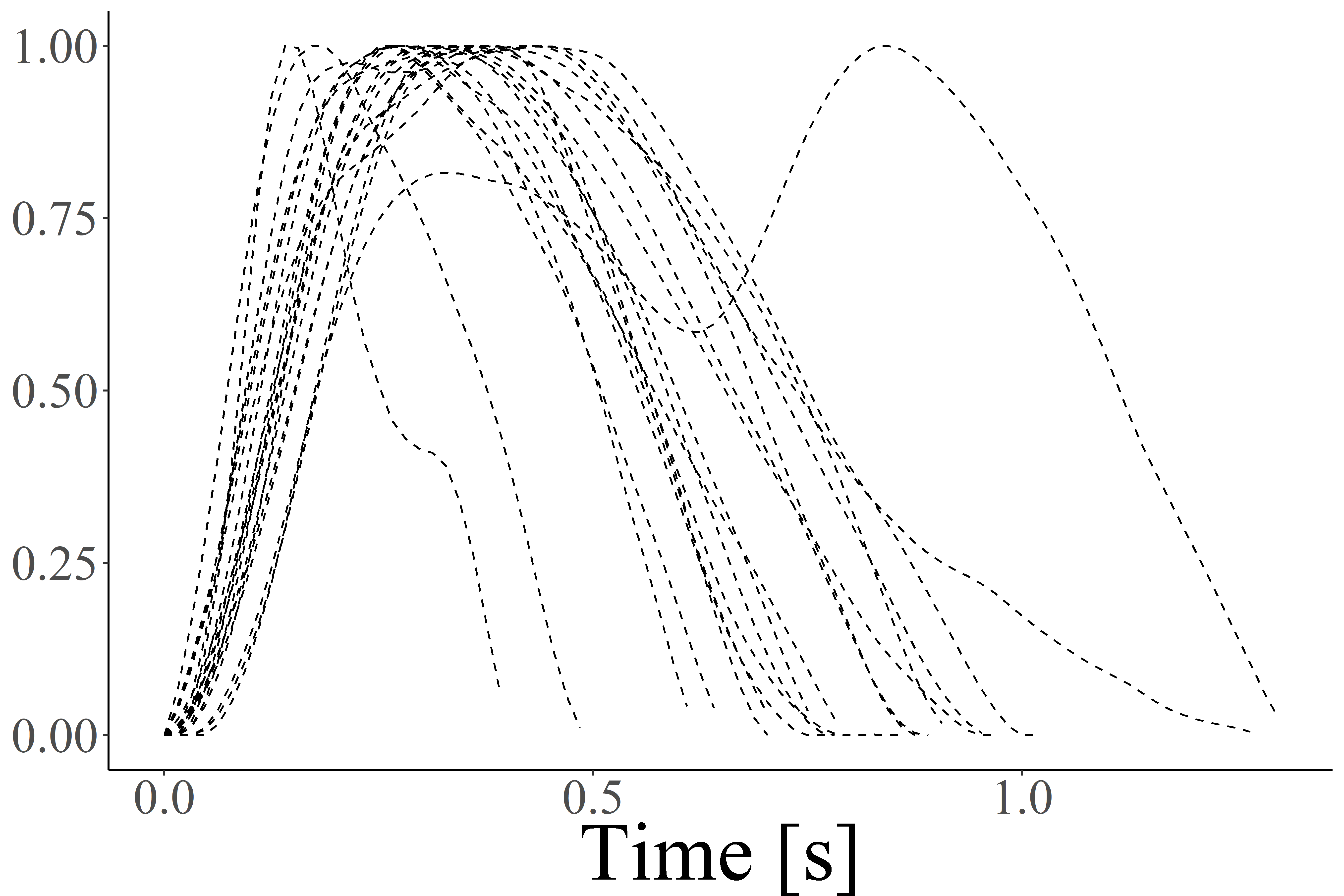}
		\subcaption{Patient 6}
	\end{subfigure}
	\begin{subfigure}[b]{0.3\textwidth}
		\centering
		\includegraphics[width=\linewidth]{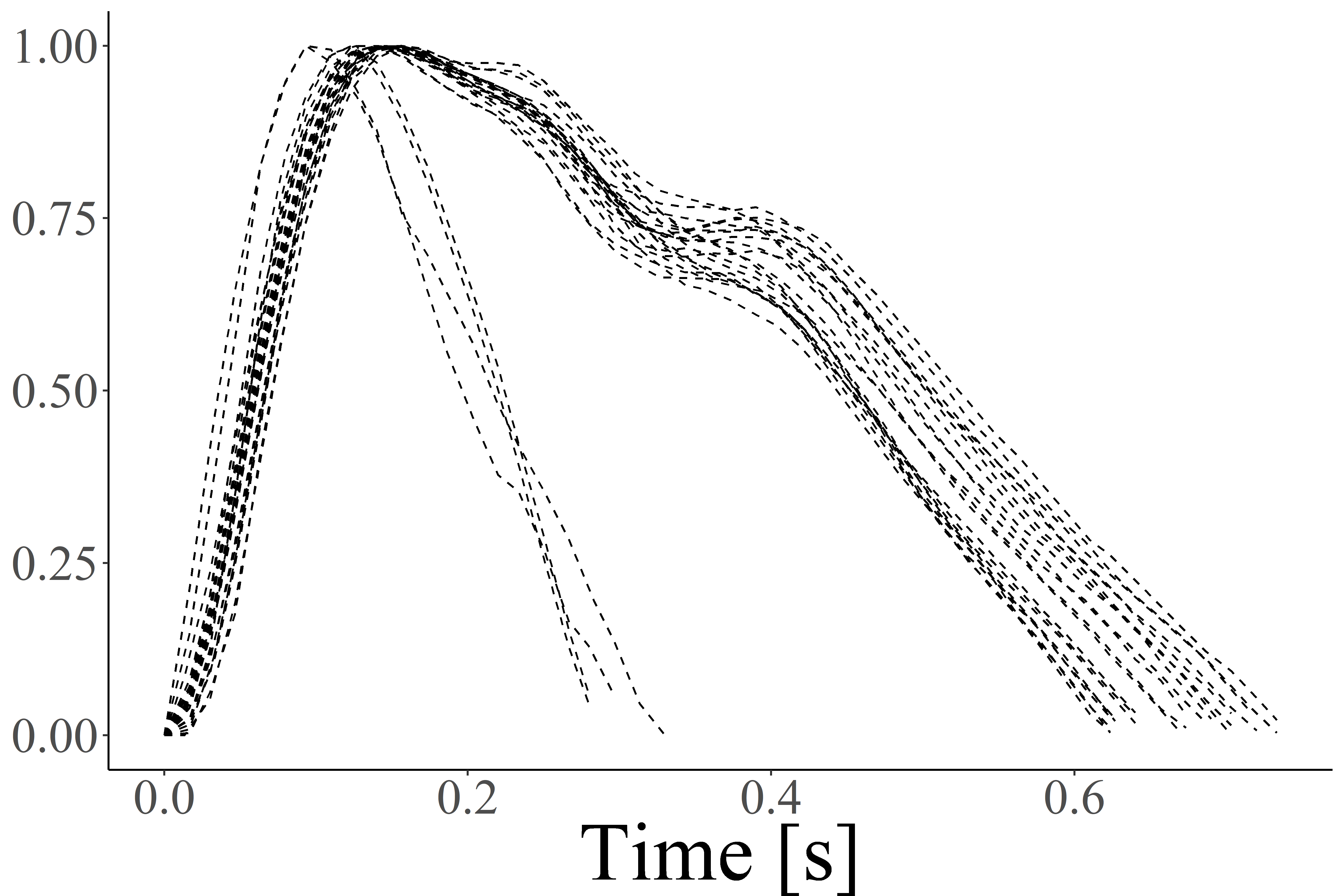}	   		
		\subcaption{Capno 32}  
	\end{subfigure}
	\caption{{\bfseries Segmentation.} Component II is segmented into pulses and overlapped to have the same first sample. Dotted line for the observations.}\label{fig:segmentation}	
\end{figure}

\begin{figure}[h!]
	\centering
	\begin{subfigure}[b]{0.3\textwidth}
		\centering
		\includegraphics[width=\linewidth]{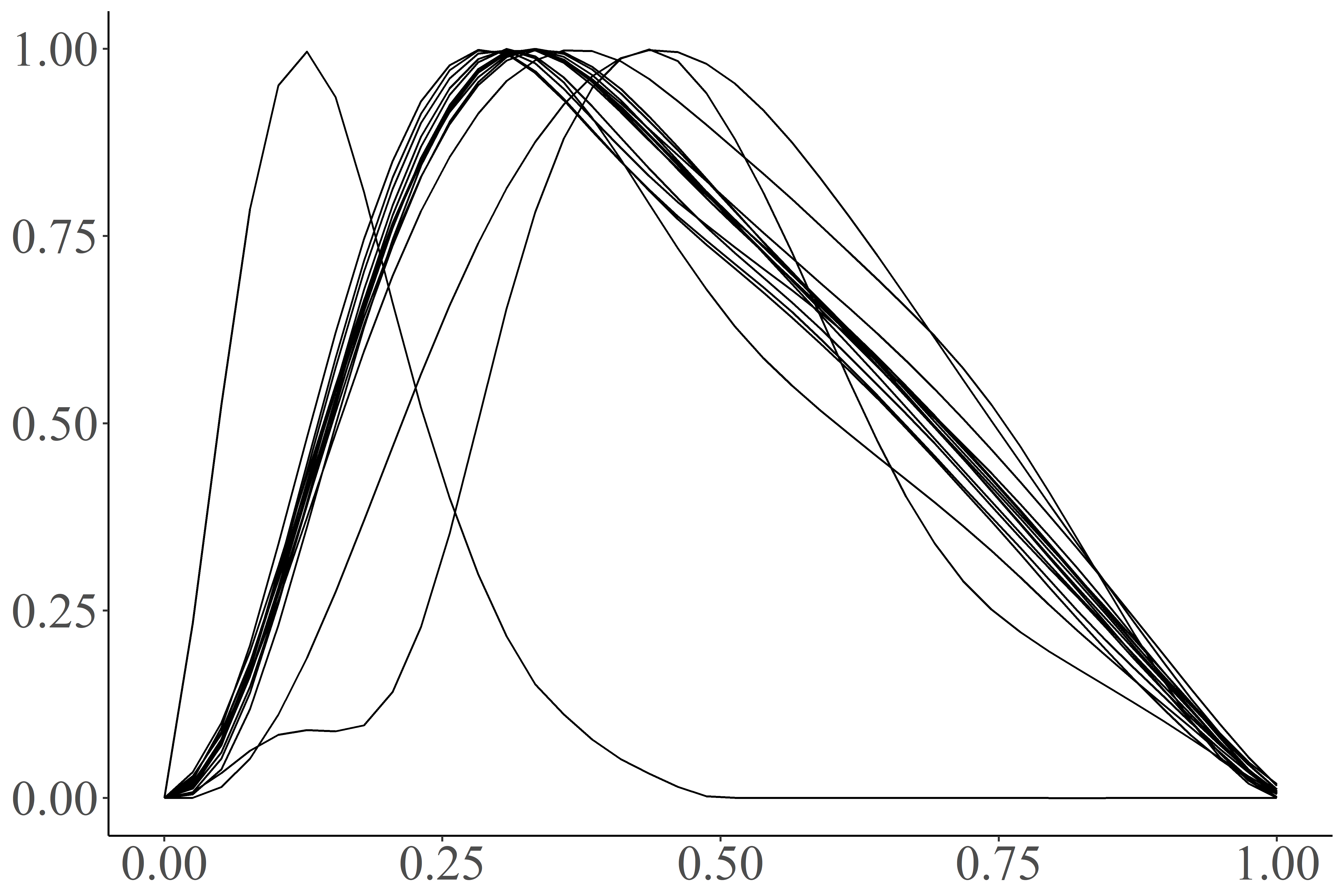}
		\subcaption{Patient 5}  
	\end{subfigure}
	\begin{subfigure}[b]{0.3\textwidth}
		\centering
		\includegraphics[width=\linewidth]{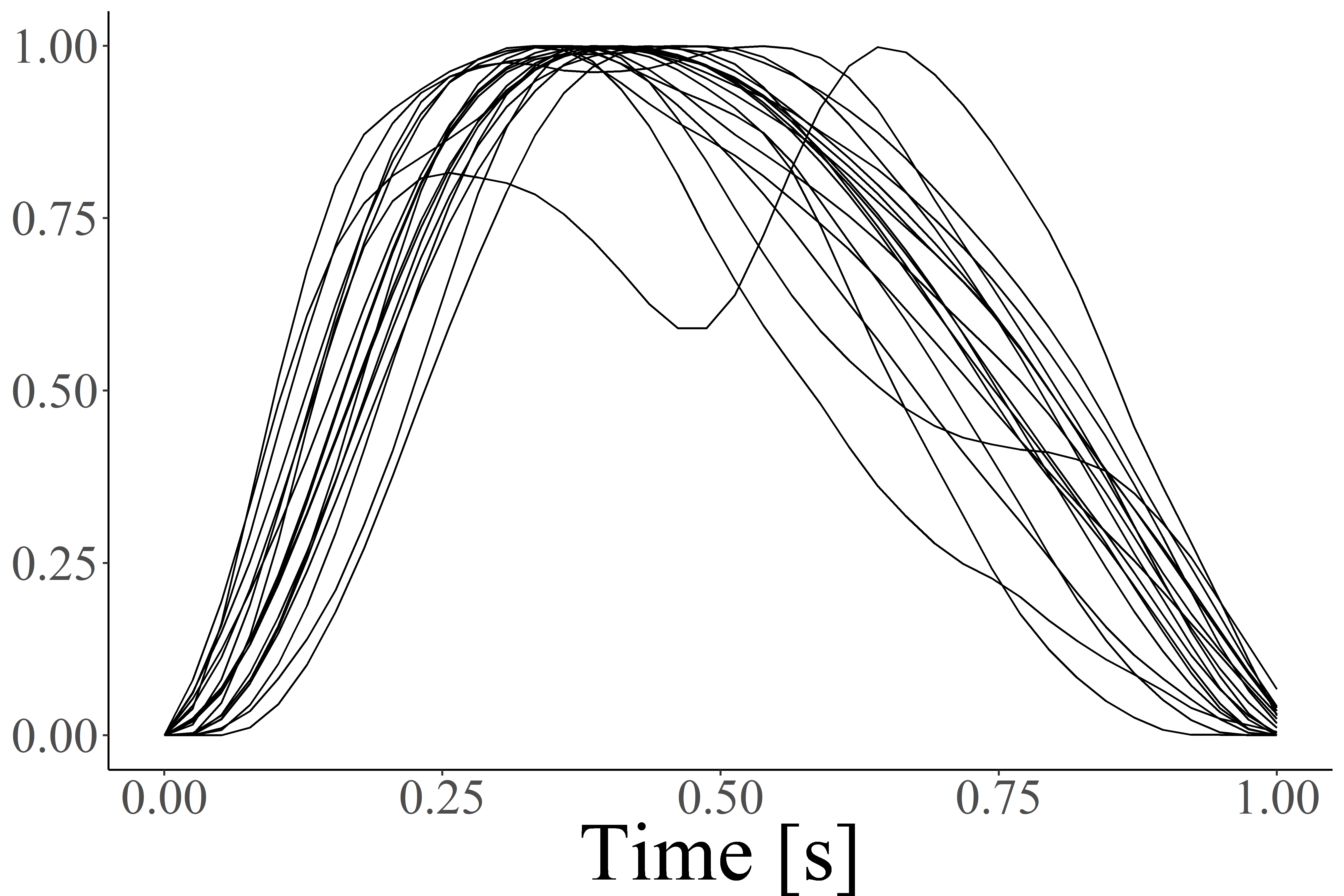}
		\subcaption{Patient 6}
	\end{subfigure}
	\begin{subfigure}[b]{0.3\textwidth}
		\centering
		\includegraphics[width=\linewidth]{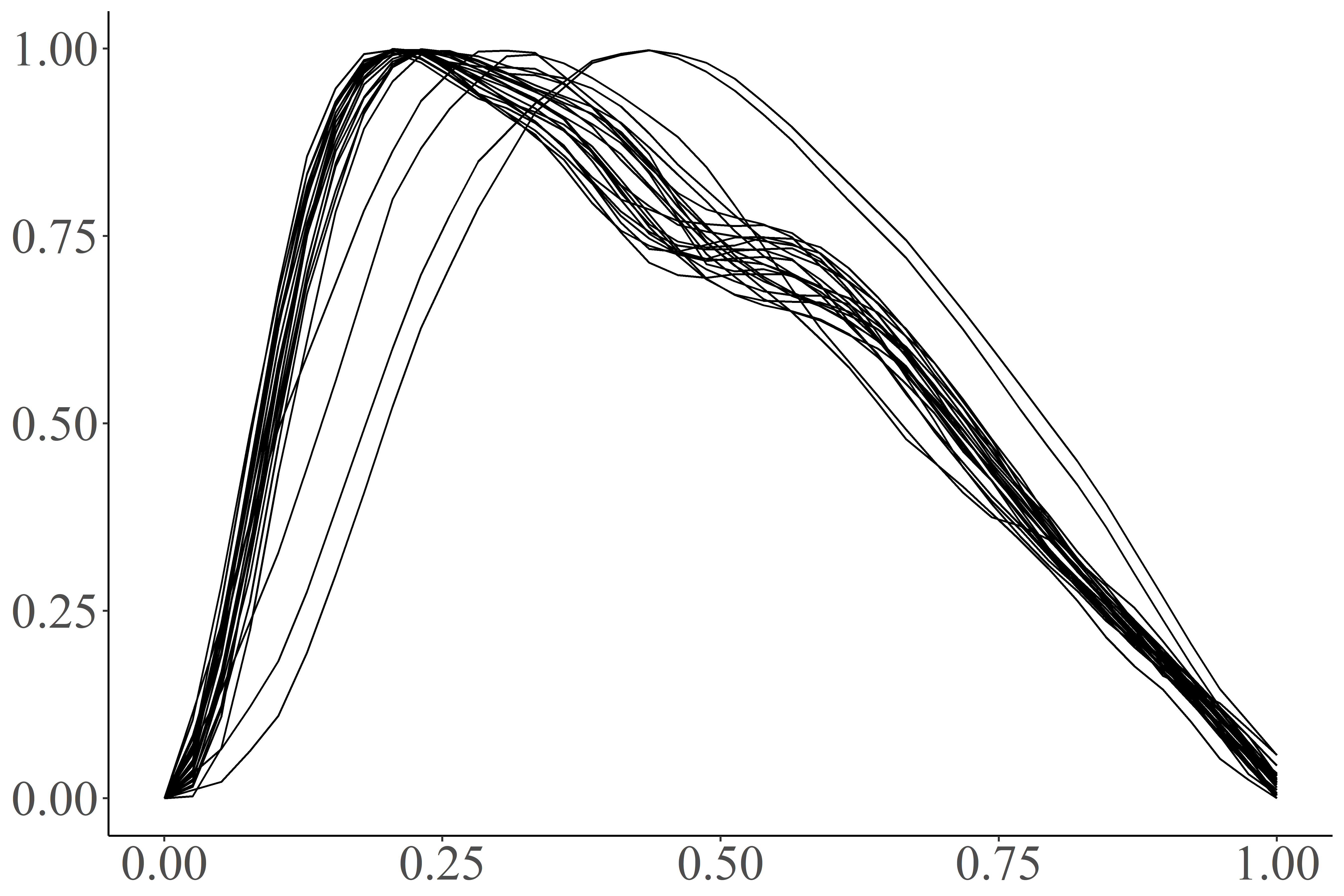}	   		
		\subcaption{Capno 32}  
	\end{subfigure}
	\caption{{\bfseries Curve registration.}  Each pulse constituting Component II is interpolated on the same $r$ points in $[0,1]$. Continuous line to represent functional objects.}\label{fig:componentII_aligned}	
\end{figure}

\begin{figure}[h!]
	\centering
	\begin{subfigure}[b]{0.3\textwidth}
		\includegraphics[width=\linewidth]{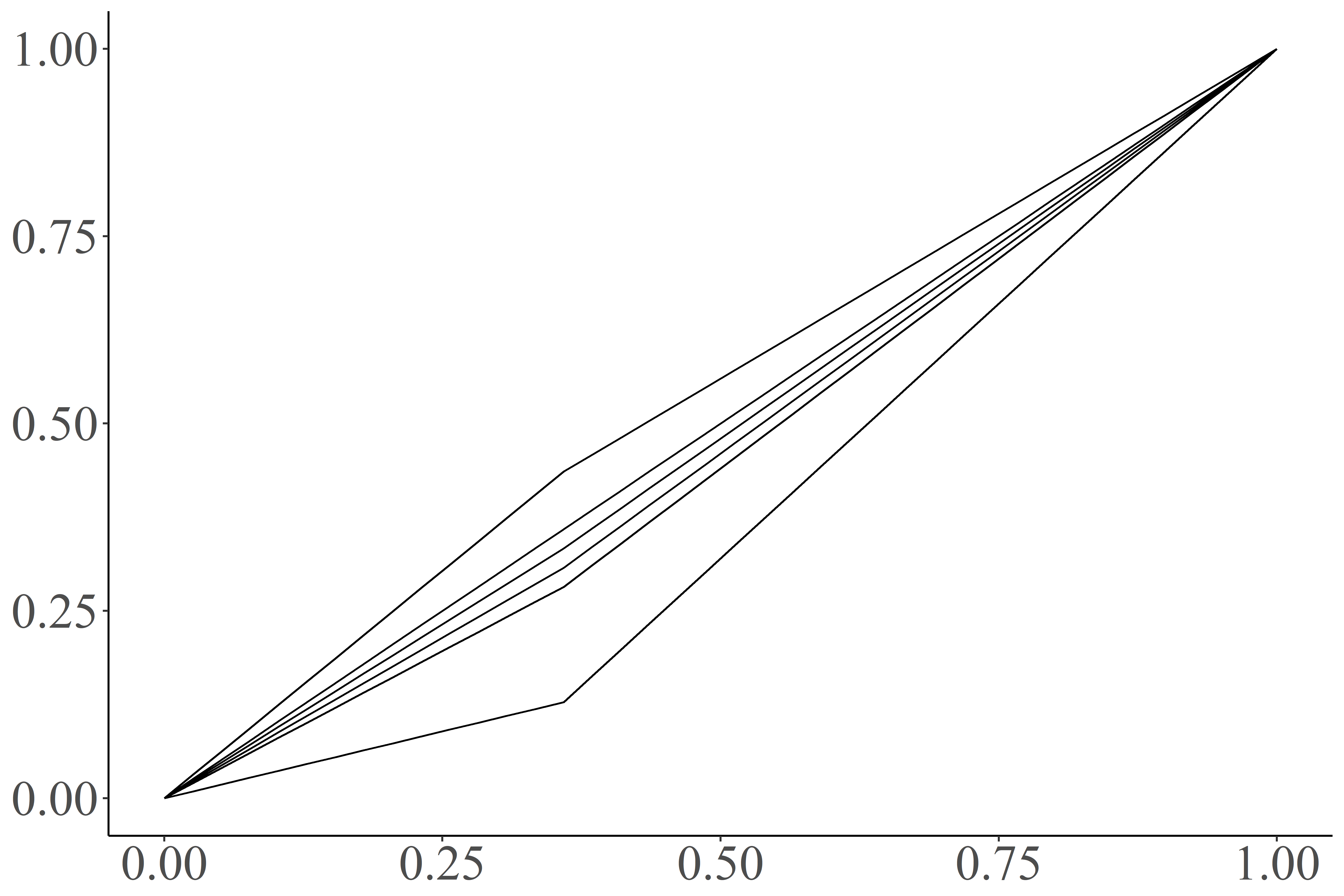}	   		
		\subcaption{Patient 5}  
	\end{subfigure}
	\begin{subfigure}[b]{0.3\textwidth}
		\includegraphics[width=\linewidth]{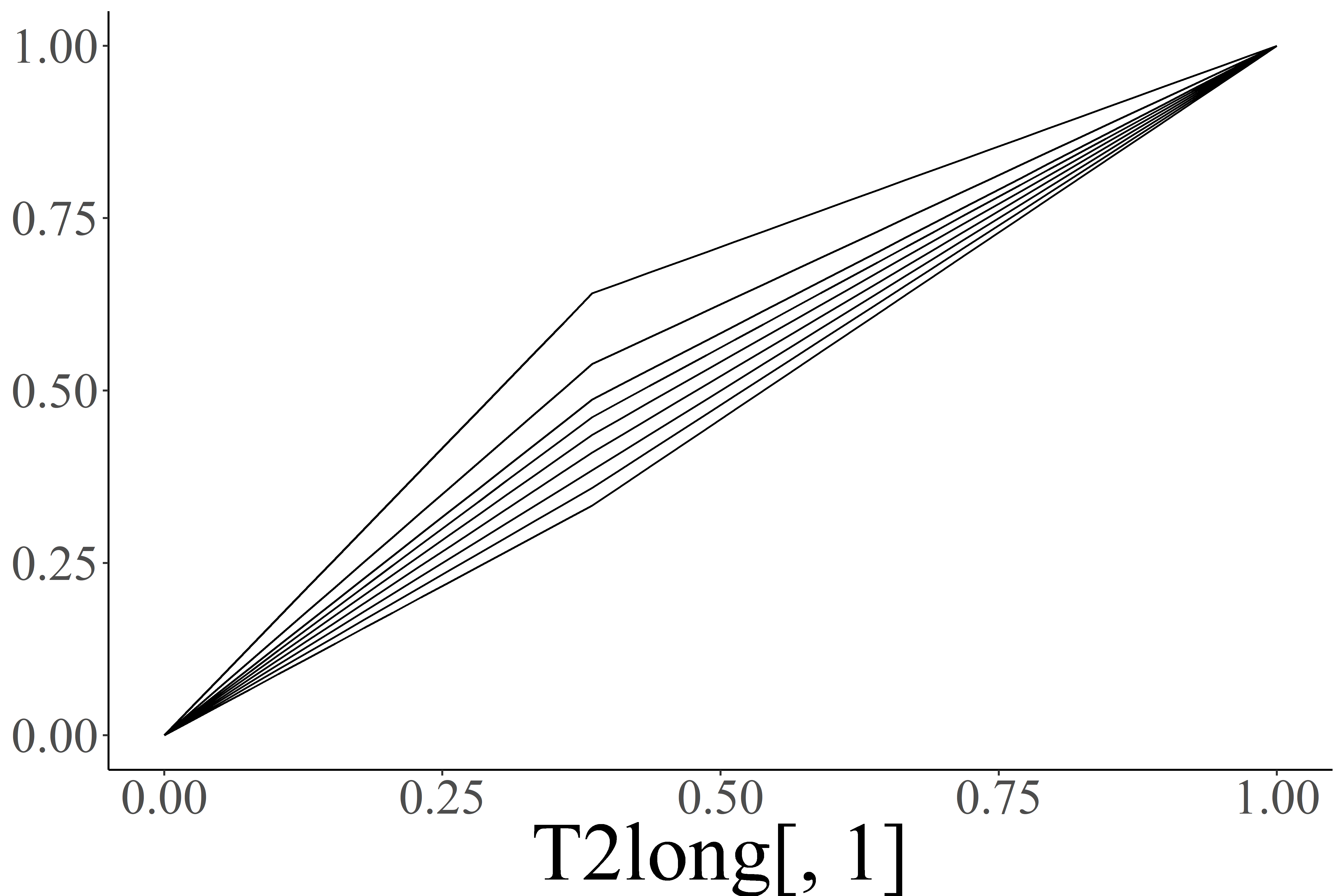}	   		
		\subcaption{Patient 6}  
	\end{subfigure}
	\begin{subfigure}[b]{0.3\textwidth}
		\includegraphics[width=\linewidth]{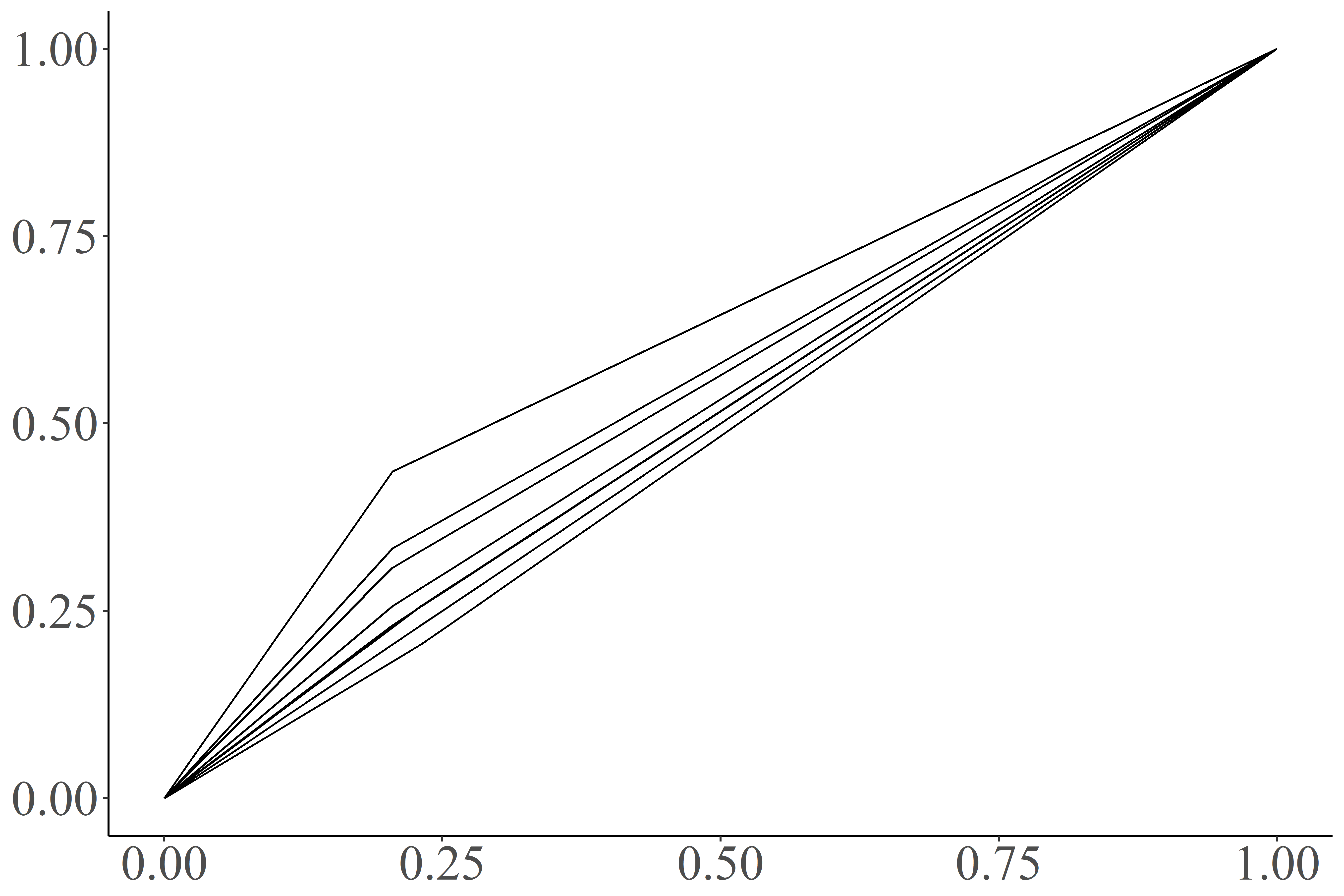}	   	
		\subcaption{Capno 32}  
	\end{subfigure}
	\caption{{\bfseries Time transformation.} Piecewise linear warping functions that lead to the pulses with aligned maxima.}\label{fig:timeTransf}	
\end{figure}

\begin{figure}[h!]	\centering
	\begin{subfigure}[b]{0.3\textwidth}
		\includegraphics[width=\linewidth]{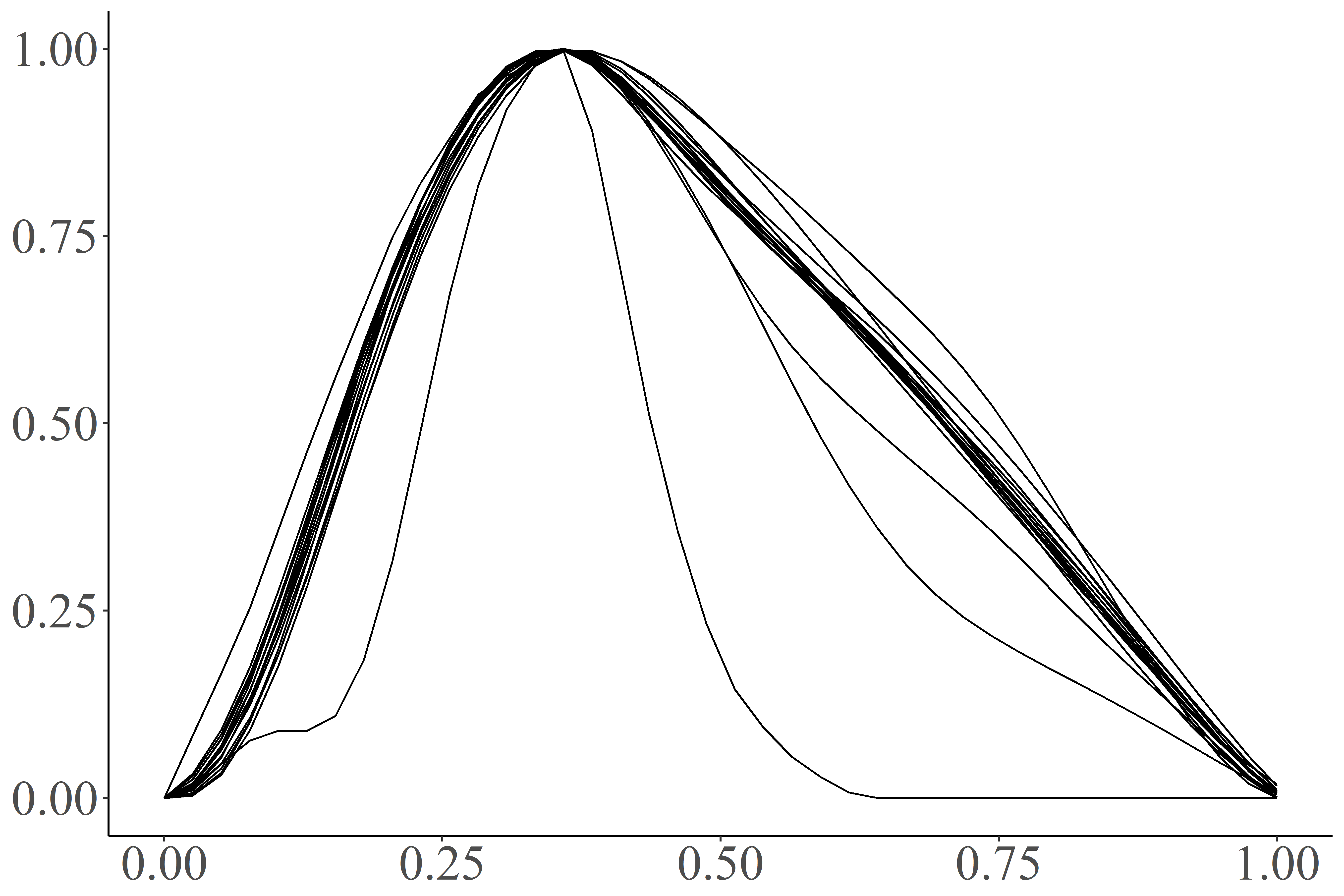}	
		\subcaption{Patient 5}  
	\end{subfigure}
	\begin{subfigure}[b]{0.3\textwidth}
		\includegraphics[width=\linewidth]{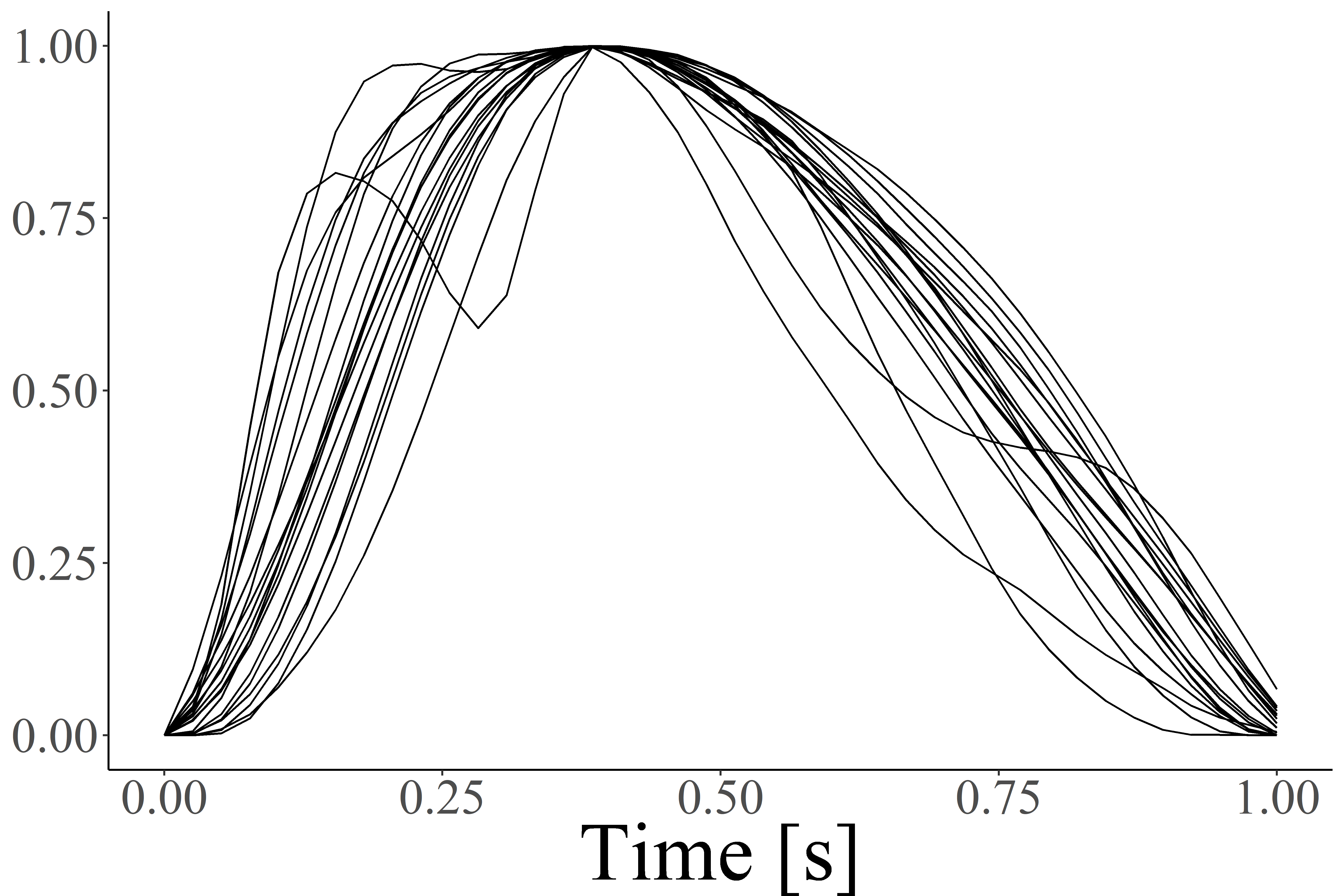}	   		
		\subcaption{Patient 6}  
	\end{subfigure}
	\begin{subfigure}[b]{0.3\textwidth}
	  \includegraphics[width=\linewidth]{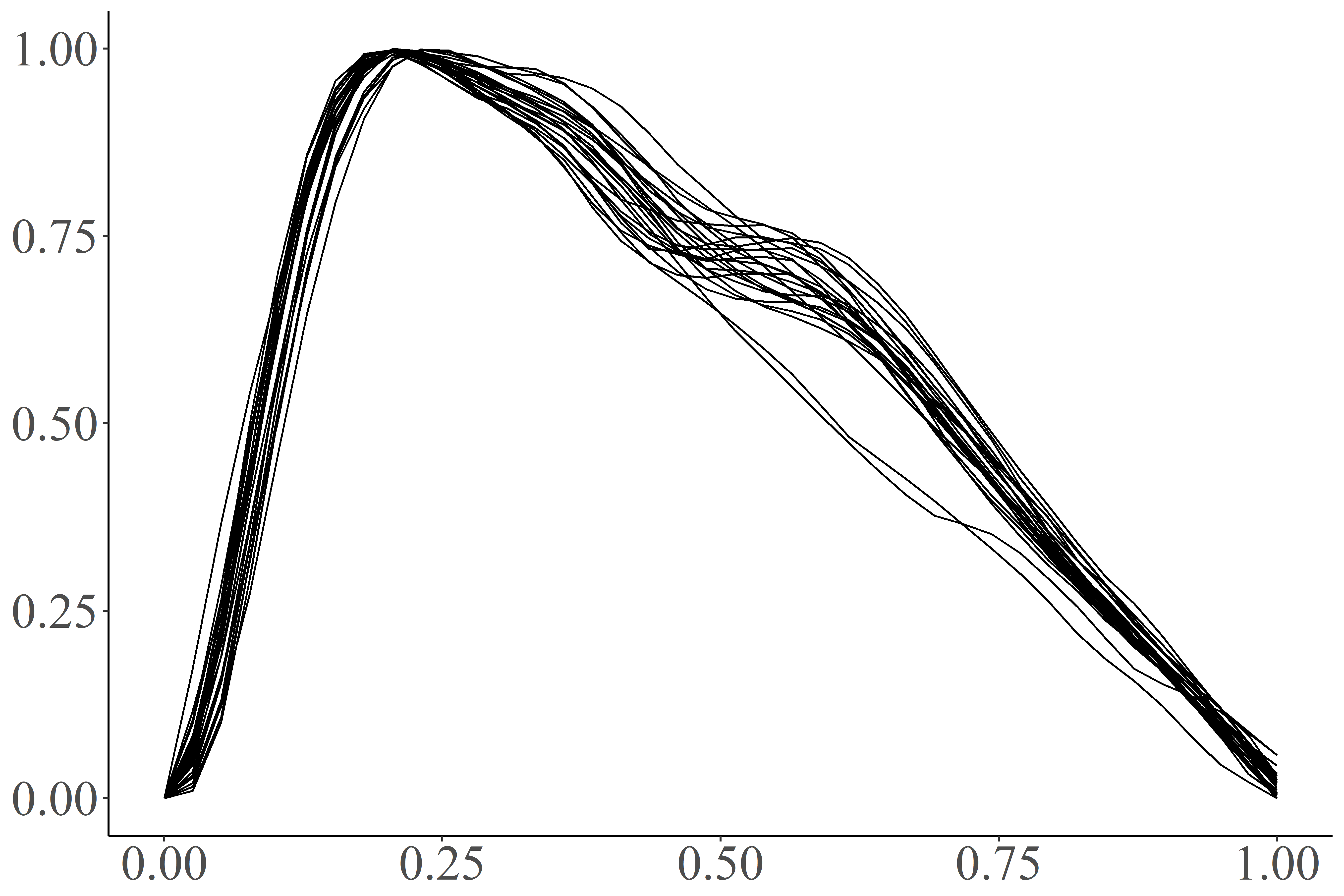}	   		
		\subcaption{Capno 32}  
	\end{subfigure}
	\caption{{\bfseries Maxima alignment.} Pulses with aligned maxima, obtained by composing Component II with the piecewise linear warping functions in Figure~\ref{fig:timeTransf}. The residual variability in the location of the maxima is due to the sequential choice of the target maximum location.}\label{fig:curveReg}	
\end{figure}

\begin{figure}[h!]
	\centering
	\begin{subfigure}[b]{0.3\textwidth}
		\includegraphics[width=\linewidth]{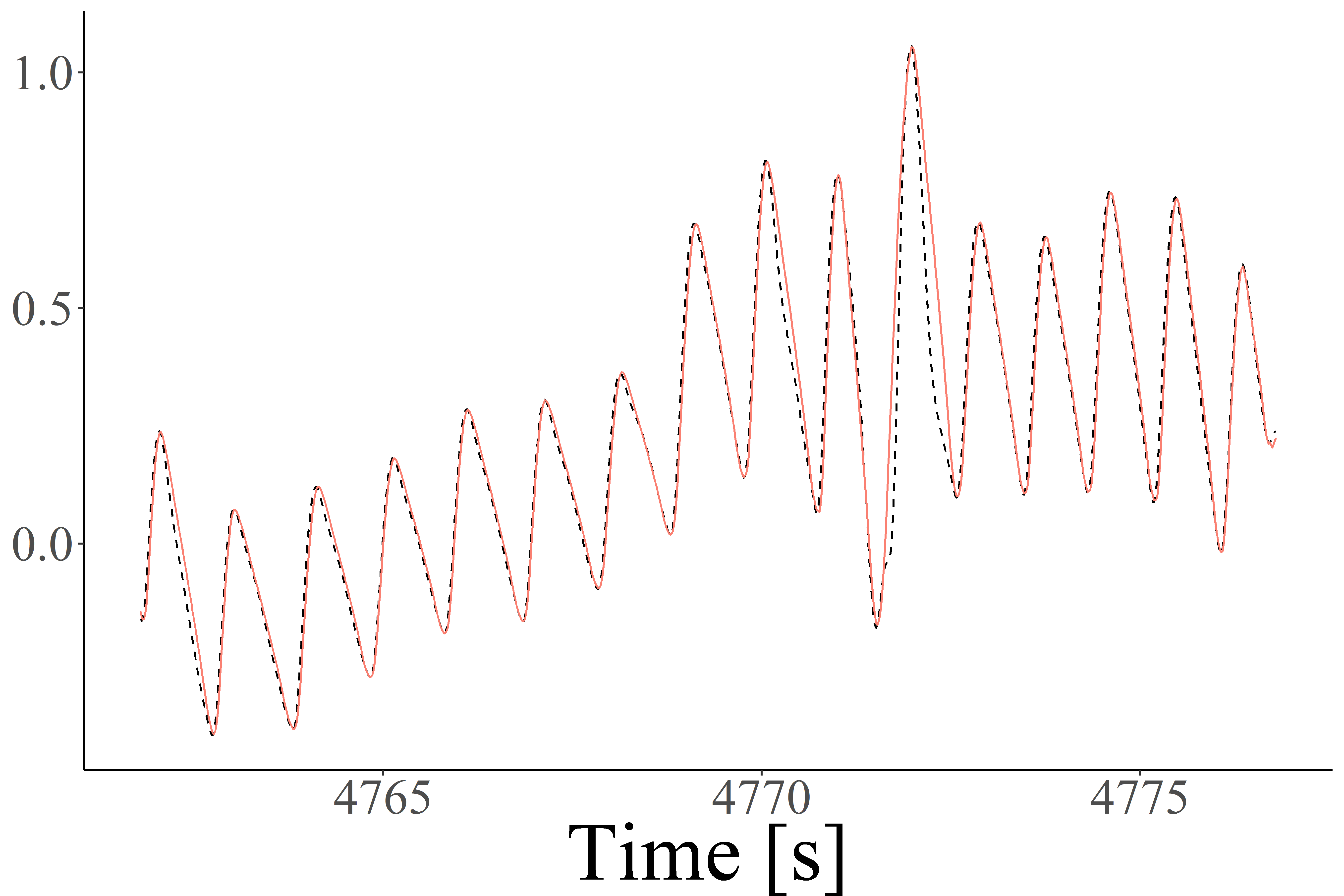}	   		
		\subcaption{Patient 5}  
	\end{subfigure}
	\begin{subfigure}[b]{0.3\textwidth}
		\includegraphics[width=\linewidth]{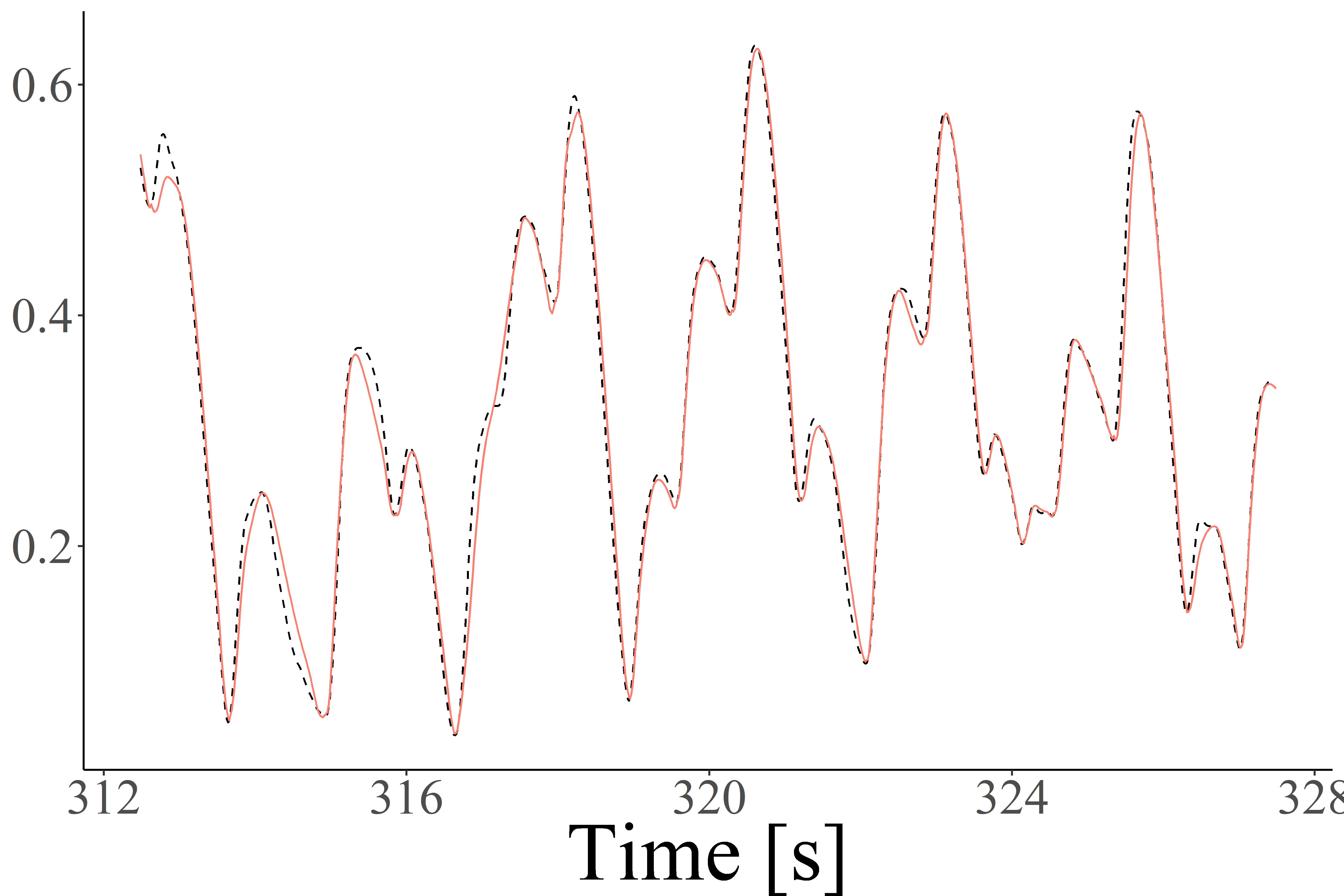}	  	
		\subcaption{Patient 6}  
	\end{subfigure}
	\begin{subfigure}[b]{0.3\textwidth}
		\includegraphics[width=\linewidth]{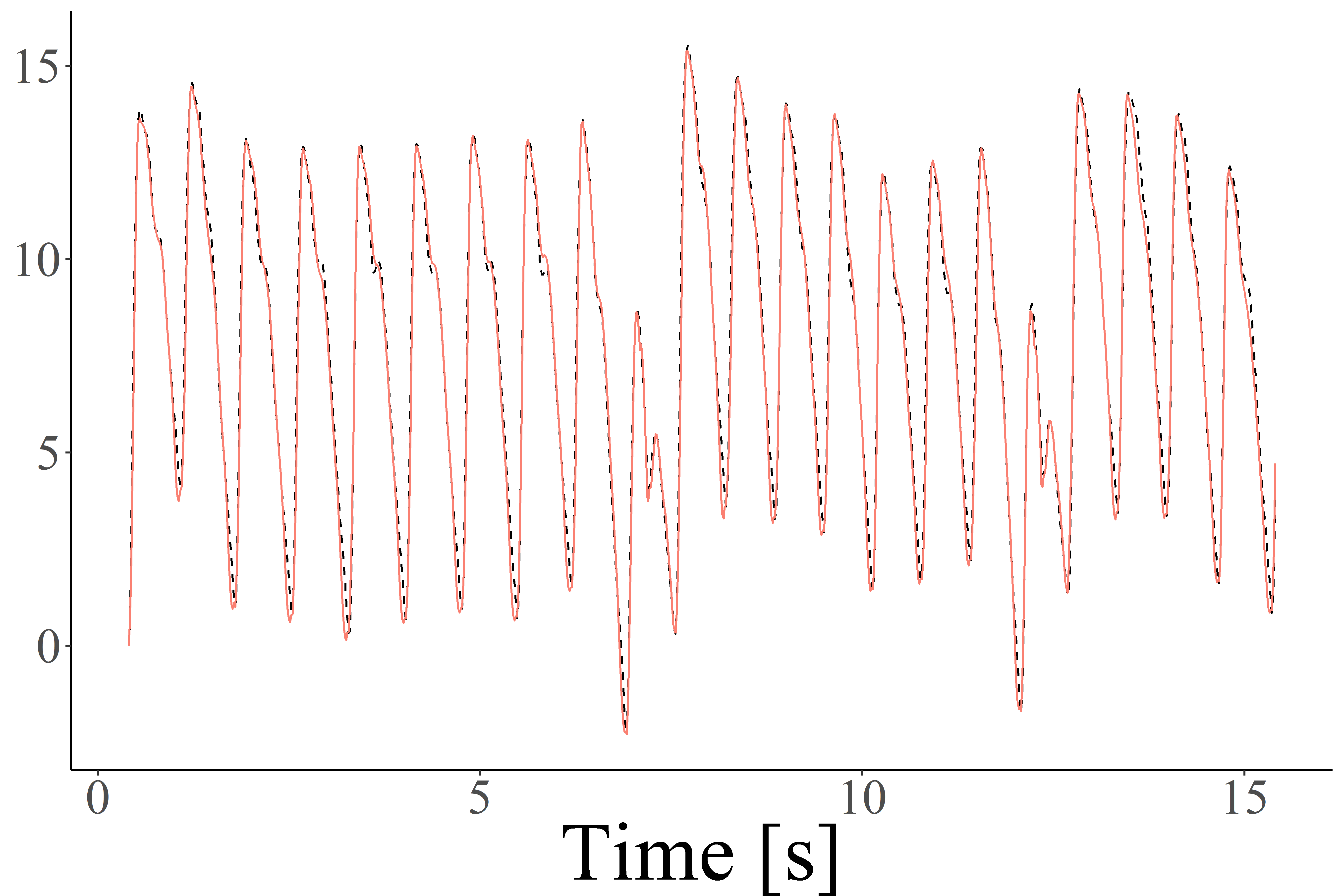}	   		
		\subcaption{Capno 32}  
	\end{subfigure}
	\caption{{\bfseries Resulting fit.} The black dotted line is the original signal, while the red continuous line is the model fit.}\label{fig:Fit}
\end{figure}

\begin{figure}[h!]
	\centering
	\begin{subfigure}[b]{0.3\textwidth}
		\includegraphics[width=\linewidth]{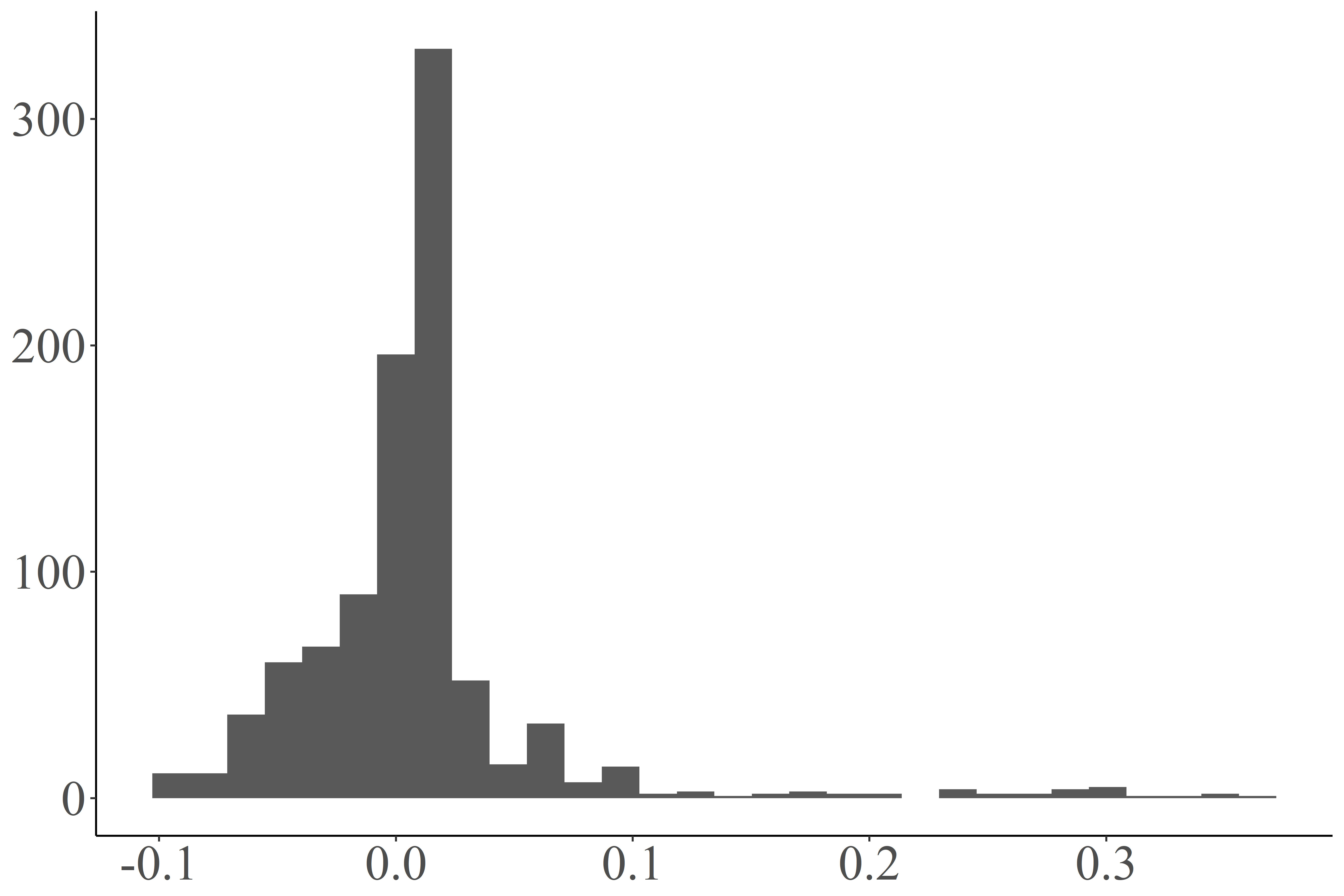}	   
		\subcaption{Patient 5}  
	\end{subfigure}
	\begin{subfigure}[b]{0.3\textwidth}
		\includegraphics[width=\linewidth]{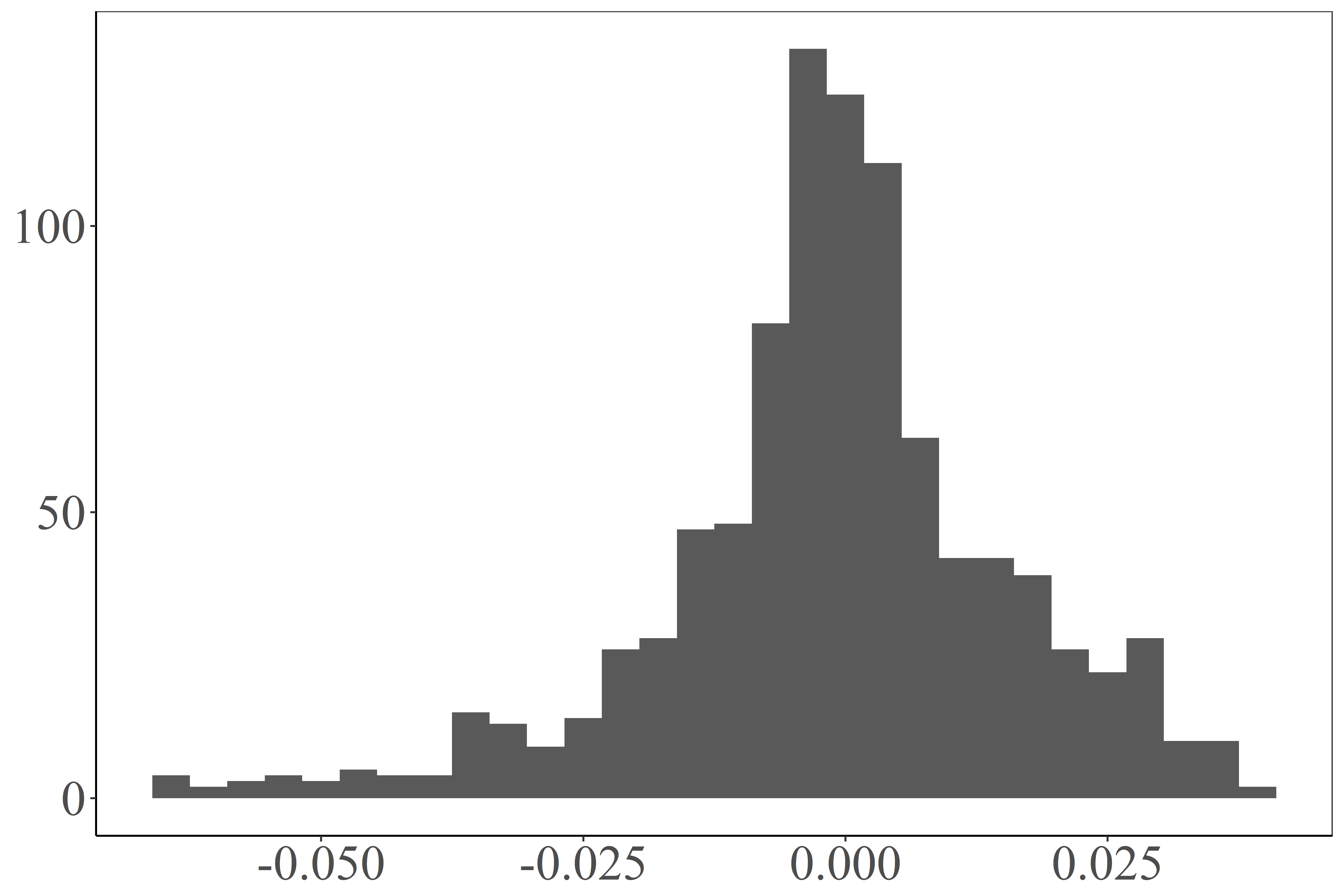}
		\subcaption{Patient 6}  
	\end{subfigure}
	\begin{subfigure}[b]{0.3\textwidth}
		\includegraphics[width=\linewidth]{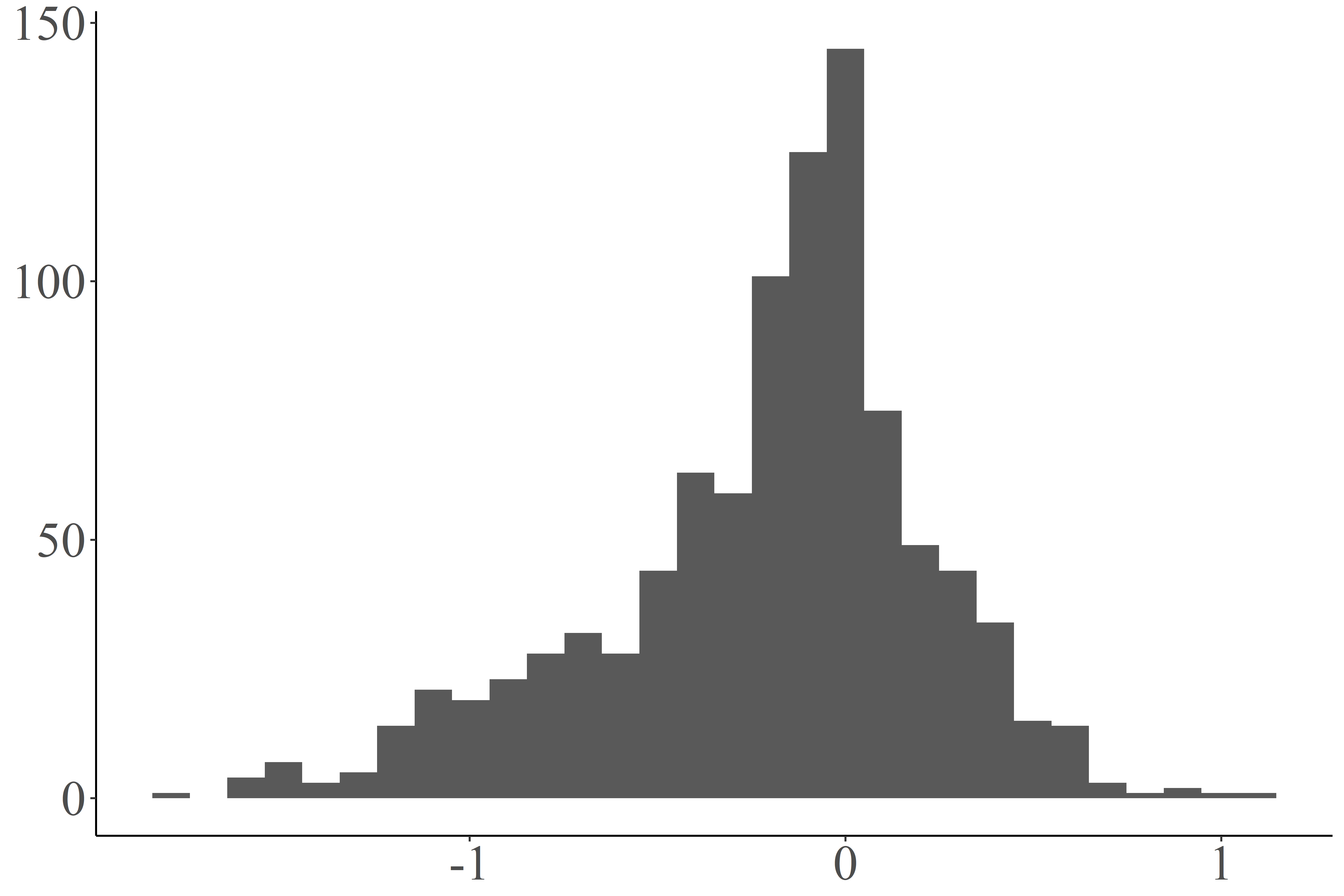}	   		
		\subcaption{Capno 32}  
	\end{subfigure}
	\caption{{\bfseries Residuals - series.} Residuals of model fit, obtained as difference between the fit and the original data.}\label{fig:Residuals}	
\end{figure}

\begin{figure}[h!]
	\centering
	\begin{subfigure}[b]{0.3\textwidth}
		\includegraphics[width=\linewidth]{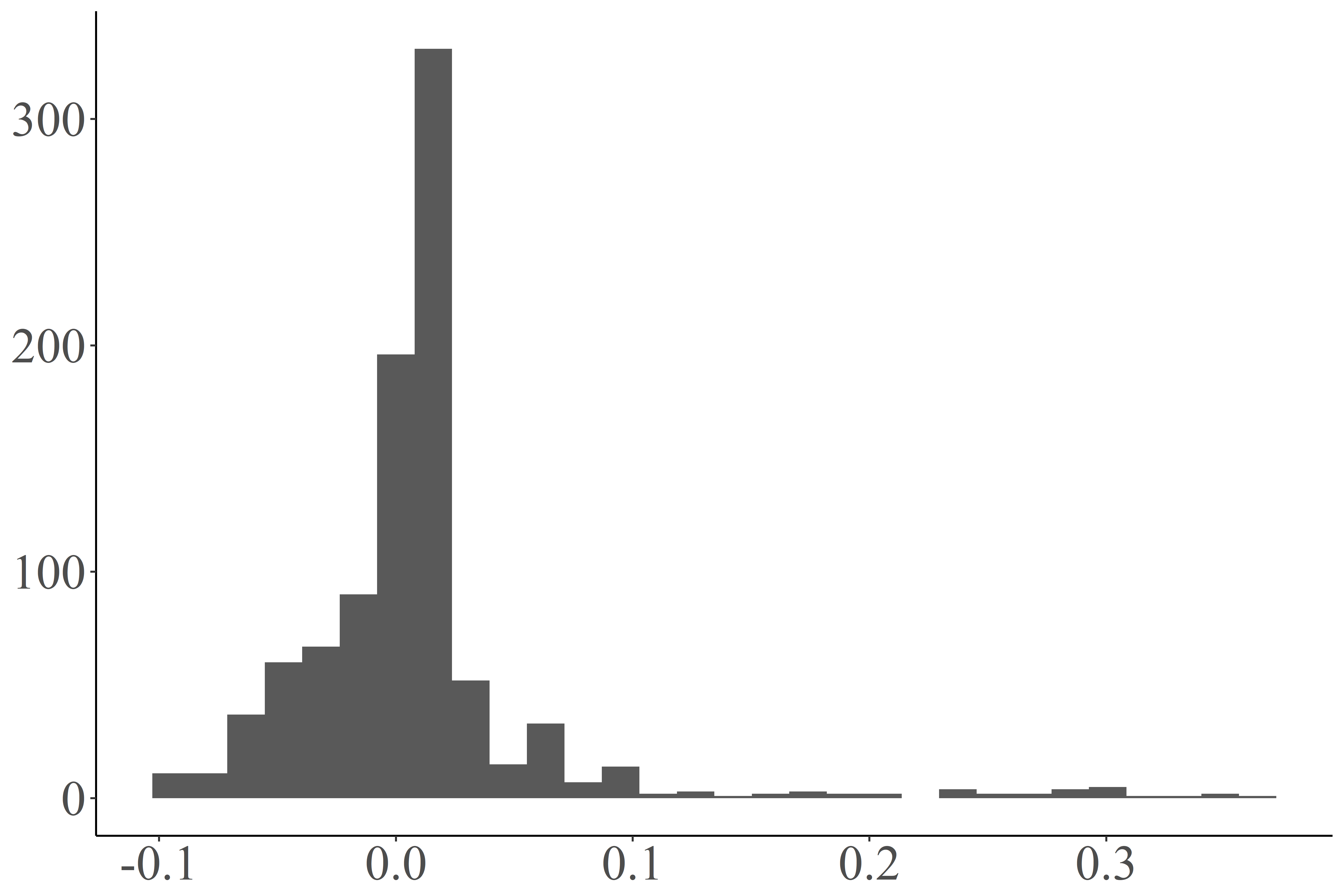}	   
		\subcaption{Patient 5}  
	\end{subfigure}
	\begin{subfigure}[b]{0.3\textwidth}
		\includegraphics[width=\linewidth]{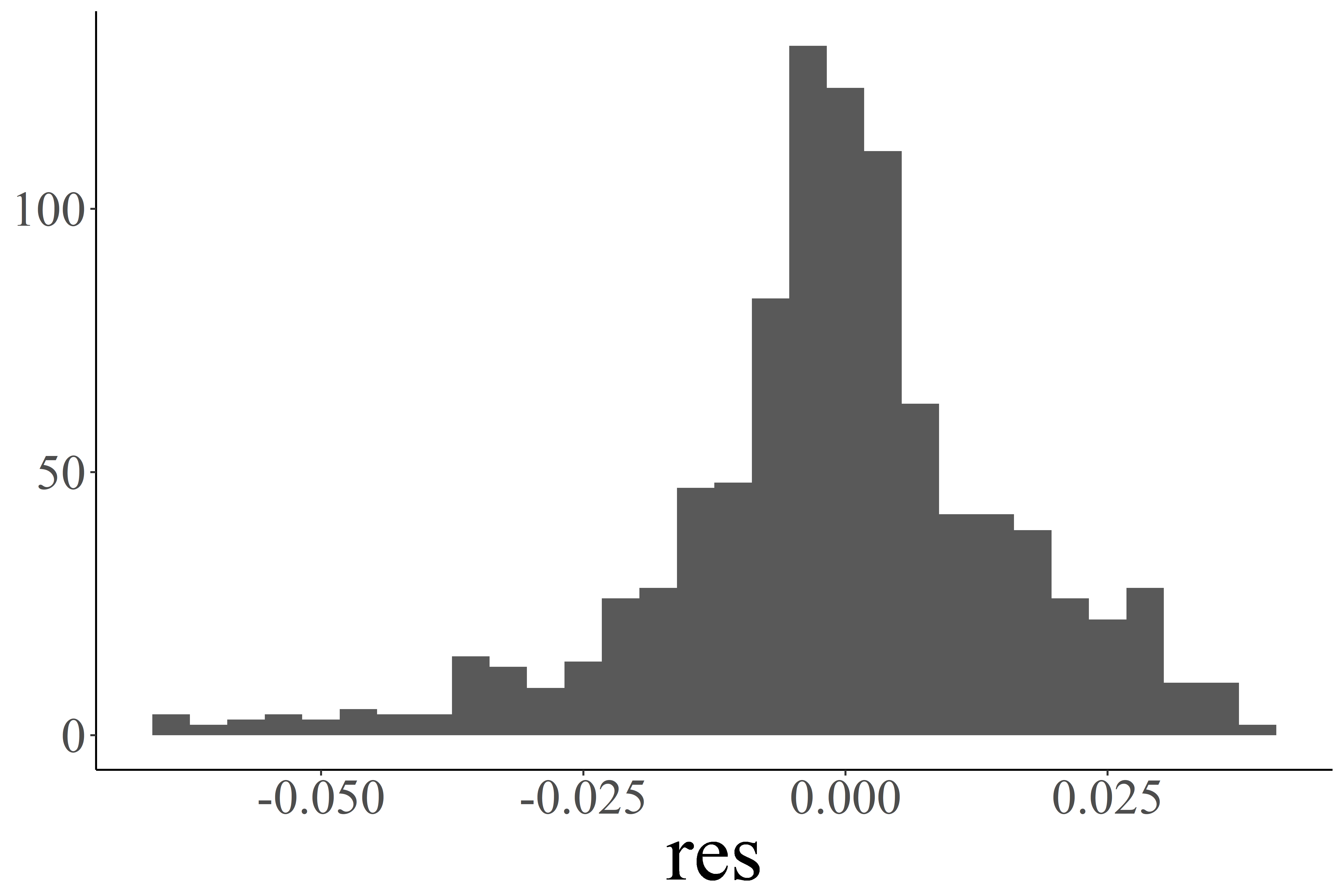}
		\subcaption{Patient 6}  
	\end{subfigure}
	\begin{subfigure}[b]{0.3\textwidth}
		\includegraphics[width=\linewidth]{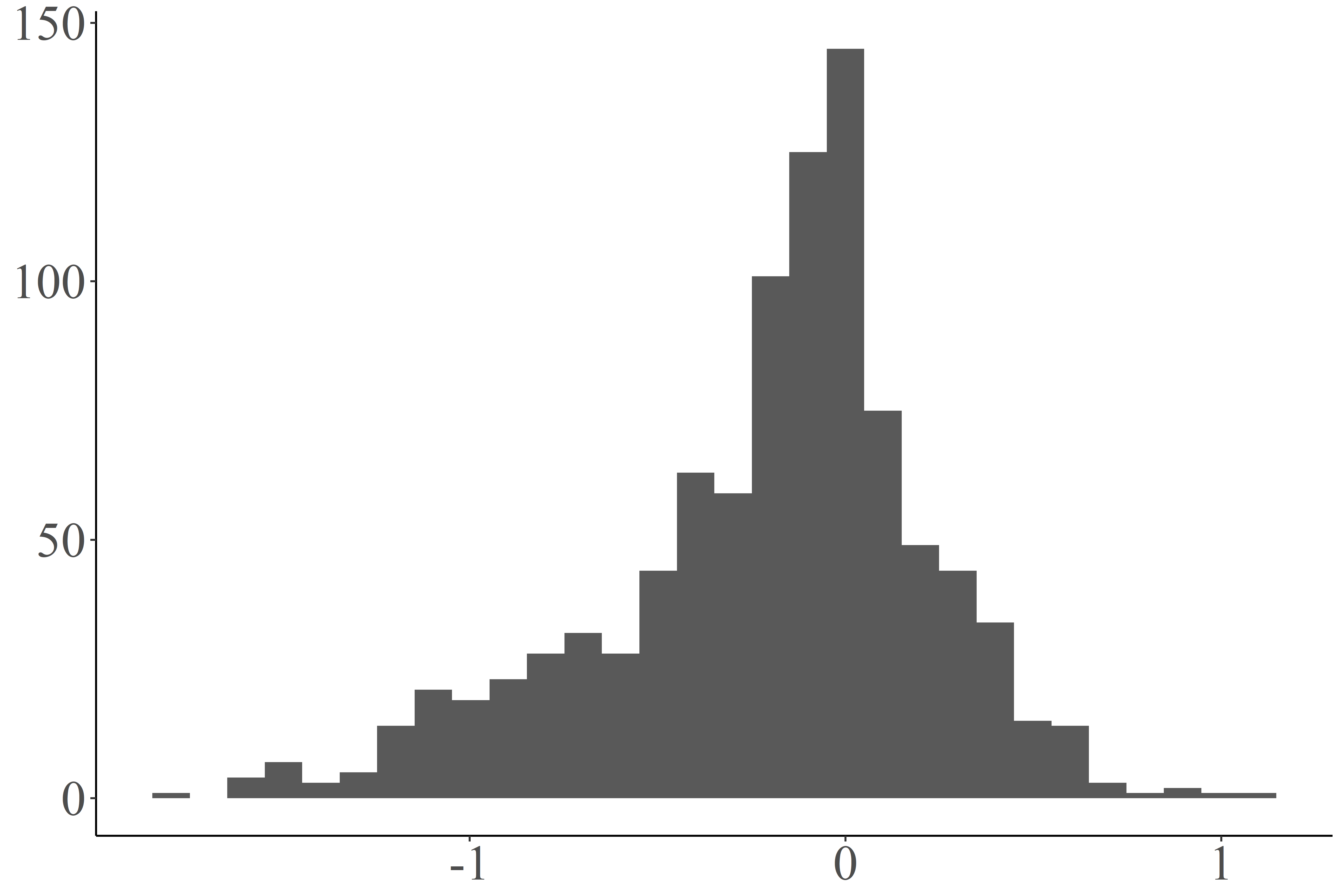}	   		
		\subcaption{Capno 32}  
	\end{subfigure}
	\caption{{\bfseries Residuals - histogram.} Histogram of the residuals from model fit, obtained as difference between the fit and the original data.}\label{fig:Residuals_hist}	
\end{figure}

\begin{figure}[h!]
	\centering
	\begin{subfigure}[b]{0.3\textwidth}
		\includegraphics[width=\linewidth]{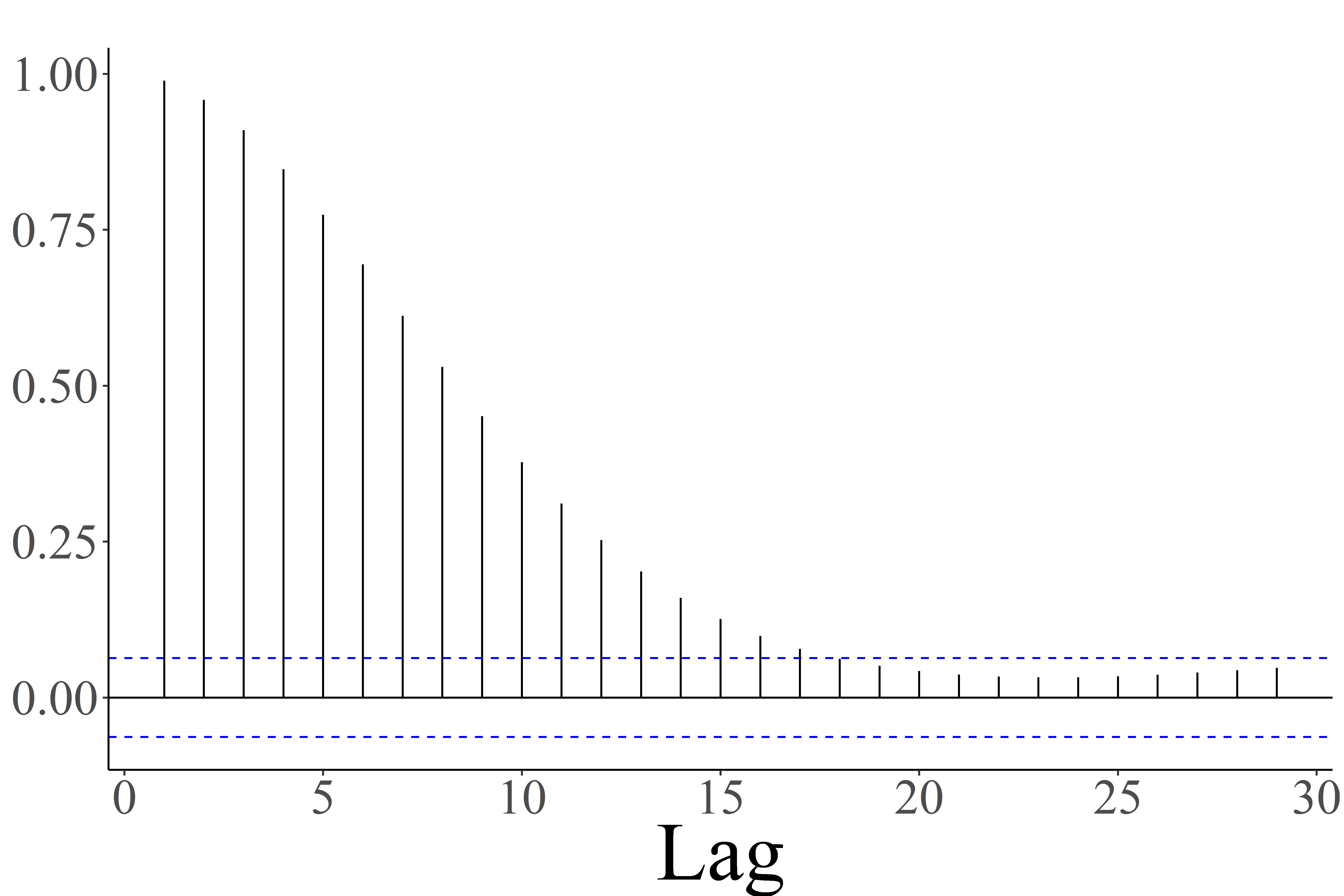}	   
		\subcaption{Patient 5}  
	\end{subfigure}
	\begin{subfigure}[b]{0.3\textwidth}
		\includegraphics[width=\linewidth]{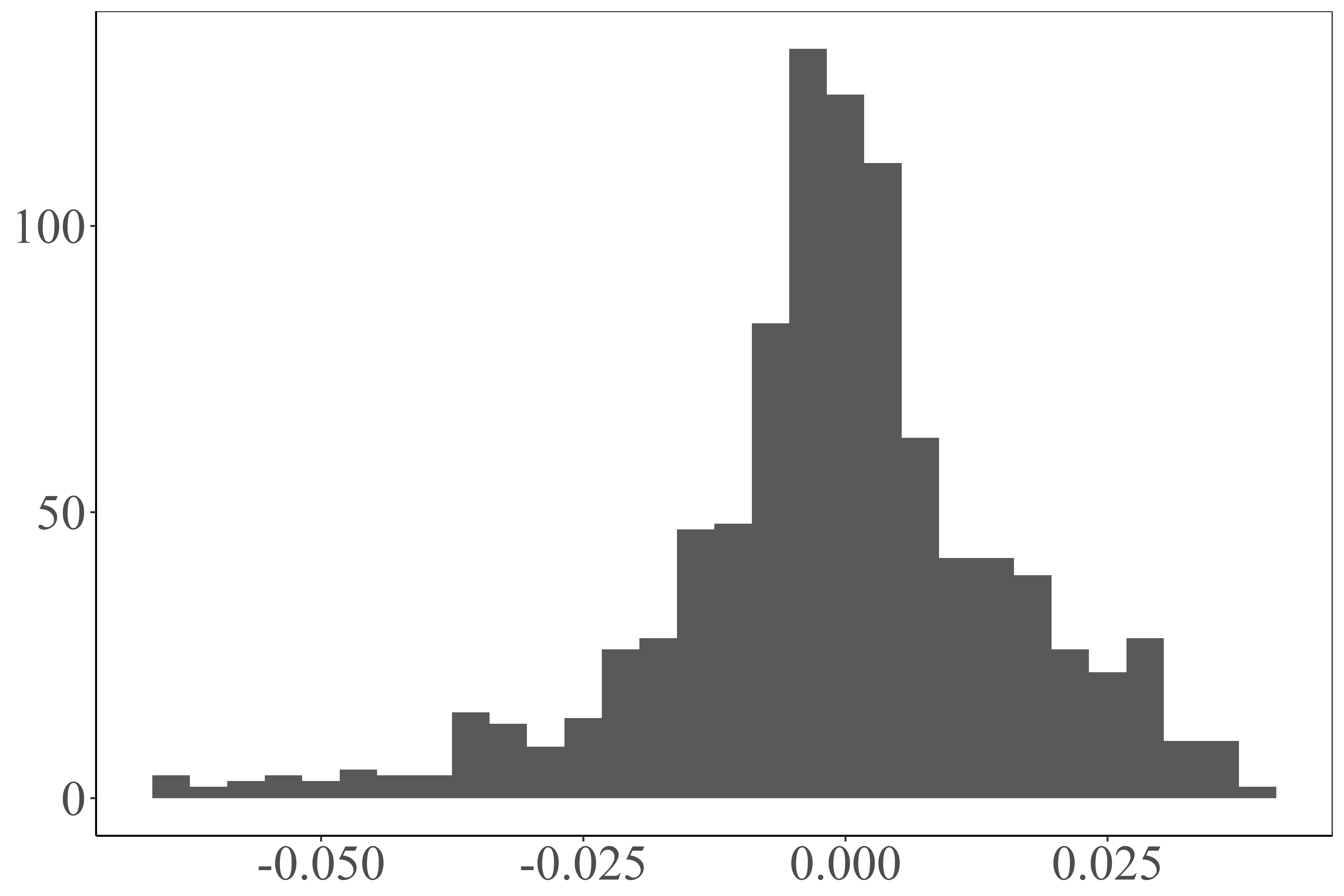}
		\subcaption{Patient 6}  
	\end{subfigure}
	\begin{subfigure}[b]{0.3\textwidth}
		\includegraphics[width=\linewidth]{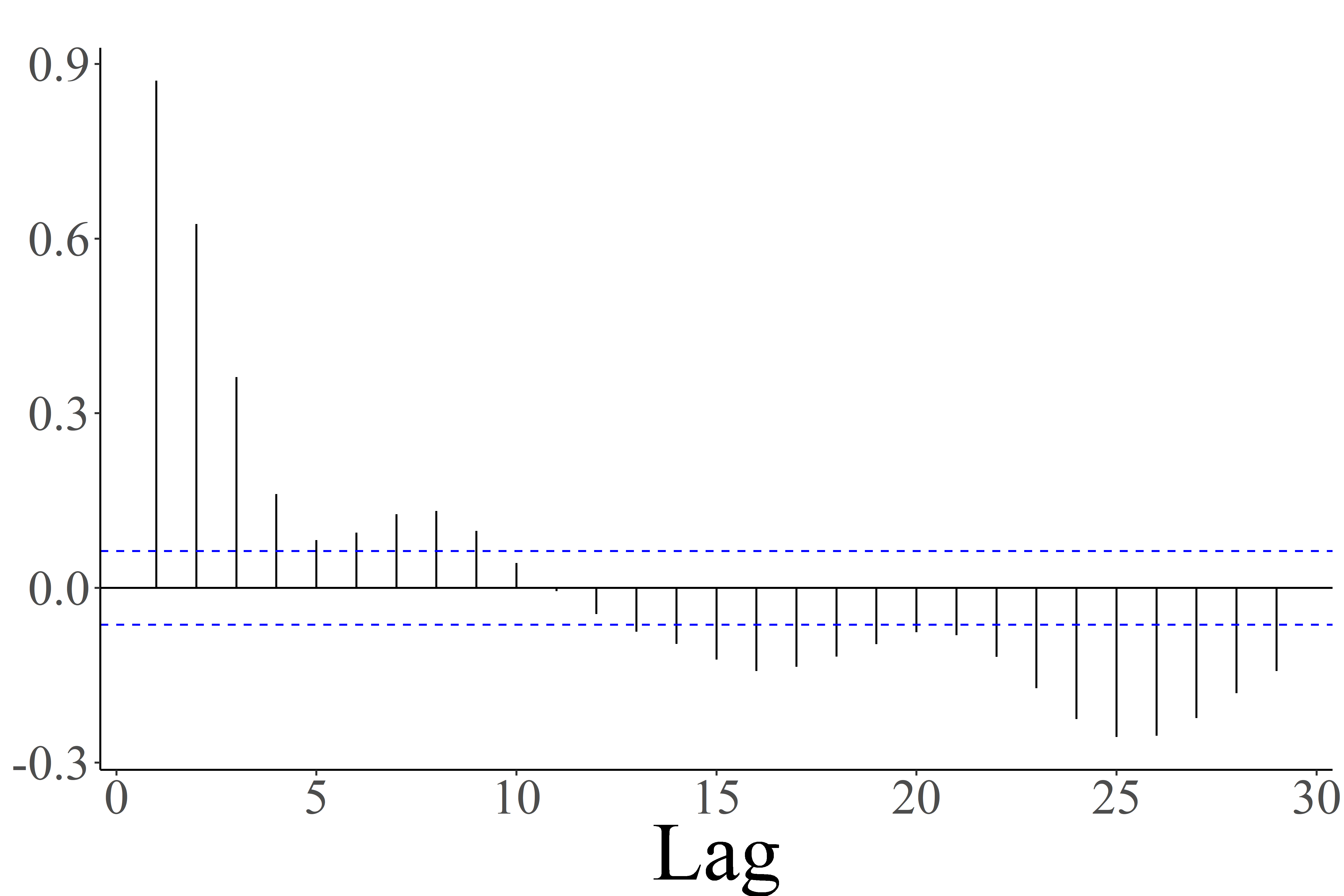}	   		
		\subcaption{Capno 32}  
	\end{subfigure}
	\caption{{\bfseries Residuals - acf.} Autocorrelation function of residuals from model fit.}\label{fig:Residuals_acf}	
\end{figure}

\begin{figure}[h!]
	\centering
	\begin{subfigure}[b]{0.3\textwidth}
		\includegraphics[width=\linewidth]{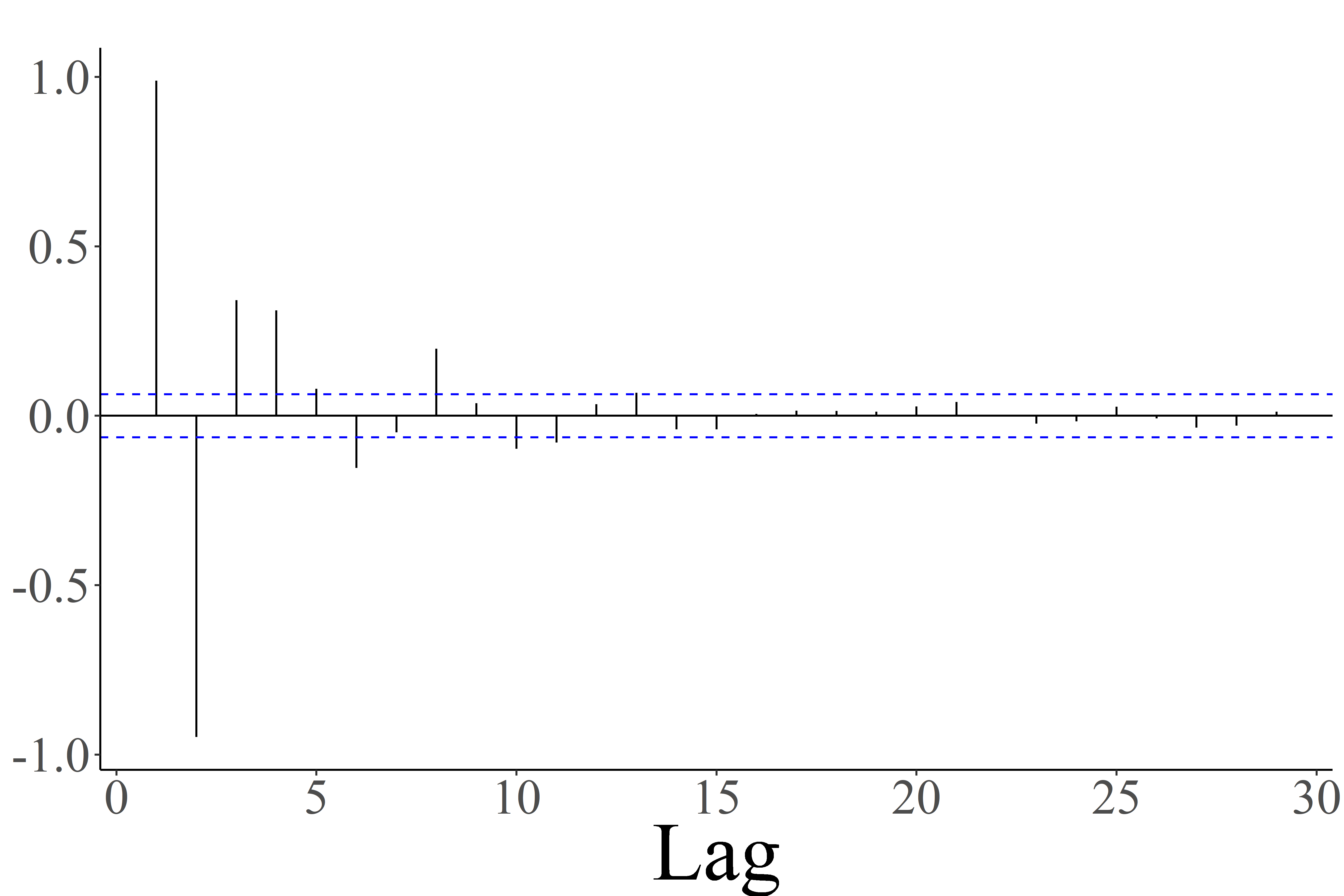}	   
		\subcaption{Patient 5}  
	\end{subfigure}
	\begin{subfigure}[b]{0.3\textwidth}
		\includegraphics[width=\linewidth]{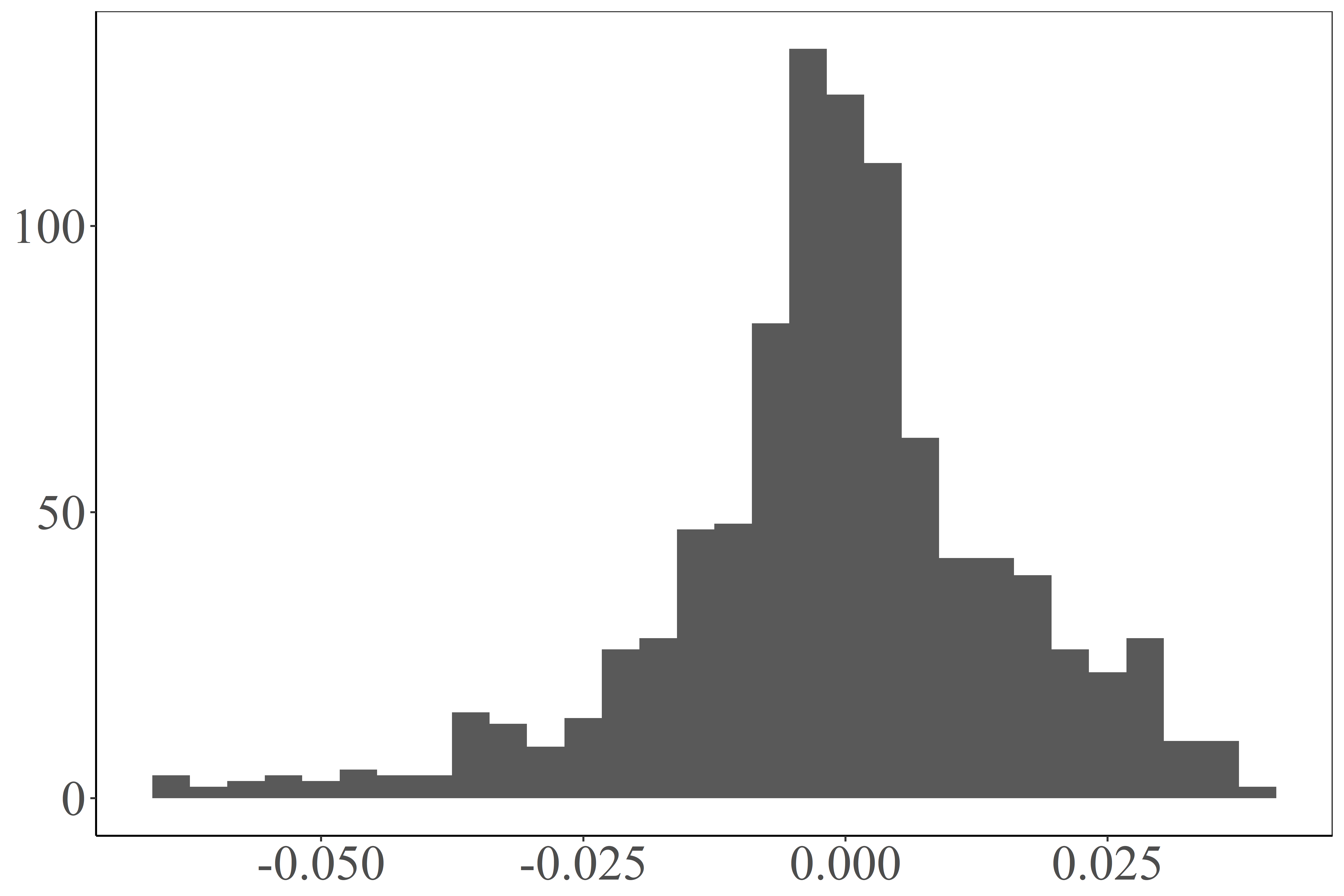}
		\subcaption{Patient 6}  
	\end{subfigure}
	\begin{subfigure}[b]{0.3\textwidth}
		\includegraphics[width=\linewidth]{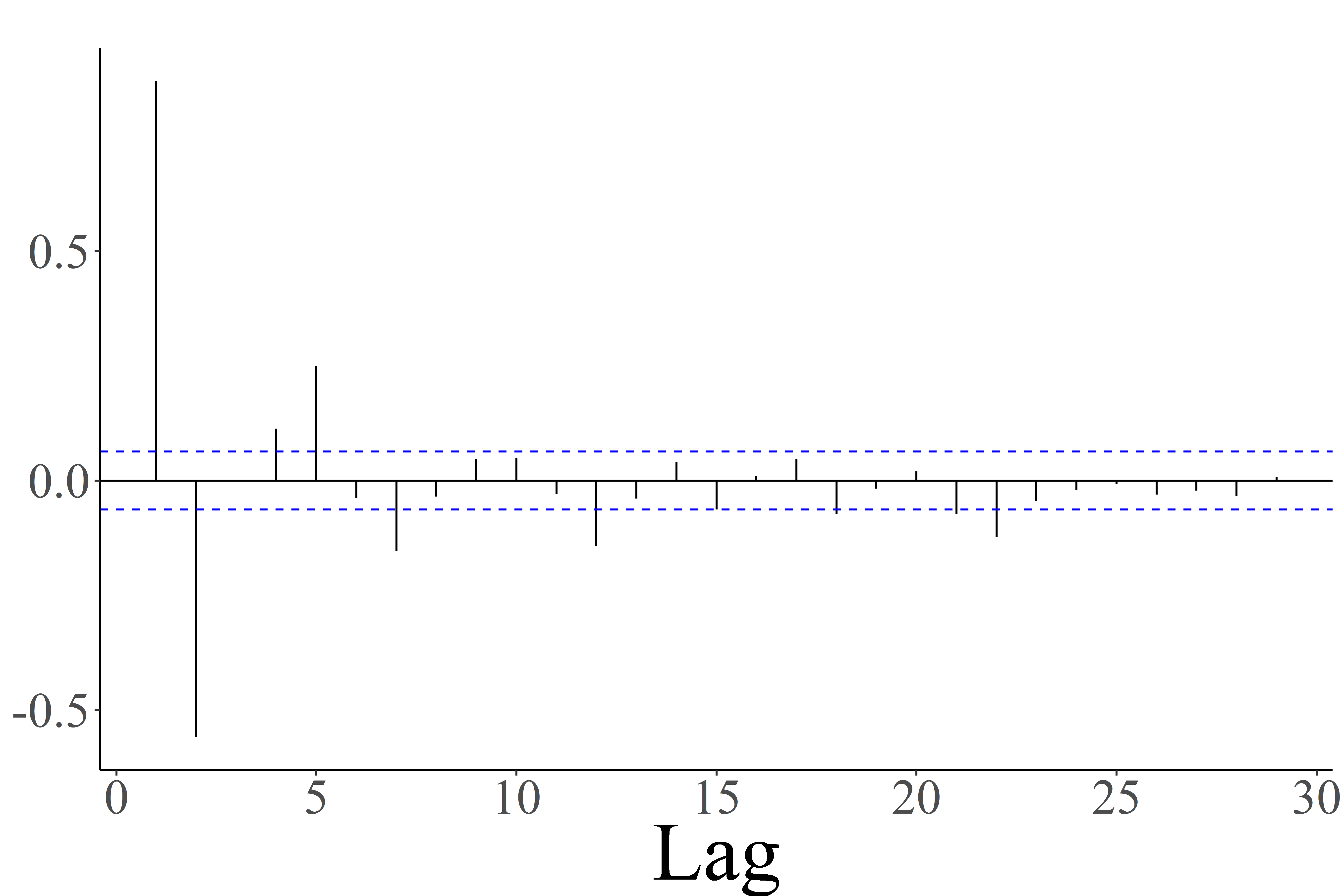}	   		
		\subcaption{Capno 32}  
	\end{subfigure}
	\caption{{\bfseries Residuals - pacf.} Partial autocorrelation function of residuals from model fit.}\label{fig:Residuals_pacf}	
\end{figure}

From Figure~\ref{fig:originalPPG} to Figure~\ref{fig:curveReg} we show the whole procedure, from the processed PPG signals, through the decomposition in Component I and II, to the curve registration (and the corresponding warping function) performed on the pulses that constitute Component II. Finally, in Figure~\ref{fig:Fit} we show the final fit against the original signals. The residuals (the difference between the fit and the processed signal) and some residual diagnostics are reported in Figure~\ref{fig:Residuals}-\ref{fig:Residuals_pacf}.
We will make use of a different notation for the original and fitted signal. Full lines will represent functional object, while dotted lines the observations. 


The residual plots in Figure~\ref{fig:Residuals} and their correlation functions (Figure~\ref{fig:Residuals_acf} and Figure~\ref{fig:Residuals_pacf}) show that residuals are still somewhat structured.
Correlations can be consequence of the simple assumptions made on the model residuals, or of the initial preprocessing procedure.
However, the magnitude of the residuals is small in absolute terms (see the histogram in Figure~\ref{fig:Residuals_hist}), and centred around zero.
The user can thus decide whether to increase model complexity/change the processing procedure, or deliberately use the simpler model while taking the consequences on the residuals into account.

The goodness of fit is also evaluated in terms of relative root mean squared error (relative RMSE) for the two datasets separately. The relative RMSE is defined as the ratio between the root mean squared error and the standard deviation of the data. It thus expresses how the residuals of the fit compare with the variability in the dataset.
Information about the internal dataset refers to the fit on the 6 minutes segments from each of the 25 patients, while information about the CapnoBase dataset refers to the whole available segments (8 minutes long) from the 9 patients.
As visible in Table~\ref{tab:relMSE}, the values are overall reasonable. One may note, however, that the statistics are increased by the fit on data from patients with rhythm irregularities and signal nuisances.

\begin{table}[h!]
	\small\sf\centering
	\caption{{\bfseries Relative RMSE.} Summaries relative to internal and CapnoBase datasets of the relative root mean square error. It is defined as the ratio between the root mean squared error and the standard deviation of the data.}\label{tab:relMSE}
	\begin{tabular}{ccccccc}
		 \hline
		Dataset  & Min    & 1st Qu. & Median  & Mean    & 3rd Qu. & Max.\\
		 \hline
		Internal & 3.54\% &7.35\%   & 11.10\% & 11.50\% & 13.57\% & 29.05\% \\
		CapnoBase& 5.53\% &10.33\%  & 12.12\% & 13.68\% & 12.57\% & 32.62\% \\    
	     \hline
	\end{tabular}
\end{table}

\subsection{Robustness}
\begin{figure}[h!]
	\centering
	\includegraphics[width=0.7\linewidth]{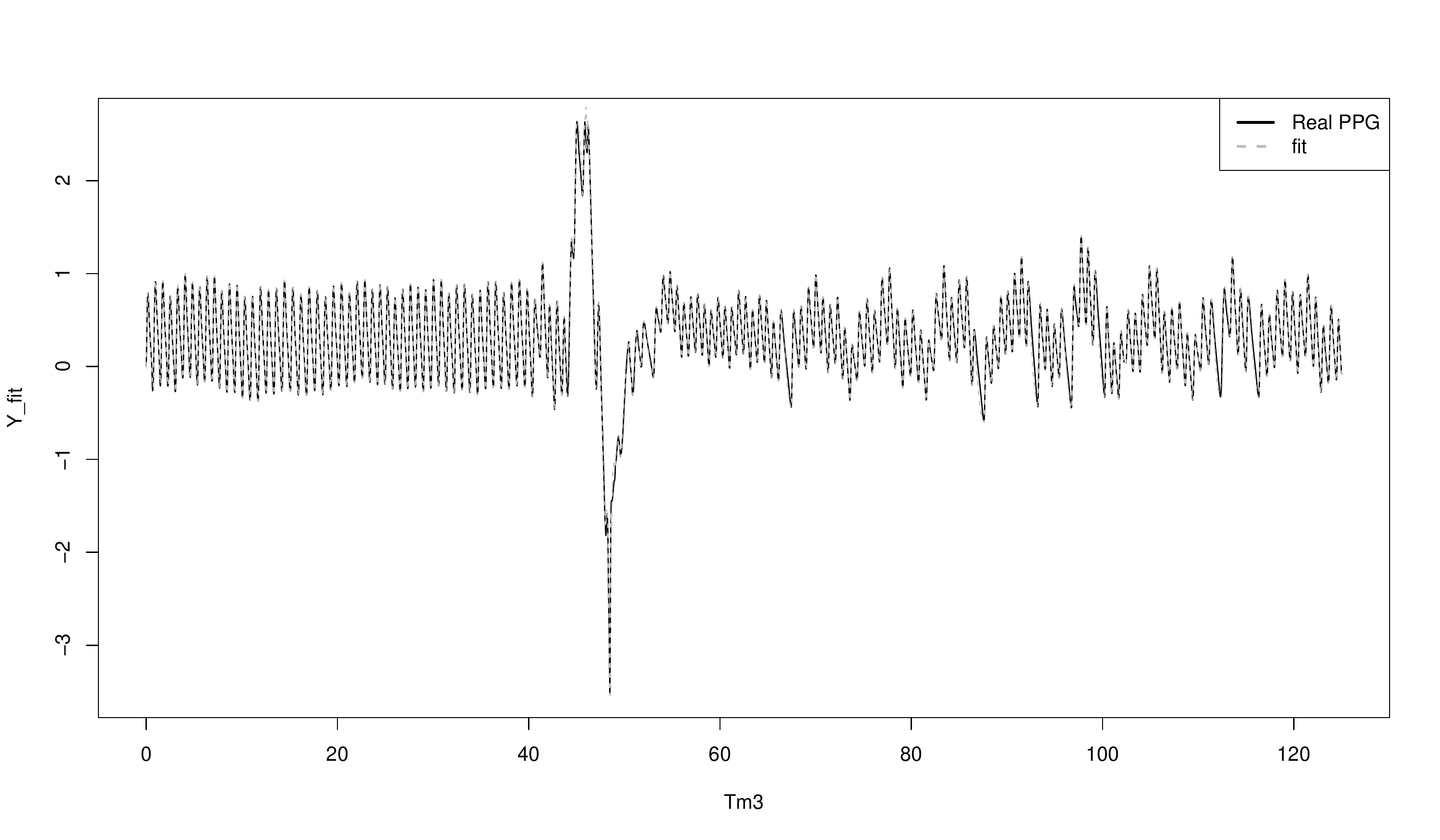}
	\caption{{\bfseries Segment of PPG signal with position change.} The long-term periodicity and the pulse shape seem to depend on wrist position.}	\label{fig:Datachangeposition}
\end{figure}

\begin{figure}[h!]
	\centering
	\begin{subfigure}[b]{0.49\textwidth}
		\centering
		\includegraphics[width=0.7\linewidth]{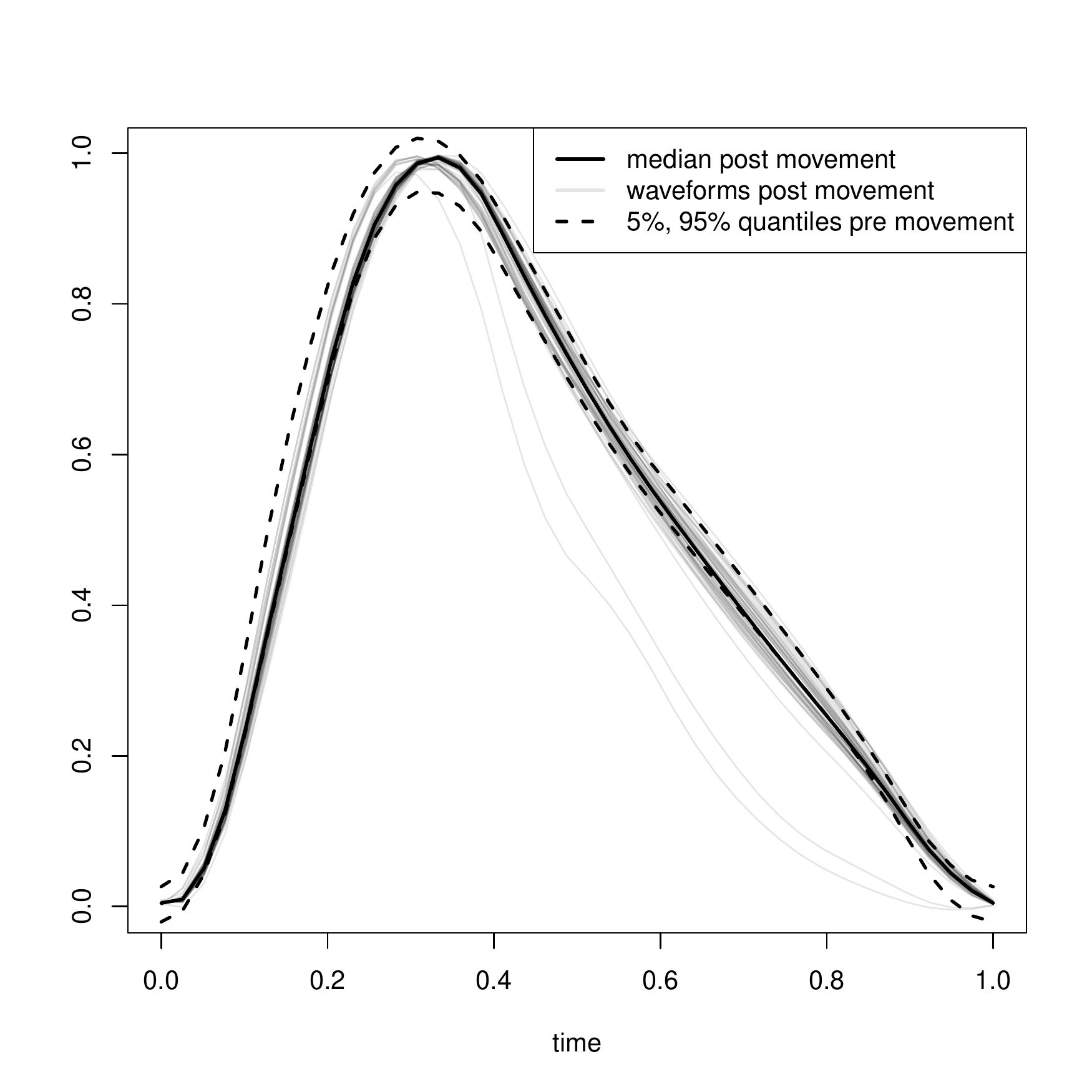}
		\subcaption{Post-movement median against 5\% and 95\% quantiles pre-movement}  
	\end{subfigure}
	\begin{subfigure}[b]{0.49\textwidth}
		\centering
		\includegraphics[width=0.7\linewidth]{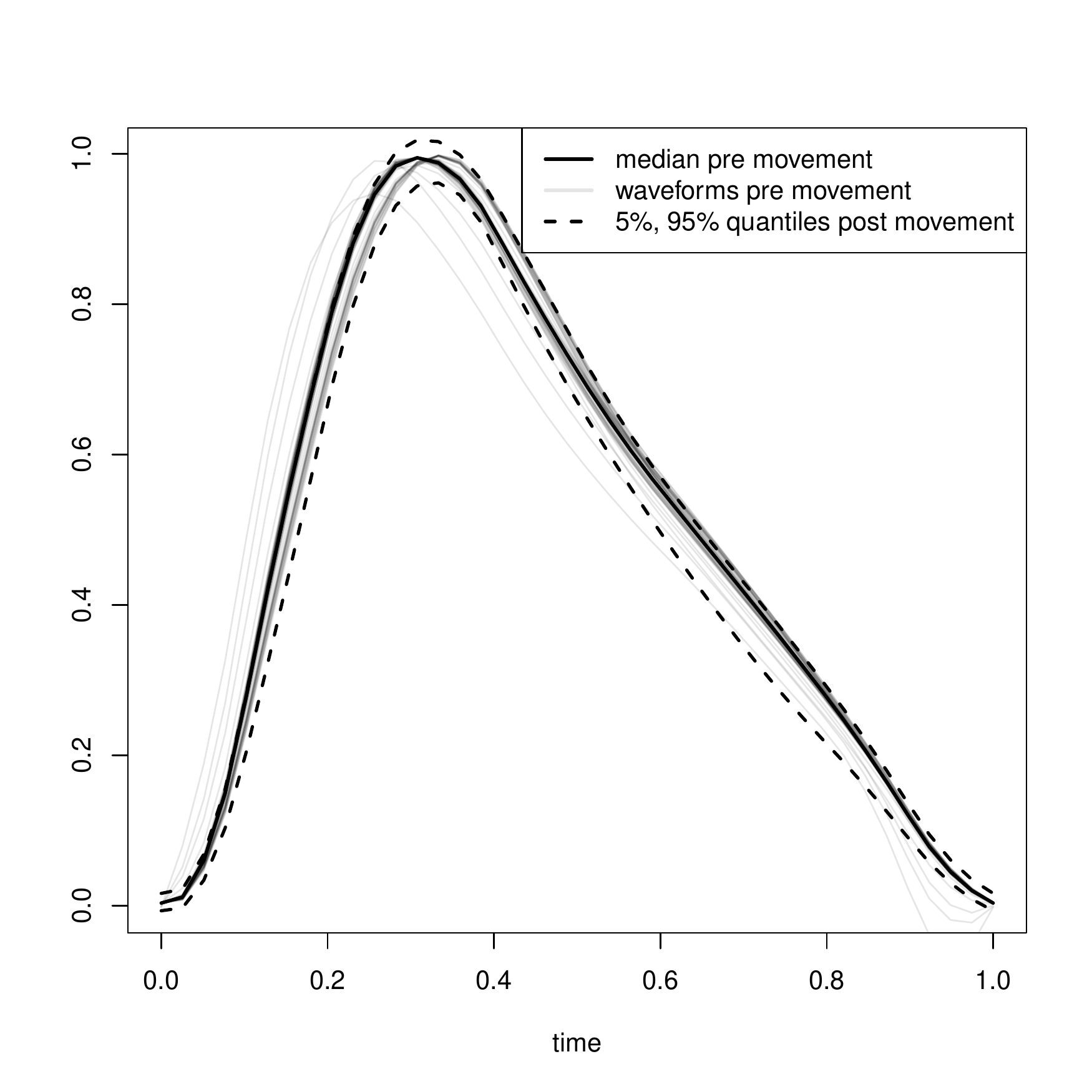}
		\subcaption{Pre-movement median against 5\% and 95\% quantile post-movement}  
	\end{subfigure}
	\caption{{\bfseries Effect of position change on model output.} Distribution of shape parameters before and after position change.
		Example illustrated from patient 1 of the internally collected dataset.}	\label{fig:PPGchangeposition}
\end{figure}

Since the focus of the present work is on modelling the shape and its characteristics, robustness of the method with respect to movement artefacts is of essential importance.
Here we show that the method can handle data with changes in wrist positions.
As visible for example from the fit in Figure~\ref{fig:Datachangeposition} (data from Patient 1 of the internally collected dataset, without AF), the model can adjust to the different amplitudes before and after movement.
More importantly, we observe that the distributional information we derive on the shape is not dependent on wrist position, and can thus be used to derive physiological insight. 
Analogous conclusions have been drawn after fitting the model to data from other patients.
In Figure~\ref{fig:PPGchangeposition} we show the median of 50 pulses after movement as obtained from the estimated parameters together with the 5\% and 95\% quantiles of the 50 pulses before perturbation, and reversely the median of the pulses before movement against the 5\% and 95\% quantile pulses after movement. 

\subsection{Inference: inter-beat-intervals}\label{res:IBIs}
In this section we show the accuracy achieved in estimating the IBIs as the distance between the local minima.
In Table~\ref{tab:IBIs} for each patient we report the correlation between the IBIs obtained through model fitting and the IBIs from ECG, when the misdetected contractions are excluded.
Furthermore, we compute the relative absolute error as the ratio between the absolute difference between the two IBIs and the true ones. Finally, we report the percentage of misaligned contractions, both missed and erroneously detected.

\begin{table}
	\small\sf\centering
	\caption{{\bfseries Accuracy in determining the IBIs as function of the fitted parameters.} Correlation and relative absolute error between IBIs from model fitting and from ECG, excluding misdetected contractions (missed and extra, reported in percentage). At the bottom mean and standard deviation of these values.}\label{tab:IBIs}	
	\begin{tabular}{cccccc}  
		 \hline
		patient & AF &  correlation & relative absolute error  &   missed (\%)  &  extra (\%)\\
		 \hline
		1  &   & 0.99 & 0.019 &   7.01 & 1.13\\
		2  & X & 0.97 & 0.091 &  19.28 & 16.42\\
		3  &   & 0.98 & 0.017 &   5.88 & 8.98\\
		4  &   & 0.97 & 0.006 &   0.31 & 0.00\\
		5  &   & 0.95 & 0.022  &   3.25 & 4.64\\
		6  & X & 0.98 & 0.071 &  15.24 & 6.47\\
		7  &   & 0.99 & 0.013 &   0.74 & 1.23 \\
		9  & X & 0.99 & 0.031 &   0.85 & 0.85\\
		10&    &0.99 &  0.012  &  4.66  & 1.86\\
		11&    & 0.99 & 0.023 & 15.89 & 24.74\\
		12& X  & 1.00 & 0.029 & 17.79 & 33.56\\
		13&    & 0.98 & 0.008& 0.75   & 0.00\\
		14&    & 0.98 & 0.011& 2.08   & 2.97\\
		15&    & 0.99 & 0.013& 3.89   & 0.24\\
		17&    & 0.98 & 0.010& 0.55   & 0.55\\
		18&    & 0.96 & 0.019& 3.69   & 4.80\\
		20& X & 0.99 &  0.050& 13.56  & 17.09\\
		22&    & 0.99 & 0.012& 4.47   & 7.67\\
		23&    & 0.97 & 0.008& 0.32   & 0.00\\
		24&    & 0.95 & 0.008& 5.62   & 11.02\\
		26&    & 0.99 & 0.009& 0.27   & 0.00\\
		27&    & 0.96 & 0.008& 0.34   & 0.34\\
		28& X & 0.98 &  0.033& 1.13   & 0.38\\
		29& X &  0.99 & 0.031& 1.94   & 1.55\\
		30& X & 1.00 &  0.040& 33.21 & 37.27 \\
		 \hline
		&& 0.98   $\pm$  0.01  & 0.027  $\pm$  0.026 & 7.70  $\pm$  10.06  & 8.16  $\pm$  11.18\\
		 \hline
	\end{tabular}
\end{table}

As visible, the correlation between the estimated and the true values is overall high, but this is influenced by the fact that we are excluding all the misspecified minima. In fact, mainly under atrial fibrillation the rates of misaligned contractions can be very large.

This example shows how IBIs can be derived from the fitted parameters. Other functionals can be foreseen that link the parameters to certain shape characteristics. Examples of informative measures include the peak-to-peak interval, the pulse amplitude, the area under the curve and the point of maximal slope.

\subsection{Detection of premature contractions}
In this section we discuss the performance of our model-based methods for detecting premature contractions.
First we report the results of the SVM approach to classify premature contractions by using the information from the fit, possibly combined with the original data. 
We show the results of this classification for various choices of predictors and kernels, both in case of two-class classification (sinus rhythm/premature contraction) in Table~\ref{tab:2class} and in case of three-class classification (sinus rhythm/premature atrial contraction/premature ventricular contraction) in Table~\ref{tab:3class}.
For the first, we report sensitivity, specificity and overall accuracy defined as
\begin{equation}
\text{accuracy} = \frac{\sum_i C_{ii}}{\sum_{ij}C_{ij}},\label{eq:accuracy}
\end{equation}
where $C$ is the confusion matrix in which the rows are the true values, and the columns are the predicted ones.
For the three-class classification, only patient 10 was included as it was the only patient showing multiple premature contractions of both types in the selected time window.
In this case we report the sensitivity of each class against the rest,
\begin{equation}
\text{sen}_i = \frac{C_{ii}}{\sum_{j}C_{ij}},
\end{equation}
and the accuracy defined in~\eqref{eq:accuracy}.

\begin{table}
	\small\sf\centering
	\caption{{\bfseries SVM two-class classification.} Average and standard deviation of out-of-sample accuracy, sensitivity and specificity of the SVM method in the classification of normal sinus rhythm against premature contractions.}\label{tab:2class}
	\begin{tabular}{ccccc}
		 \hline
		kernel & predictor & accuracy (\%) & sensitivity (\%) & specificity (\%)\\
		 \hline
		linear & $(\bm Y_{i}$, $\bm\epsilon_i$, $\bm X_i)$  &	97.88  $\pm$  2.36 & 50.04  $\pm$  36.18  & 99.15  $\pm$  0.82 \\
		linear & $\bm X_i$                          &  98.15  $\pm$  2.25 & 42.86  $\pm$  40.63  & 99.51  $\pm$  0.64 \\		
		linear & $\bm\epsilon_i$                           &  97.49  $\pm$ 3.01 & 15.79  $\pm$  30.19 & 99.69  $\pm$  0.64 \\
		linear & $\bm Y_{i}$                           &  97.44  $\pm$  3.08 & 15.94  $\pm$  30.59 & 99.64  $\pm$  0.78 \\ 
		 \hline
	\end{tabular}

\end{table}

\begin{table}
	\small \sf\centering
	\caption{{\bfseries SVM three-class classification.} Accuracy and sensitivity in the detection of normal sinus rhythm (sin), premature atrial and ventricular contractions (pac and pvc respectively) against the rest. Results refer to patient 10 from the internally collected dataset. Under predictor, between brackets: next refers to using the shape parameters relative to the next pulse to predict the label of the current pulse, and (paired, next) means using the features from the next two pulses to predict the label of the current pulse.}\label{tab:3class}
	\begin{tabular}{cccccc}
		 \hline
		kernel & predictor & acc (\%) & sen sin (\%) & sen pac (\%) & sen pvc (\%)\\
		 \hline
		polyn & $\bm X_i$ (next)         & 96.23 & 97.68 & 50.00 & 71.43\\	
		polyn & $\bm X_i$ (paired, next) & 95.27 & 98.34 & 50.00 & 35.71\\	
		linear     & $\bm X_i$ (paired, next) & 94.95 & 98.34 & 100	 & 21.43\\
		
		 \hline	
	\end{tabular}
\end{table}
The use of shape parameters ($\bm X_i$) as input is shown to provide good accuracy for both the two- and three-class classification. Furthermore, predictions can be improved by combining this information with the residuals from the fit ($\bm \epsilon_i$) and the data itself ($\bm Y_i$) in the two-class classification (cf.~Table~\ref{tab:2class}).

\begin{figure}[h!]
	\centering
	\begin{subfigure}[b]{0.3\textwidth}
		\centering
		\includegraphics[width=\linewidth]{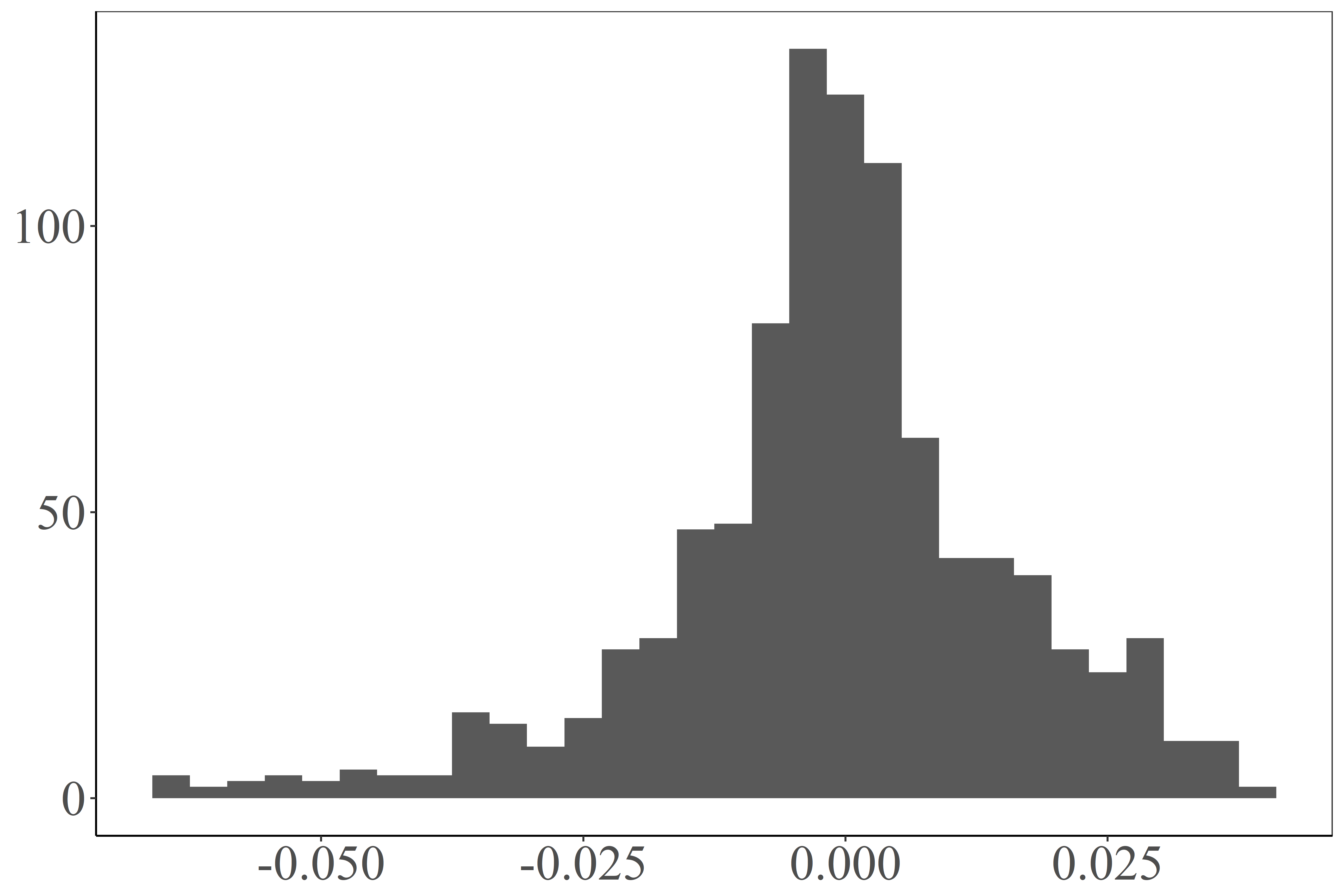}
		\subcaption{$\lambda$}  		
	\end{subfigure}
	\begin{subfigure}[b]{0.3\textwidth}
		\centering
		\includegraphics[width=\linewidth]{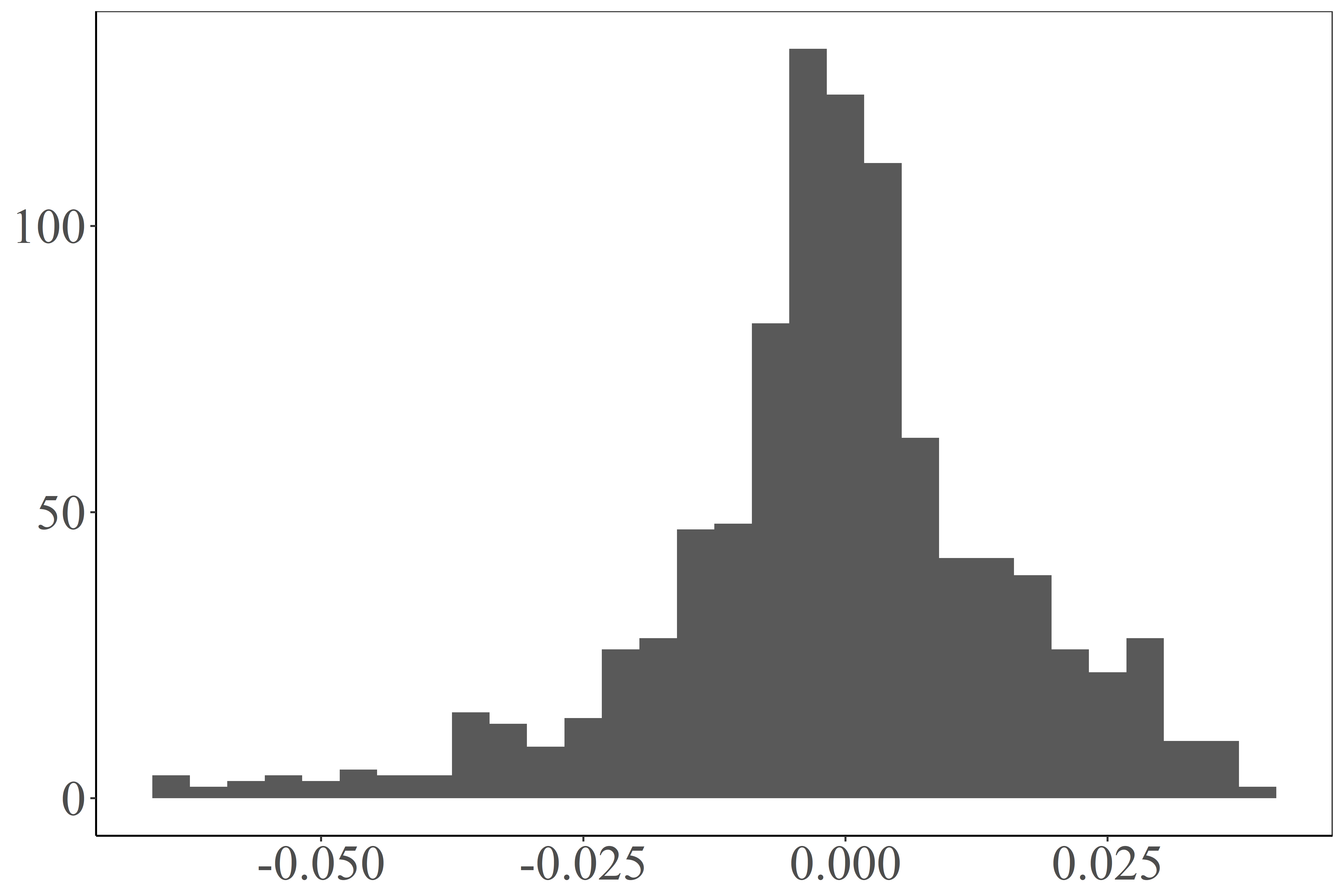}
		\subcaption{$\rho$}
	\end{subfigure}
	\begin{subfigure}[b]{0.3\textwidth}
		\centering
		\includegraphics[width=\linewidth]{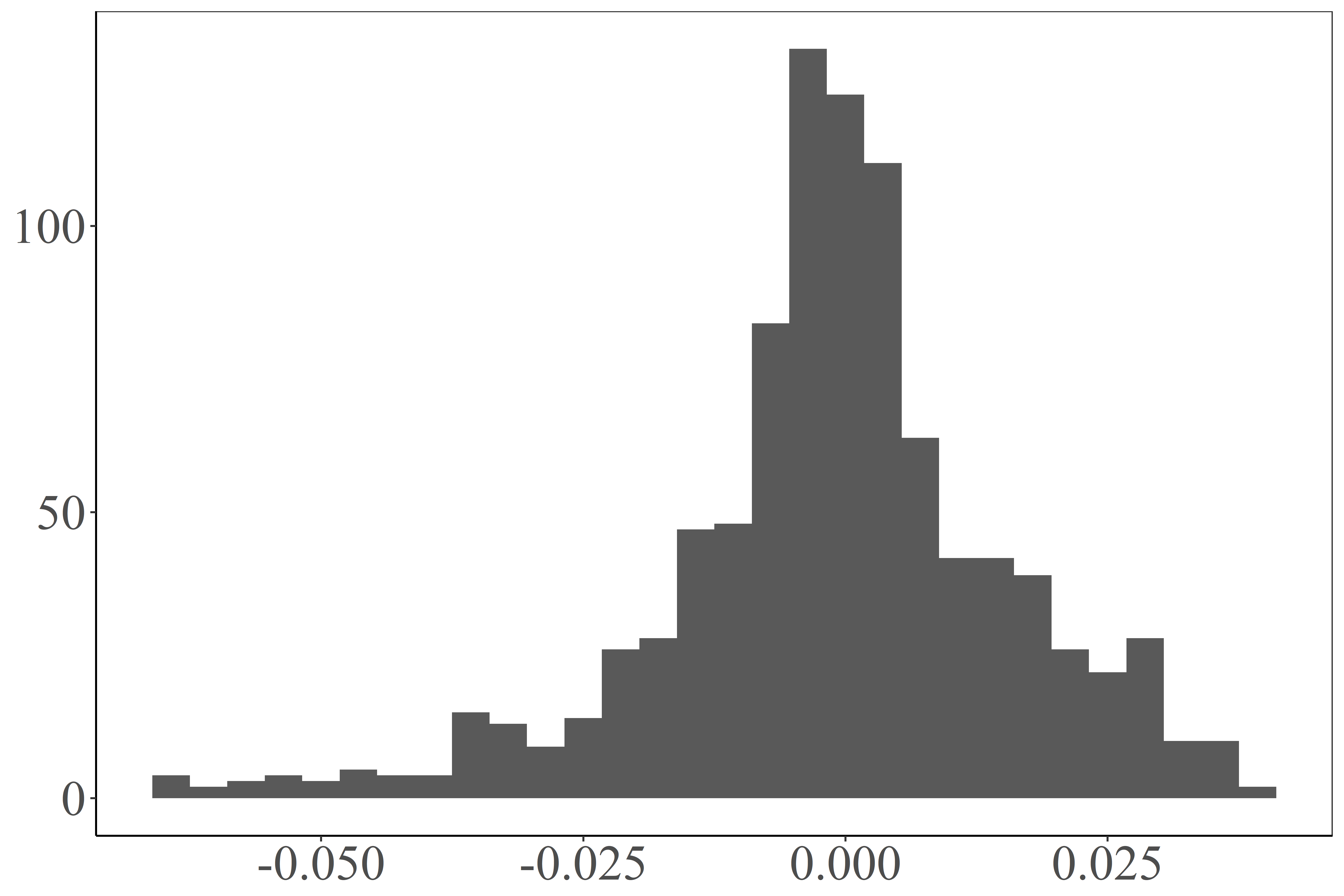}
		\subcaption{$\pi$}  		
	\end{subfigure}		
	\caption{{\bfseries  Histogram of the optimal values for the parameters.}}\label{fig:histParam}	
\end{figure}

In the following we show the performance of the proposed threshold method for the detection of premature contractions.
The procedure involves optimization of three parameters, namely the window length for moving average smoothing $\lambda$, the radius of the interval to be considered for the detection of local maxima $\rho$ and the threshold $\pi$ for the detection of premature contractions. 
Parameter optimization with leave-one-out cross-validation from a grid of possible values led to an optimal set of parameters.
For each left-out patient, a certain combination of $\lambda,\ \rho$ and $\pi$ was found, and the values are shown in the histogram in Figure~\ref{fig:histParam}.

As visible, these values are highly concentrated. We choose thus to fix these values to the medians $\bar{\lambda}=$45, $\bar{\rho}=$55, $\bar{\pi}=$0.051 for the remainder of the analysis.
The results from the detection of premature contractions with the threshold method with these median values for parameters are shown in Table~\ref{tab:thresholdmethod}. Note that the procedure does not distinguish between premature atrial and ventricular contractions.
\begin{table}[h!]
	\small \sf\centering
	\textbf{\caption{{\bfseries  Detection accuracy of threshold procedure on the internal dataset when the parameter values are fixed to the median values.} 
			For each patient we report the number of premature atrial and ventricular contractions (pac and pvc respectively), accuracy, sensitivity and specificity of the detection method.
			In the last row, the average and standard deviation of accuracy, sensitivity and specificity.}\label{tab:thresholdmethod}}
	\begin{tabular}{cccccc}
		 \hline
		patient & n. pac & n. pvc & accuracy (\%) & sensitivity (\%) & specificity (\%) \\
		 \hline
		1	    & 44      & 0                & 98.19     & 100.00        & 97.98	\\
		3       & 13      & 0                & 71.12     & 100.00        & 69.90  \\
		4       & 1       & 0                & 99.69     & 100.00       & 99.69  \\
		5       & 3       & 1                & 92.79     & 100.00        & 92.72  \\
		7       & 41      & 0                & 90.12     & 2.44        &	100.00\\
		10      & 2       & 16               & 91.59     & 88.89       &91.75\\
		11      & 1       & 0                & 73.89    & 100.00       & 73.82\\
		13      & 0       & 2                & 100.00     & 100.00       & 100.00\\
		14      & 2       & 0                & 95.24    & 100.00        & 95.21\\
		15      & 0       & 17               & 98.29     & 93.75        &	98.48\\
		17      & 0       & 1                & 97.22     & 100.00       & 97.21\\
		18      & 0       & 2                & 96.86     & 100.00       & 96.85\\
		22      & 25      & 0                & 76.92     & 64.00        &78.05\\
		23      & 2       & 0                & 93.91     & 50.00        & 94.19\\
		24      & 3       & 0                & 90.69     & 66.67        &90.85\\
		26      & 22      & 0                & 95.38     & 31.82        &99.42\\
		27      &2        & 0                & 85.03    & 100.00      & 84.93\\
		 \hline
		& && 	91.00  $\pm$  8.77 &   82.21  $\pm$  28.55 & 	91.83  $\pm$  9.24\\
		 \hline
	\end{tabular}
\end{table}

The detection procedure performs reasonably well on all the patients, without marked differences in the detection of premature atrial or ventricular contractions. Exceptions are patients $7$ and $26$ that show long sequences of premature contractions without alternation to regular contractions.
The premature contractions detected by the threshold method are then classified into the two classes via SVM.
Among the possible choices of kernel functions and predictors, we found the linear kernel combined with a predictor including data $(\bm Y_{i})$, fitted shape parameters ($\bm X_i$) and residuals ($\bm\epsilon_i$) to deliver the best results. 
The average out-of-sample accuracy (obtained via cross-validation) is 56.01\% $\pm$ 31.40\%, the sensitivity for sinus rhythm versus the other classes is 71.92\% $\pm$ 29.78\%, while it is 36.31\% $\pm$ 40.95\% and 29.44\% $\pm$ 36.54\% for premature atrial and ventricular contractions, respectively.
For reference, we report the results from the classification of the premature contractions as detected by the ECG. 
Although these results do not directly compare with the results obtained on the detected premature contractions (where there is the addition of false positives), it is interesting to see the information contained in the predictors about the two types of premature contractions. The overall accuracy is 73.29\% $\pm$ 38.52\%. The sensitivity for premature atrial contractions is 79.82\% $\pm$ 36.15\% while this value for premature ventricular contractions is 54.17\% $\pm$ 46.58\%. This is achieved with sigmoid kernel and a combination of data, residuals and shape parameters as predictor $(\bm Y_{i}$, $\bm\epsilon_i$, $\bm X_i)$.

\section{Discussion}\label{sec:PPGDiscussion}
In this paper we have proposed a new model for photoplethysmography signals.
Motivated by the dataset in hand, including patients with various comorbidities and having different medications, we aimed at a flexible yet robust model.
This goal is achieved by combining curve registration and spline smoothing. 
Indeed, results show that the model can fit data measured with finger and wrist probes, from patients with very disparate conditions (including atrial fibrillation and premature contractions).
The temporal information is embedded together with the relationship between parameters via a state space representation, and estimation is performed applying an extended Kalman filter.
The algorithm is modular, and each step can be replaced to adapt the approach to the particular case study in hand.
The method is also shown to be robust to changes in wrist position, feature that is particularly desirable to make inference on the estimated parameters, since data from weareables is easily corrupted by such disruptions.

Residual diagnostics show patterns. The structure contained therein is due to the fact that residuals follow pulse shapes, and thus for each pulse, they lie all either above or below the observed signal. 
Furthermore, from the correlation plots we observe that residuals are not independent. Additional error structure and more spline parameters 
could greatly improve the fit and consequently the residual pattern. However, we avoid this because we want the results to be comparable with methods existing in the literature, and we want to use the lack of fit for clinical applications.

In particular, we show how residuals can be used to detect premature atrial and ventricular contractions.
Furthermore, by combining the information from the fit (in particular the shape parameters), we are also able to classify them into the two types.
With support vector machine we achieve 97.88\% $\pm$ 2.36\% accuracy in the classification of regular and premature contractions, by using a combination of shape parameters, residuals and data as predictors. Including only the shape parameters would lead to comparable results, while this would not hold in case of including only the residuals from the fit or the data.
We achieve furthermore 96.23\% accuracy in distinguishing normal sinus rhythm (sensitivity 97.68\%), atrial (sensitivity 50.00\%) and ventricular (sensitivity 71.43\%) premature contractions, although the data in hand allows this analysis on one patient only. It is interesting to see that in this case different combinations of the shape parameters alone are the best predictors.
Here the sensitivity in detecting premature ventricular contractions against the other rhythms is higher than the one for premature atrial contractions, reflecting an unbalanced classification problem.
These results strengthen the importance of morphology analyses in general, and in particular the relevance of the information contained in the shape parameters obtained from the model fit.
SVM is a standard approach, and can provide directly three-class classification. 
Furthermore, being patient-specific, it just needs some historical data from the patient before being able to provide insight.
We propose also a threshold method as an alternative to the SVM approach. 
This procedure seems to be universal (parameters are stable across the considered group of patients) and provides a visualization of the signal/fit mismatch. 
Furthermore, different objective functions can be chosen in order to maximize specificity or the sensitivity, according to the goal.
We optimized the two with equal weights, reaching 82.21\% $\pm$ 28.55\% sensitivity and 91.83\% $\pm$ 9.24\% specificity, although with a relatively high rate of false positives.
It is interesting to note that there is no significant difference in the detection accuracy of the two types of premature contractions. 

These results on the classification of premature atrial vs ventricular contractions are new and promising.
No beat-by-beat benchmark exists for the detection of premature atrial and ventricular contractions, and for their classification.
On the other hand, a direct comparison for the detection of premature ventricular contractions alone with published results has not been possible, since the data used is often openly available, but the labelling is performed by the authors and labels are not reported.

Our results anticipate the richness of information contained in the waveform characteristics, and the insight we can get by improving our understanding of it.

\section{Future work}\label{sec:FutureWork}
Our results are an indication that shape analysis is a very promising direction for future research.
The studies should continue further to improve the understanding of signal morphology, and to use this for relevant clinical applications.

In the current implementation, segmentation into pulses is performed on the basis of detected local minima. Since the largest first derivative in the rising slope of the pulses has been shown to be the most accurate fiducial point in comparison with ECG~\cite{Nano2017}, this could be considered for improvement. Other methods for pulse delineation exist, such as~\cite{Aboy2005}.
Further research could also address the structure in the residuals, either increasing model complexity, or proposing alternative processing procedures.
Modelling choices can be tuned according to the specific aim without changing the structure of the algorithm, and the fit can be improved as much as desired, minding the trade-off with overfitting.

We have shown that the model can be fitted to two very different datasets.
Since the dataset in hand is highly heterogeneous, we expect that the approaches proposed for the detection of premature contractions will be valid on other datasets, but further studies will have to prove this concept.
We also foresee that detection and classification of premature contractions can be improved, for example by combining existing methods for the detection of premature ventricular contractions with our model-based approach, able to detect also atrial contractions, and to distinguish the type of premature contraction.
Having more data from patients with such arrhythmias will lead to the enhancement of the current method in the unobtrusive detection and classification of premature contractions.

In fact, the dataset in hand was limited in terms of health condition of the patients included. In the future, it would be interesting to apply the method to data from other categories of patients: data from healthy individuals or with other comorbidities, segments showing onset of atrial fibrillation.
Inference from the distribution of the fitted parameters will enable patient-specific analyses, disease detection and classification of individuals into categories.

\bibliographystyle{SageV}
\bibliography{PPGrefs}

\begin{thebibliography}{10}
\providecommand{\url}[1]{\texttt{#1}}
\providecommand{\urlprefix}{URL }
\expandafter\ifx\csname urlstyle\endcsname\relax
  \providecommand{\doi}[1]{DOI:\discretionary{}{}{}#1}\else
  \providecommand{\doi}{DOI:\discretionary{}{}{}\begingroup
  \urlstyle{rm}\Url}\fi
\providecommand{\eprint}[2][]{\url{#2}}

\bibitem{Allen2007}
Allen J.
\newblock Photoplethysmography and its application in clinical physiological
  measurement.
\newblock \emph{Physiological measurement} 2007; 28(3): R1.

\bibitem{Shelley2007}
Shelley KH.
\newblock Photoplethysmography: beyond the calculation of arterial oxygen
  saturation and heart rate.
\newblock \emph{Anesthesia \& Analgesia} 2007; 105(6): S31--S36.

\bibitem{Peper2007}
Peper E, Harvey R, Lin IM et~al.
\newblock Is there more to blood volume pulse than heart rate variability,
  respiratory sinus arrhythmia, and cardiorespiratory synchrony?
\newblock \emph{Biofeedback} 2007; 35(2).

\bibitem{Elgendi2012}
Elgendi M.
\newblock On the analysis of fingertip photoplethysmogram signals.
\newblock \emph{Current cardiology reviews} 2012; 8(1): 14--25.

\bibitem{Charlton2018}
Charlton PH, Birrenkott DA, Bonnici T et~al.
\newblock Breathing rate estimation from the electrocardiogram and
  photoplethysmogram: A review.
\newblock \emph{IEEE reviews in biomedical engineering} 2018; 11: 2--20.

\bibitem{Shariati2005}
Shariati NH and Zahedi E.
\newblock Comparison of selected parametric models for analysis of the
  photoplethysmographic signal.
\newblock In \emph{Computers, Communications, \& Signal Processing with Special
  Track on Biomedical Engineering, 2005. CCSP 2005. 1st International
  Conference on}. IEEE, pp. 169--172.

\bibitem{Rubins2008}
Rubins U.
\newblock Finger and ear photoplethysmogram waveform analysis by fitting with
  gaussians.
\newblock \emph{Medical \& biological engineering \& computing} 2008; 46(12):
  1271--1276.

\bibitem{Couceiro2012}
Couceiro R, Carvalho P, Paiva RP et~al.
\newblock Multi-gaussian fitting for the assessment of left ventricular
  ejection time from the photoplethysmogram.
\newblock In \emph{2012 Annual International Conference of the IEEE Engineering
  in Medicine and Biology Society}. IEEE, pp. 3951--3954.

\bibitem{Wang2013}
Wang L, Xu L, Feng S et~al.
\newblock Multi-gaussian fitting for pulse waveform using weighted least
  squares and multi-criteria decision making method.
\newblock \emph{Computers in biology and medicine} 2013; 43(11): 1661--1672.

\bibitem{Martin2013}
Mart{\'\i}n-Mart{\'\i}nez D, Casaseca-de-la Higuera P,
  Mart{\'\i}n-Fern{\'a}ndez M et~al.
\newblock Stochastic modeling of the {PPG} signal: a synthesis-by-analysis
  approach with applications.
\newblock \emph{IEEE Transactions on Biomedical Engineering} 2013; 60(9):
  2432--2441.

\bibitem{Baruch2014}
Baruch MC, Kalantari K, Gerdt DW et~al.
\newblock Validation of the pulse decomposition analysis algorithm using
  central arterial blood pressure.
\newblock \emph{Biomedical engineering online} 2014; 13(1): 96.

\bibitem{Martin2015}
Mart{\'\i}n-Mart{\'\i}nez D, Domingues A, Casaseca-de-la Higuera P et~al.
\newblock {PPG} beat reconstruction based on shape models and probabilistic
  templates for signals acquired with conventional smartphones.
\newblock In \emph{Iberian Conference on Pattern Recognition and Image
  Analysis}. Springer, pp. 595--602.

\bibitem{Luo2016}
Luo Y, Li W, Rao W et~al.
\newblock A new modeling method of photoplethysmography signal based on
  lognormal basis.
\newblock In \emph{International Conference on Internet and Distributed
  Computing Systems}. Springer, pp. 12--21.

\bibitem{He2017}
He D, Wang L, Fan X et~al.
\newblock A new mathematical model of wrist pulse waveforms characterizes
  patients with cardiovascular disease--a pilot study.
\newblock \emph{Medical Engineering and Physics} 2017; 48: 142--149.

\bibitem{Elgendi2012a}
Elgendi M.
\newblock Standard terminologies for photoplethysmogram signals.
\newblock \emph{Current cardiology reviews} 2012; 8(3): 215--219.

\bibitem{Solovsenko2017}
Solo{\v{s}}enko A, Petr{\.e}nas A, Marozas V et~al.
\newblock Modeling of the photoplethysmogram during atrial fibrillation.
\newblock \emph{Computers in biology and medicine} 2017; 81: 130--138.

\bibitem{Addison2017}
Addison PS.
\newblock Respiratory effort from the photoplethysmogram.
\newblock \emph{Medical Engineering and Physcs} 2017; 41: 9--18.

\bibitem{Mason2013}
Mason GR and Criley JM.
\newblock Method and system for detection of respiratory variation in
  plethysmographic oximetry, 2013.
\newblock US Patent 8,465,434.

\bibitem{Karlen2013}
Karlen W, Raman S, Ansermino JM et~al.
\newblock Multiparameter respiratory rate estimation from the
  photoplethysmogram.
\newblock \emph{IEEE transactions on bio-medical engineering} 2013; 60(7):
  1946--53.

\bibitem{Suzuki2009}
Suzuki T, Kameyama Ki and Tamura T.
\newblock Development of the irregular pulse detection method in daily life
  using wearable photoplethysmographic sensor.
\newblock In \emph{Engineering in Medicine and Biology Society, 2009. EMBC
  2009. Annual International Conference of the IEEE}. IEEE, pp. 6080--6083.

\bibitem{Gil2013}
Gil E, Laguna P, Mart{\'\i}nez JP et~al.
\newblock Heart rate turbulence analysis based on photoplethysmography.
\newblock \emph{IEEE Transactions on Biomedical Engineering} 2013; 60(11):
  3149--3155.

\bibitem{Solovsenko2015}
Solo{\v{s}}enko A, Petr{\.e}nas A and Marozas V.
\newblock Photoplethysmography-based method for automatic detection of
  premature ventricular contractions.
\newblock \emph{IEEE transactions on biomedical circuits and systems} 2015;
  9(5): 662--669.

\bibitem{Yousefi2018}
Yousefi MR, Khezri M, Bagheri R et~al.
\newblock Automatic detection of premature ventricular contraction based on
  photoplethysmography using chaotic features and high order statistics.
\newblock In \emph{2018 IEEE International Symposium on Medical Measurements
  and Applications (MeMeA)}. IEEE, pp. 1--5.

\bibitem{Polania2015}
Polania LF, Mestha LK, Huang DT et~al.
\newblock Method for classifying cardiac arrhythmias using
  photoplethysmography.
\newblock In \emph{Engineering in Medicine and Biology Society (EMBC), 2015
  37th Annual International Conference of the IEEE}. IEEE, pp. 6574--6577.

\bibitem{Papini2018}
Papini GB, Fonseca P, Eerik{\"a}inen LM et~al.
\newblock Sinus or not: a new beat detection algorithm based on a pulse
  morphology quality index to extract normal sinus rhythm beats from wrist-worn
  photoplethysmography recordings.
\newblock \emph{Physiological measurement} 2018; 39(11): 115007.

\bibitem{Karlen2010}
Karlen W, Turner M, Cooke E et~al.
\newblock Capnobase: Signal database and tools to collect, share and annotate
  respiratory signals.
\newblock In \emph{Annual Meeting of the Society for Technology in Anesthesia
  (STA)}. West Palm Beach.

\bibitem{Nano2017}
Nano M, Papini G, Fonseca P et~al.
\newblock Comparing inter beat and inter pulse intervals from ecg and ppg
  signals.
\newblock In \emph{11th Biomedica Summit, May 9-10, 2017, Eindhoven,
  Netherlands}.

\bibitem{Aboy2005}
Aboy M, McNames J, Thong T et~al.
\newblock An automatic beat detection algorithm for pressure signals.
\newblock \emph{IEEE Transactions on Biomedical Engineering} 2005; 52(10):
  1662--1670.

\end{thebibliography}

\section*{Acknowledgments}
This research was performed within the framework of the strategic joint research program on Data Science between TU/e and Philips Electronics Nederland B.V. 
We would like to thank Evan Friedland for providing and making openly available the function {\itshape inflect} to find local extrema.

\end{document}